\documentclass[12pt]{article}
\pdfoutput=1
\usepackage{pstricks}
\usepackage{color}
\usepackage{amssymb,amsmath,bm,bbm}
\usepackage{epsf}
\usepackage{epsfig}
\usepackage{afterpage}
\usepackage{longtable}
\usepackage{cite}
\usepackage{latexsym,mathrsfs,dsfont}
\usepackage{graphics}
\usepackage{url}
\usepackage{paralist}
\usepackage{bbold}

\newcommand{\vcb}{|V_{cb}|}

\newcommand{\be}{\begin{equation}}
\newcommand{\ee}{\end{equation}}
\newcommand{\bi}{\begin{itemize}}
\newcommand{\ei}{\end{itemize}}
\newcommand{\ba}{\begin{array}}
\newcommand{\ea}{\end{array}}
\newcommand{\bea}{\begin{eqnarray}}
\newcommand{\eea}{\end{eqnarray}}

\newcommand{\nn}{\nonumber}

\usepackage{fancyhdr}
\advance \headheight by 3.0truept       

\begin{document}

\begin{flushright} {BARI-TH/18-715}\end{flushright}

\medskip

\begin{center}
{\Large  {Scrutinizing $\bar B \to D^*(D \pi) \ell^- \bar \nu_\ell$ and $\bar B \to D^*(D \gamma) \ell^- \bar \nu_\ell$ \\ \vspace*{0.4cm}in  search of new physics footprints}}
\\[1.0 cm]
{\large P.~Colangelo and F.~De~Fazio 
 \\[0.5 cm]}
{\small
Istituto Nazionale di Fisica Nucleare, Sezione di Bari, Via Orabona 4,
I-70126 Bari, Italy}
\end{center}

\vskip0.5cm

\begin{abstract}
\noindent
Besides being  important  to determine Standard Model parameters such as the CKM matrix elements $|V_{cb}|$ and $|V_{ub}|$, semileptonic $B$ decays  seem also promising  to reveal new physics (NP) phenomena, in particular in connection with the possibility of uncovering lepton flavour universality (LFU) violating  effects. 
In this view, it could be natural to connect the tensions in the inclusive versus exclusive determinations  of $|V_{cb}|$ to the anomalies  in  the ratios $R(D^{(*)})$ of decay rates into $\tau$ vs $\mu, e$.
However, the question has been raised about the role of the  parametrization of the hadronic $B \to D^{(*)}$ form factors in exclusive $B$ decay modes.
We focus on the fully differential angular distributions of $\bar B \to D^* \ell^-{\bar \nu}_\ell$ with $D^* \to D \pi$ or $D^* \to D \gamma$,  the latter mode being  important in the case of $B_s \to D_s^*$ decays. We  show that the angular coefficients in the distributions can be used  to scrutinize the role of the form factor parametrization and to  pin down deviations from SM. 
As an example of a NP scenario, we include  a tensor operator  in the $b \to c$ semileptonic effective Hamiltonian, and discuss how the angular coefficients allow to construct  observables sensitive to this structure,  also defining  ratios useful to test LFU.
\end{abstract}

\thispagestyle{empty}

\newpage
\setcounter{page}{1}


\section{Introduction}
Despite the lack  of new physics (NP) signals in direct searches at colliders, there are hints of physics beyond the Standard Model (SM)  in  a few anomalies in the  flavour sector, with   observables in  tension with   the SM predictions. In particular,  tree-level semileptonic $B$ decays  unexpectedly point to violation of  lepton flavour universality (LFU), 
since the measured
 ratios $R(D^{(*)})=\displaystyle\frac{{\cal B}(B \to D^{(*)} \tau \bar \nu_\tau)}{{\cal B}(B \to D^{(*)} \ell \bar \nu_\ell)}$    reveal an anomalous deviation of semitauonic $B$ modes with respect to  $\mu$ and $e$ ones. 
The HFLAV  averages  \cite{Amhis:2016xyh}  of BaBar \cite{Lees:2012xj,Lees:2013uzd}, Belle \cite{Huschle:2015rga,Sato:2016svk,Hirose:2016wfn} and LHCb \cite{Aaij:2015yra} Collaboration measurements,
\be
 R(D)=0.403 \pm 0.040\pm0.024  \,\,\,\, , \hskip 1 cm R(D^*)=0.310\pm 0.015\pm 0.008 \,\,\,\, ,
 \label{hfag}
 \ee
  compared to the first SM predictions 
$R(D)=0.296 \pm 0.016$, $R(D^*)=0.252\pm 0.003$ \cite{Fajfer:2012vx} and to the updated ones $R(D)=0.300 \pm 0.008$ \cite{Aoki:2016frl} and $R(D^*)=0.260 \pm0.008$ \cite{Bigi:2017jbd}\footnote{Further recent SM calculations of $R(D^*)$ can be found in \cite{Jaiswal:2017rve,Bernlochner:2017jka}.},
show  a deviation  at a global 3.9$\sigma$ level. In the case of $R(D^*)$, the recent Belle result    $R(D^*)=0.270 \pm 0.035({\rm stat})\pm^{0.028}_{0.025}({\rm syst})$ \cite{Hirose:2017dxl} reduces the average  in (\ref{hfag}).
 The LHCb  measurement  $R(J/\psi)=\displaystyle\frac{{\cal B}(B_c^+ \to J/\psi \tau^+  \nu_\tau)}{{\cal B}(B_c^+ \to J/\psi \mu^+ \nu_\mu)}=0.71 \pm 0.17(\rm stat) \pm 0.18 (\rm syst)$  \cite{Aaij:2017tyk} is also  slightly above  the range of existing  predictions  within SM, but for this mode the theoretical error still needs to be  precisely assessed \cite{Tran:2018kuv}. In  SM  the couplings of charged leptons  to  gauge bosons  are  lepton-flavour independent,  and LFU is  only broken by the Yukawa interaction, hence, evidences of LFU violation in $b$-hadron semileptonic modes signal  physics beyond  SM.

There are other puzzles  affecting semileptonic heavy meson  decays, in particular the tension  between  the determinations of  the CKM matrix elements $|V_{cb}|$ and $|V_{ub}|$ from inclusive and exclusive $B$ modes. 
Focusing on $\vcb$,  precise determinations are obtained from the  exclusive $B \to D^*\ell \bar \nu_\ell$ and  $B \to D \ell \bar \nu_\ell$ decays and from the inclusive $B \to X_c \ell \bar \nu_\ell$ mode.
In  $B \to D^*$ the procedure to determine $\vcb$ is based on  the extrapolation of  the dilepton invariant mass  spectrum  up to the maximum value, using as an input hadronic  form factors at this kinematical point  computed by lattice QCD.  The  FLAG  averages  $\vcb_{\rm excl}^{D^*}=(39.27 \pm 0.56_{\rm th} \pm 0.49_{\rm exp}) \times 10^{-3}$ and  $\vcb_{\rm excl}^D=(40.85 \pm 0.98)\times 10^{-3}$ \cite{Aoki:2016frl} have to be compared to  $\vcb_{\rm incl}=(42.46 \pm 0.88)\times 10^{-3}$ obtained in the kinetic scheme  \cite{Amhis:2016xyh}.

Considering the two sets of anomalies,  the  past viewpoint  was to invoke NP in the ratios $R(D^{(*)})$, and to attribute the inclusive/exclusive tensions in $|V_{cb}|$   and $|V_{ub}|$ to  some underlying assumptions, namely  the uncertainty  in the quark-hadron duality ansatz adopted for the inclusive measurement. Recent studies   for $\vcb$ have focused, instead,  on the errors involved in the analysis of 
the   $B \to D^* \ell \bar \nu_\ell$ spectrum at the maximum  dilepton invariant mass.  
The procedure  usually adopted in the experimental determinations was based  on the Caprini-Lellouch-Neubert (CLN) parametrization of the  $B \to D^*$  form factors  \cite{Caprini:1997mu}, which  uses   heavy quark (HQ) symmetry relations with 
the inclusion of radiative and  $1/m_Q$ corrections.  On the other hand,  the deconvoluted fully differential $\bar B^0 \to D^{*+} \ell^- \bar \nu_\ell$   decay distribution   measured by Belle \cite{Abdesselam:2017kjf}
 has been fitted   adopting the Boyd-Grinstein-Lebed (BGL) parametrization of the form factors \cite{Boyd:1994tt,Boyd:1995cf,Boyd:1995sq},   resulting  in a value for $\vcb$  compatible with the inclusive one \cite{Bigi:2017njr,Grinstein:2017nlq,Jaiswal:2017rve}.  Although the outcome refers to a single  data set,  the question has been  raised if the  form factor parametrization provides a solution of the $\vcb$ anomaly. 

The idea that a common  explanation could be found for the $R(D^{(*)})$ and $\vcb$ anomalies,  invoking NP,  has also been put forward  \cite{Colangelo:2016ymy}. 
 As an example, adding a  tensor operator  to the SM effective $b \to c$ semileptonic  Hamiltonian, weighted by a  complex lepton-flavour dependent parameter $\epsilon_T^\ell$,  it has been shown that  a difference of $\epsilon_T^\tau$ with respect to $\epsilon_T^{\mu,e}$ could account for the $R(D^{(*)})$ anomaly,  considering $\epsilon_T^\tau \neq 0$ and $\epsilon_T^\mu=\epsilon_T^e=0$  \cite{Biancofiore:2013ki}. Relaxing the latter assumption,   inclusive and exclusive $B$ semileptonic decays with  $\mu$ and $e$  have been scrutinized showing that, for  $\epsilon_T^\mu \neq 0$ and  $\epsilon_T^e \neq 0$, it is also possible to pin down a region in the parameter space $(Re (\epsilon_T^\ell), \, Im (\epsilon_T^\ell), \, \vcb)$ where the inclusive  ${\cal B}(B^- \to X_c^0 \ell^- \bar \nu_\ell)$ and  exclusive ${\cal B}(B^- \to  D^{(*)0} \ell^- \bar \nu_\ell)$ branching fractions, as well as the spectrum  ${d{\cal B}(B^- \to  D^{*0} \ell^- \bar \nu_\ell)}/{dq^2}$ close to maximum $q^2$ are recovered   \cite{Colangelo:2016ymy}.

Here we reconsider the two issues, the role of the form factor parametrization and the possibility of  non SM effects.  We focus on   $B \to D^* \ell \bar \nu_\ell$  in the case of both light $\mu, e$  and heavy $\tau$  lepton, with the $D^*$ decaying  to $D \pi$ or  $D \gamma$. The latter mode is particularly  relevant for $B_s \to D^*_s$ transitions.
We  express  the fully differential decay rate in  $\bar B \to D^*(D \pi) \ell^- \bar \nu_\ell$ and  $\bar B \to D^*(D \gamma) \ell^- \bar \nu_\ell$ in terms of angular coefficient functions, 
 and show how the analysis of the two modes may shed light on the  form factor parametrization.
We also reconsider the NP model in \cite{Biancofiore:2013ki,Colangelo:2016ymy} and study the modified  angular coefficients, proposing a set of sensitive observables. Other  investigations focusing on the angular distributions have been carried out  in   \cite{Duraisamy:2013kcw,Bhattacharya:2015ida,Bardhan:2016uhr,Alonso:2016gym,Becirevic:2016hea,Ligeti:2016npd,Alok:2016qyh,Ivanov:2017mrj,Alonso:2017ktd,Jung:2018lfu}.
 In particular, differential distributions including subsequent $\tau$ decay have been studied in \cite{Alonso:2016gym,Ligeti:2016npd,Ivanov:2017mrj,Alonso:2017ktd}.
A tensor structure appears, for example, in the effective Hamiltonian of leptoquark models, in  variants of which  it is possible to accommodate a few $B$  anomalies \cite{Bauer:2015knc,Becirevic:2016yqi,Crivellin:2017zlb}. Attempts for  a combined explanation of  the anomalies in  NP  frameworks can  be found in \cite{Buttazzo:2017ixm}, while the role of  ew corrections has been studied in \cite{Feruglio:2017rjo}.

A comment is in order, concerning the differences between the modes with  light and $\tau$ leptons. The final state with $\tau$  necessarily contains at least one neutrino, making  the full reconstruction of $\tau$ kinematics challenging.  BaBar, Belle and the first LHCb studies of semileptonic $B$ decays to $\tau$  exploited $\tau$ decay to light leptons, with a  final state involving two neutrinos. For this reason, the amplitude $\tau \to \nu X$, with $X = \ell \nu$ has been coherently included  in  analyses as, e.g., in  \cite{Ligeti:2016npd}. 
An  interesting path has been followed in the recent LHCb study  which exploits three-prong $\tau$ decay to the visible  $\pi^+ \pi^- \pi^+ $  final state \cite{Aaij:2017deq}. The topology of this mode allows the precise reconstruction of the $\tau$ decay vertex well separated from the $B$ vertex due to the $\tau$ lifetime. This improves the discrimination from the background and, due to the presence of a single neutrino in the final state, would allow the determination of  the complete kinematics of the decay (up to  two two-fold ambiguities). However,  $\tau$ decays  with both three prong $\pi^+ \pi^- \pi^+ $ and four prong $\pi^+ \pi^- \pi^+ \pi^0$ pions   enter in the signal,  and these two modes are treated on the basis of their known reconstruction efficiencies, so that the measurement of  the kinematic variables  in semitauonic $B$ decays  still represents an experimental challenge.

This is the plan of the paper. After having set the stage for the calculation,  in section \ref{sec:NP} we discuss  the fully angular distributions and the properties of the angular coefficient functions. Results in SM are presented in section \ref{sec:FF},  where the effects  of the form factor parametrization, in particular CLN vs BGL, are investigated.
In section \ref{sec:NPcase} we compare the angular coefficients  in SM and  in the NP  model with the  tensor operator. A set of observables is considered in section \ref{sec:obs}, and  ratios useful to  test  LFU are scrutinized.  Our  conclusions are presented in the last section.

\section{Setting the stage}
\begin{figure}[t]
\begin{center}
\includegraphics[width = 0.6\textwidth]{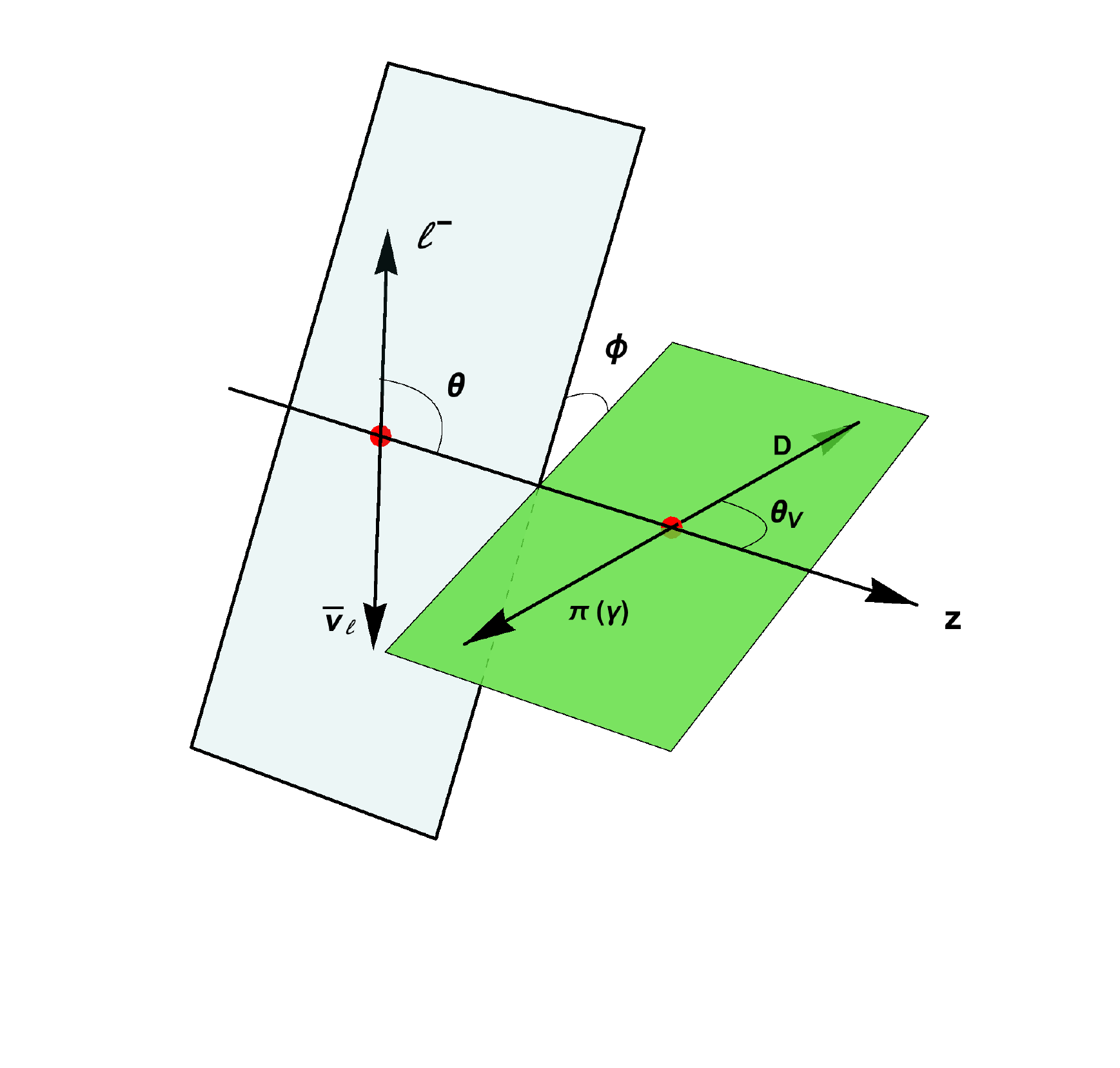} \vspace*{-1.5cm}
    \caption{ Kinematics of  ${\bar B} \to D^* (\to D F)  \ell^-  {\bar \nu}_\ell$ }\label{fig:piani}
\end{center}
\end{figure}

We consider  ${\bar B}(p_B) \to D^*(p_{D^*},\,\epsilon) \ell^- (k_1) {\bar \nu}_\ell(k_2)$, where ${\bar B} \to D^*$ denotes either ${\bar B}^0 \to D^{*+}$ or $B^- \to D^{*0}$, followed by the decay $D^*(p_{D^*},\,\epsilon) \to D(p_D) F(p_F)$ with $F=\pi$ or $\gamma$.
For the  kinematics we adopt the convention for angles and momenta   as in figure \ref{fig:piani}, with   lepton-pair momentum  $q=k_1+k_2=p_B-p_{D^*}$.
In the derivation, we extend to NP  the procedure  in \cite{Korner:1989qb,Gilman:1989uy} for  $F=\pi$, considering also the case $F=\gamma$.

The amplitude  of the process 
\be
{\cal A}_{TOT}({\bar B} \to D^* (\to D F) \ell^-  {\bar \nu}_\ell)={\cal A}({\bar B} \to D^* \ell^-  {\bar \nu}_\ell)\,\frac{i}{p_{D^*}^2-m_{D^*}^2+i m_{D^*}\Gamma(D^*)} \, {\cal A}(D^* \to D F) \label{atot}
\ee
involves three  factors.
To describe   ${\bar B} \to D^* \ell^-  {\bar \nu}_\ell$ we focus on the effective Hamiltonian 
 \be
H_{eff}= {G_F \over \sqrt{2}}V_{cb} \left[ {\bar c} \gamma_\mu (1-\gamma_5) b \, {\bar \ell} \gamma^\mu (1-\gamma_5) {\nu}_\ell + \epsilon_T^\ell \, {\bar c} \sigma_{\mu \nu} (1-\gamma_5) b \, {\bar \ell} \sigma^{\mu \nu} (1-\gamma_5) { \nu}_\ell \right] + h.c.\,\,\, , \label{heff}
\end{equation}
consisting in the Standard Model  term and in  a new physics term with a tensor operator weighted by a lepton-flavour dependent complex parameter $\epsilon_T^\ell$.
\footnote{We only consider the tensor operator, although other operators could be produced by  ew renormalization-group evolution   \cite{Gonzalez-Alonso:2017iyc}.}
This allows to write
\be
{\cal A}({\bar B} \to D^* \ell^-  {\bar \nu}_\ell)={G_F \over \sqrt{2}}V_{cb} \left[H^{SM}_\mu L^{SM \, \mu}+\epsilon_T^\ell H^{NP}_{\mu \nu} L^{NP \, \mu \nu}\right] 
\ee
in terms of the quark current matrix elements
\bea
H^{SM}_\mu(m)&=& \langle D^*(p_{D^*},\epsilon(m))|{\bar c} \gamma_\mu(1-\gamma_5) b| {\bar B}(p_B) \rangle =\epsilon^{*\alpha}(m) T^{SM}_{\mu \alpha}
\label{HSM} \\
H^{NP}_{\mu \nu} (m)&=& \langle D^*(p_{D^*},\epsilon(m))|{\bar c} \sigma_{\mu \nu}(1-\gamma_5) b| {\bar B}(p_B) \rangle =
\epsilon^{*\alpha}(m) T^{NP}_{\mu \nu \alpha} \label{HNP} 
\eea
and  of the lepton currents
\bea
L^{SM \, \mu}&=&  {\bar \ell} \gamma^\mu (1-\gamma_5) { \nu}_\ell 
\label{leptSM} \\
 L^{NP \, \mu \nu} &=&{\bar \ell} \sigma^{\mu \nu} (1-\gamma_5) { \nu}_\ell  .  \label{leptNP}
 \eea
In (\ref{HSM}) and (\ref{HNP})  the   index $m$ of the $D^*$ polarization vector $\epsilon$ runs over $m=\pm,0$.
In the lepton-pair rest-frame (LRF), with the $D^*$ three-momentum  along the positive $z$-axis,  one has:
\bea
p_B&=&(E_B,0,0,|{\vec p}_{D^*}|) \,\,\, , \hskip 1cm
p_{D^*}= (E_{D^*},0,0,|{\vec p}_{D^*}|) \,\,\, , \hskip 1cm
q=(\sqrt{q^2},0,0,0) \,\,\, ,  \nn \\
\epsilon_\pm &=&\frac{1}{\sqrt{2}}(0,1, \mp i,0) \,\,\, , 
\hskip 1.4 cm
\epsilon_0 = \frac{1}{m_{D^*}}(|{\vec p}_{D^*}|,0,0, E_{D^*}) \,\,\, ,
\eea
with  $|{\vec p}_{D^*}|=\displaystyle{\frac{\lambda^{1/2}(m_B^2,m_{D^*}^2,q^2)}{2 \sqrt{q^2}}}$ and $E_{D^*}=\displaystyle{\frac{m_B^2-m_{D^*}^2-q^2}{2\sqrt{q^2}}}$,   $\lambda$ being the triangular function.
The orientation of the lepton momenta is fixed by  the angles $\theta$ and $ \phi$ as in figure \ref{fig:piani}, so that
\bea
k_1&=&(k_1^0,|{\vec k}_1| \sin\theta \cos \phi, |{\vec k}_1| \sin\theta \sin \phi, |{\vec k}_1| \cos\theta )
\nn \\
k_2&=&(k_2^0,-|{\vec k}_1| \sin\theta \cos \phi,- |{\vec k}_1| \sin\theta \sin \phi, -|{\vec k}_1| \cos\theta ) \,\,.
\eea
In terms of the $D^*$ polarization indices one can write
\be
|{\cal A}({\bar B} \to D^* \ell^-  {\bar \nu}_\ell) (m,n)|^2={G_F^2 \over 2}|V_{cb}|^2 \left[{\cal H}^{SM}(m,n)+{\cal H}^{NP}(m,n) +{\cal H}^{INT}(m,n) \right] \,\,\, , 
\ee
where
\bea
{\cal H}^{SM}(m,n) &=& H^{SM}_\mu(m) (H^{SM})^\dagger_{\mu^\prime}(n) {\cal L}^{SM \, \mu \mu^\prime} \,\,,
\label{asmsq} \\
{\cal H}^{NP}(m,n) &=&|\epsilon_T|^2 \left[ H^{NP}_{\mu \nu}(m)(H^{NP})^\dagger_{\mu^\prime \nu^\prime}(n) {\cal L}^{NP \, \mu \nu \mu^\prime \nu^\prime}\right] \,\,,
\label{anpsq} \\
{\cal H}^{INT}(m,n)  &=&  \epsilon_T H^{SM}_\mu(m) (H^{NP})^\dagger_{\mu^\prime \nu^\prime}(n){\cal L}_1^{INT \, \mu  \mu^\prime \nu^\prime} 
+\epsilon_T^* H^{NP}_{\mu \nu}(m)(H^{SM})^\dagger_{\mu^\prime}(n){\cal L}_2^{INT \, \mu \nu \mu^\prime} ,\,\,\,\,\,  
\label{aint} \eea
in terms of the quantities in (\ref{HSM}),(\ref{HNP}) and
\bea
{\cal L}^{SM \, \mu \mu^\prime}  &=& L^{SM \, \mu} (L^{SM \, \mu^\prime})^\dagger \,\,,
\nn \\
{\cal L}^{NP \, \mu \nu \mu^\prime \nu^\prime} &=& L^{NP \, \mu \nu}(L^{NP \,  \mu^\prime \nu^\prime})^\dagger \nn \\
{\cal L}_1^{INT \, \mu  \mu^\prime \nu^\prime} &=&  L^{SM \, \mu} (L^{NP \,  \mu^\prime \nu^\prime})^\dagger  \\
{\cal L}_2^{INT \, \mu \nu \mu^\prime} &=&  L^{NP \, \mu \nu}  (L^{SM \, \mu^\prime})^\dagger \nn
\,\,. \eea

As for  the  $D^*$ propagator,   the narrow-width approximation can be used for the state produced nearly on-shell \cite{Uhlemann:2008pm},
\be
\frac{1}{(p_{D^*}^2-m_{D^*}^2)+ m_{D^*}^2\Gamma(D^*)^2}=\frac{\pi}{m_{D^*}\Gamma(D^*)}\, \delta(p_{D^*}^2-m_{D^*}^2)\,.
\label{nwa}
\ee
On the other hand, the $D^* \to D F$  amplitude can be written as
\be
{\cal A}(D^* \to D F)=g_{D^* D F} \, \left( \epsilon \cdot Q \right) \label{D*DFamp}
\ee
where $Q=p_D$ for $F=\pi$,  and  $Q_\beta=i \, \epsilon_{ \alpha \beta \sigma \tau}\eta^{* \alpha} p_{D^*}^\sigma p_D^\tau$  for $F=\gamma$, with  $\eta$  the photon polarization vector. 
One can get rid of the coupling $g_{D^{*}DF}$ considering
\be
\Gamma(D^* \to DF)=g_{D^{*}DF}^2 \frac{|{\vec p}_D|}{24 \pi m_{D^*}^4}\left[(p_{D^*} \cdot Q)^2-Q^2 m_{D^*}^2 \right] \label{gammaD*DF}
\,,
\ee
with  $|{\vec p}_D|=\displaystyle{\frac{\lambda^{1/2}(m_{D^*}^2,m_D^2,m_F^2)}{2m_{D^*}}}$  the $D$ three-momentum in the $D^*$ rest frame ($D^*$RF).
In particular,    one has  $\left[(p_{D^*} \cdot Q)^2-Q^2 m_{D^*}^2 \right]=m_{D^*}^2 |{\vec p}_D|^2$ for $F=\pi$ and
$2 m_{D^*}^4 |{\vec p}_D|^2$ for $F=\gamma$.
Specifying the $D^*$ polarization indices,  one can write
\be
|{\cal A}(D^* \to D F)|^2(m,n)=\Gamma(D^* \to DF)\frac{24 \pi m_{D^*}^2}{|{\vec p}_D|^3}  \,F_F(m,n),
\ee
with
\be
F_F(m,n)=c_F \left[\epsilon(m) \cdot Q \right]\left[\epsilon(n) \cdot Q\right]^\dagger 
\label{Ftensor}
\ee
and the constant   $c_\pi=1$  for $F=\pi$, and     $c_\gamma=1/(2m_{D^*}^2)$  for $F=\gamma$.
The ($3\times 3$)  $F_F(m,n)$ matrices in (\ref{Ftensor}) involve the angle $\theta_V$:
\begin{equation}
F_\pi=\frac{|{\vec p}_D|^2}{2} \left(\begin{array}{ccc}
 \sin^2 \theta_V  &  \sin^2 \theta_V &\frac{1}{\sqrt{2}} \sin 2\theta_V \\
 \sin^2 \theta_V  &  \sin^2 \theta_V &\frac{1}{\sqrt{2}} \sin 2\theta_V \\
\frac{1}{\sqrt{2}} \sin 2\theta_V &\frac{1}{\sqrt{2}} \sin 2\theta_V & 2 \cos^2 \theta_V
\end{array}\right) \label{Fpimat}
\end{equation}
\begin{equation}
F_\gamma=\frac{|{\vec p}_D|^2}{4 } \left(\begin{array}{ccc}
\frac{3+\cos 2 \theta_V}{2} & -\sin^2 \theta_V & -\frac{1}{\sqrt{2}} \sin 2 \theta_V \\
-\sin^2 \theta_V &\frac{3+\cos 2 \theta_V}{2} &  -\frac{1}{\sqrt{2}} \sin 2 \theta_V \\
 -\frac{1}{\sqrt{2}} \sin 2 \theta_V &  -\frac{1}{\sqrt{2}} \sin 2 \theta_V & 2 \sin^2 \theta_V 
\end{array}\right) \,\,\, . \label{Fgammamat}
\end{equation}
Collecting the various terms in Eq.~(\ref{atot}) we obtain
\bea
&&|{\cal A}_{TOT}({\bar B} \to D^*(\to D F) \ell^-  {\bar \nu}_\ell)|^2=G_F^2|V_{cb}|^2 \frac{12 \pi^2 m_{D^*}}{|{\vec p}_D|^3} {\cal B}(D^* \to DF) \delta(p_{D^*}^2-m_{D^*}^2) \hspace*{2.5cm}\nn \\
&&\hspace*{3.5cm}
\times \left\{ Tr \left[({\cal H}^{SM})^T \cdot F_F\right]+Tr \left[({\cal H}^{NP})^T \cdot F_F\right]+Tr \left[({\cal H}^{INT})^T \cdot F_F\right] \right\},    \label{atotfin}
\eea
where the trace is carried out over the indices $(m,n)$, ordered as $(1,2,3)=(+,-,0)$, and $T$ meaning the transpose.
The expression of the fully differential decay distribution can be worked out considering the four-body phase-space recalled  in  appendix \ref{kinematics}: 
\bea
&&\frac{d^4  \Gamma (\bar B \to D^*( \to D F) \ell^- \bar \nu_\ell)}{dq^2 \,d\cos \theta \,d\phi \, d \cos \theta_V}=
\displaystyle{\frac{3G_F^2 |V_{cb}|^2 {\cal B}(D^* \to D F)}{128(2\pi)^4m_B^2}} \frac{|{\vec p}_{D^*}|_{BRF}}{|{\vec p_D}|_{D^*RF}^2} \left(1-  \frac{ m_\ell^2}{q^2}\right)  \hspace*{2.5cm} \nn \\
&&\hspace*{3.5cm}\times\left\{ Tr \left[({\cal H}^{SM})^T \cdot F_F\right]+Tr \left[({\cal H}^{NP})^T \cdot F_F\right]+Tr \left[({\cal H}^{INT})^T \cdot F_F\right] \right\}.  \label{ris}
\eea

The hadronic matrix elements  (\ref{HSM}),(\ref{HNP}) can be parametrized in terms of  form factors. We use the  definition
\bea
\langle D^*(p_{D^*},\epsilon)|{\bar c} \gamma_\mu(1-\gamma_5) b| {\bar B}(p_B) \rangle &=&
- {2 V(q^2) \over m_B+m_{D^*}} i \epsilon_{\mu \nu \alpha \beta} \epsilon^{*\nu}  p_B^\alpha p_{D^*}^\beta \nn \\
&-&\Big\{ (m_B+m_{D^*}) \left[ \epsilon^*_\mu -{(\epsilon^* \cdot q) \over q^2} q_\mu \right] A_1(q^2) \nn\\
&-& {(\epsilon^* \cdot q) \over  m_B+m_{D^*}} \left[ (p_B+p_{D^*})_\mu -{m_B^2-m_{D^*}^2 \over q^2} q_\mu \right] A_2(q^2) \nn \\
&+& (\epsilon^* \cdot q){2 m_{D^*} \over q^2} q_\mu A_0(q^2) \Big\}  \label{FF-D*-mio}
\eea
(with the condition  $\displaystyle A_0(0)= \frac{m_B + m_{D^*}}{2 m_{D^*}} A_1(0) -  \frac{m_B - m_{D^*}}{2 m_{D^*}}  A_2(0)$) and
\bea
\langle D^*(p_{D^*},\epsilon)|{\bar c} \sigma_{\mu \nu}(1-\gamma_5) b| {\bar B}(p_B) \rangle &=&
T_0(q^2) {\epsilon^* \cdot q \over (m_B+ m_{D^*})^2} \epsilon_{\mu \nu \alpha \beta} p_B^\alpha p_{D^*}^\beta+
T_1(q^2) \epsilon_{\mu \nu \alpha \beta} p^\alpha_B \epsilon^{*\beta} \nn \\ 
&+&T_2(q^2) \epsilon_{\mu \nu \alpha \beta} p_{D^*}^\alpha \epsilon^{*\beta}\nn \\
&+&i \, \Big[ T_3(q^2) (\epsilon^*_\mu p_{B \nu} -\epsilon^*_\nu p_{B \mu})+T_4(q^2) (\epsilon^*_\mu p_{D^* \nu} -\epsilon^*_\nu p_{D^*\mu}) \nn \\
&+&T_5(q^2) {\epsilon^* \cdot q \over (m_B+ m_{D^*})^2}(p_{B \mu} p_{D^* \nu} -p_{B \nu} p_{D^* \mu})\Big]  \,\,.  \label{mat-tensor-Dstar}
\eea
We also define
$\tilde T_0=T_0-T_5$, $\tilde T_1=T_1+T_3$ and $\tilde T_2=T_2+T_4$. 
In appendix \ref{appA}  we describe  other  matrix element parametrizations.  
In  SM one can relate the helicity amplitudes for the  $D^*$ polarization states  to  the polarizations of the virtual $W(q,{\bar \epsilon})$. 
 In the LRF one writes 
\be
{\bar \epsilon}_\pm =\frac{1}{\sqrt{2}}(0,1, \pm i,0) \,\,\, ,
\hskip 0.8cm 
{\bar \epsilon}_0 =(0,0,0, 1) \,\,\, , 
\hskip 0.8cm 
{\bar \epsilon}_t =(1,0,0, 0)\,\, .
\ee
This allows to define the amplitudes
 \bea
 H_m&=&{\bar \epsilon}_m^{*\mu}\epsilon_{m}^{*\alpha}T_{\mu \alpha}\, \, \hskip 1cm \,\,(m=0,\pm) \nn \\
 H_t&=&{\bar \epsilon}_t^{*\mu}\epsilon_{0}^{*\alpha}T_{\mu \alpha}\, \,\, \hskip 1cm\,\, (m=t) \,\,\, ,
 \eea
which can be expressed  in terms of the form factors  in (\ref{FF-D*-mio}):
\bea
H_0 &=&\frac{(m_B+m_{D^*})^2(m_B^2-m_{D^*}^2-q^2) A_1(q^2)-\lambda(m_B^2,\,m_{D^*}^2,\,q^2) A_2(q^2)}{2m_{D^*}(m_B+m_{D^*}) \sqrt{q^2}} \nn \\
H_\pm&=& \frac{(m_B+m_{D^*})^2 A_1(q^2)\mp\sqrt{\lambda(m_B^2,\,m_{D^*}^2,\,q^2)}V(q^2)}{m_B+m_{D^*}}  \label{Hamp}\\
H_t&=& -\frac{\sqrt{\lambda(m_B^2,\,m_{D^*}^2,\,q^2)}}{\sqrt{q^2}} \,A_0(q^2) \,\,\, . \nn
\eea
All the entries in  ${\cal H}^{SM}(m,n)$ can be written in terms of $H_\pm$, $H_0$ and $H_t$.

\section{Angular  decomposition of the  fully differential decay distribution}\label{sec:NP}
The fully differential decay distribution for the chain process $\bar B \to D^* (\to DF) \ell^- {\bar \nu_\ell}$,  with $F=\pi$ and $F=\gamma$, can be worked out in terms of the angles in fig.~\ref{fig:piani}.
For $F=\pi$ it can be expressed as\footnote{In principle, other two structures $I_8 \sin 2\theta_V \sin 2 \theta \sin \phi +I_9 \sin^2 \theta_V \sin^2 \theta \sin 2 \phi$ could be present in these decompositions. We do not include them, since they are absent in SM and in the NP model considered here.}
\bea
\frac{d^4 \Gamma (\bar B \to D^*( \to D \pi) \ell^- \bar \nu_\ell)}{dq^2 \,d\cos \theta \,d\phi \,d \cos \theta_V} 
&=&{\cal N_\pi}|{\vec p}_{D^*}| \left(1-  \frac{ m_\ell^2}{q^2}\right)^2 \Big\{I_{1s}^\pi \,\sin^2 \theta_V+I_{1c}^\pi \,\cos^2\theta_V \nn \\
&+&\left(I_{2s}^\pi \,\sin^2 \theta_V+I_{2c}^\pi \,\cos^2 \theta_V\right) \cos 2\theta \nn  \\ 
&+&I_3^\pi \,\sin^2 \theta_V \sin^2 \theta  \cos 2 \phi +I_4^\pi \, \sin 2\theta_V \sin 2\theta \cos  \phi \label{angularpi} \\  
&+&I_5^\pi \, \sin 2 \theta_V \sin \theta \cos  \phi +\left(I_{6s}^\pi \,\sin^2 \theta_V+I_{6c}^\pi \,\cos^2\theta_V\right)\cos \theta \nn \\
&+& I_7^\pi \sin 2 \theta_V \sin \theta \sin  \phi
  \Big\}\,\,, \nn
\eea
with ${\cal N}_F=\displaystyle{\frac{3G_F^2 |V_{cb}|^2 {\cal B}(D^* \to D F)}{128(2\pi)^4m_B^2}}$.
For $F=\gamma$ we adopt the decomposition
\bea
\frac{d^4 \Gamma (\bar B \to D^*( \to D \gamma) \ell^- \bar \nu_\ell)}{dq^2 \,d\cos \theta \,d\phi \,d\cos \theta_V}  
&=&{\cal N_\gamma}|{\vec p}_{D^*}|\left(1- \frac{ m_\ell^2}{q^2}\right)^2 \Big\{I_{1s}^\gamma \,\sin^2 \theta_V+I_{1c}^\gamma \,(3+\cos 2\theta_V )\nn \\ 
&+& (I_{2s}^\gamma \,\sin^2 \theta_V+I_{2c}^\gamma \,(3+\cos 2\theta_V )) \cos 2\theta \nn \\   
&+&I_3^\gamma \,\sin^2 \theta_V \sin^2 \theta  \cos  2 \phi +I_4^\gamma \sin 2 \theta_V \sin 2\theta \cos  \phi  \label{angulargamma} \\ 
&+&I_5^\gamma \, \sin 2 \theta_V  \sin \theta \cos  \phi +(I_{6s}^\gamma \, \sin^2 \theta_V+I_{6c}^\gamma \,(3+\cos 2\theta_V ))\cos \theta
 \nn \\
&+& I_7^\gamma \sin 2 \theta_V \sin \theta \sin  \phi \Big\}  \,\,\, . \nn
\eea
In  the Standard Model the  coefficients of the angular terms are related to the helicity amplitudes (\ref{Hamp}):
\bea
&&
I_{1s}^\pi=\frac{1}{2}(H_+^2+H_-^2)(m_\ell^2+3q^2) \,\,\, ,  \hskip 1 cm I_{1c}^\pi=2(2m_\ell^2 H_t^2+H_0^2(m_\ell^2+q^2)) \,\,\, , 
\nn \\
&&
I_{2s}^\pi=\frac{1}{2}(H_+^2+H_-^2)(q^2-m_\ell^2) \,\,\, ,  \hskip 1.1 cm I_{2c}^\pi=2 H_0^2(m_\ell^2-q^2) \,\,\, , 
\nn \\
&&
I_{3}^\pi=2H_+ H_-(m_\ell^2-q^2)  \,\,\, , \hskip 2.1 cm I_{4}^\pi= H_0(H_+ +H_-)(m_\ell^2-q^2) \,\,\, , 
\nn \\
&&
I_5^\pi=-2 (H_+ + H_-) H_t m_\ell^2 - 2 H_0 (H_+ - H_-) q^2 \,\,\, , 
\label{angpi} \\
&&I_{6s}^\pi= 2(H_+^2-H_-^2)q^2  \,\,\, , \hskip 2.5 cm I_{6c}^\pi=-8H_0 H_t m_\ell^2 \,\,\, , 
\nn \\
&&I_7^\pi=0 \,\,\, , \nn
\eea
and
\bea
&&
I_{1s}^\gamma=2 m_\ell^2 H_t^2+H_0^2 (m_\ell^2+q^2) \,\,\, , \hskip 1 cm I_{1c}^\gamma=\frac{1}{8}(H_+^2+H_-^2)(m_\ell^2+3q^2) \,\,\, , \nn \\
&&
I_{2s}^\gamma= H_0^2 (m_\ell^2-q^2) \,\,\, , \hskip 3 cm I_{2c}^\gamma=\frac{1}{8}(H_+^2+H_-^2)(q^2-m_\ell^2) \,\,\, , \nn \\
&&
I_{3}^\gamma=-H_+ H_- (m_\ell^2-q^2) \,\,\, , \hskip 2.2 cm I_{4}^\gamma= -\frac{1}{2} H_0 (H_+ +H_-) (m_\ell^2-q^2) \,\,\, , \nn \\
&&
I_5^\gamma= (H_+ + H_-) H_t m_\ell^2 +  H_0 (H_+ - H_-) q^2 \,\,\, , \label{anggamma} \\
&&I_{6s}^\gamma=-4 H_0 H_t m_\ell^2  \,\,\, , \hskip 3.5 cm I_{6c}^\gamma=\frac{1}{2}(H_+^2-H_-^2) q^2 \,\,\, , 
 \nn \\
&&I_7^\gamma=0 \,\,\, . \nn
\eea
Hence, the coefficients   in  $D \pi$ and $D \gamma$ angular distributions obey the relations, for all $q^2$,
\be
\frac{I_{1s}^\pi}{4 I_{1c}^\gamma}=\frac{I_{1c}^\pi}{2I_{1s}^\gamma}=\frac{I_{2s}^\pi}{4 I_{2c}^\gamma}=\frac{I_{2c}^\pi}{2I_{2s}^\gamma}=\frac{I_{6s}^\pi}{4 I_{6c}^\gamma}=\frac{I_{6c}^\pi}{2I_{6s}^\gamma}=-\frac{I_{3}^\pi}{2I_{3}^\gamma}=-\frac{I_{4}^\pi}{2I_{4}^\gamma}=-\frac{I_{5}^\pi}{2I_{5}^\gamma} =1 \,\, . \label{relationspigamma}
\ee
Integrated  distributions are written in terms of the angular coefficients. In particular, the $q^2$ distributions read: 
\bea
\frac{d\Gamma}{dq^2}\Big|_{F=\pi}&=&{\cal N}_\pi |{\vec p}_{D^*}|\left(1-  \frac{ m_\ell^2}{q^2}\right)^2 \frac{8}{9} \pi \left( 6 I_{1s}^\pi+3 I_{1c}^\pi-2I_{2s}^\pi-I_{2c}^\pi \right) \,\,\, ,
\label{dgdq2pi} \\
\frac{d\Gamma}{dq^2}\Big|_{F=\gamma}&=&{\cal N}_\gamma |{\vec p}_{D^*}|\left(1-  \frac{ m_\ell^2}{q^2}\right)^2 \frac{16}{9} \pi \left( 3 I_{1s}^\gamma+12 I_{1c}^\gamma-I_{2s}^\gamma-4I_{2c}^\gamma \right) \,\,\, .
\label{dgdq2gamma}
\eea
The angular coefficients encode information on the form factors, and vice-versa.
Their fit from the experimental fully differential decay distribution allows to reconstruct the form factors,  with  a  possible  comparison of measurements to theory determinations. 
Considering  the $D \pi$ mode one has
\bea
A_1(q^2)&=&\frac{1}{4(m_B+m_{D^*})}\left\{ \sqrt{\frac{4 I_{1s}^\pi}{m_\ell^2+3 q^2}-\frac{ I_{6s}^\pi}{ q^2}}+\sqrt{\frac{4 I_{1s}^\pi}{m_\ell^2+3 q^2}+\frac{ I_{6s}^\pi}{ q^2}} \right\} ,
\nn \\
A_2(q^2)&=&\frac{(m_B+m_{D^*})}{4 \lambda(m_B^2,\,m_{D^*}^2,\,q^2)}\Bigg\{(m_B^2-m_{D^*}^2-q^2)\Bigg[ \sqrt{\frac{4 I_{1s}^\pi}{m_\ell^2+3 q^2}-\frac{ I_{6s}^\pi}{ q^2}}+\sqrt{\frac{4 I_{1s}^\pi}{m_\ell^2+3 q^2}+\frac{ I_{6s}^\pi}{ q^2}} \, \Bigg] 
\nn \\
&-& 4\sqrt{2}m_{D^*}\sqrt{q^2}\sqrt{-\frac{I_{2c}^\pi}{q^2-m_\ell^2}}\, \Bigg\} ,
\nn \\
V(q^2)&=&\frac{(m_B+m_{D^*})}{4 \lambda^{1/2}(m_B^2,\,m_{D^*}^2,\,q^2)}\left\{ \sqrt{\frac{4 I_{1s}^\pi}{m_\ell^2+3 q^2}-\frac{ I_{6s}^\pi}{ q^2}} -\sqrt{\frac{4 I_{1s}^\pi}{m_\ell^2+3 q^2}+\frac{ I_{6s}^\pi}{ q^2}} \right\} ,
\label{Vsolpi}
\\
A_0(q^2)&=&\frac{1}{2}\frac{\sqrt{q^2}}{ \lambda^{1/2}(m_B^2,\,m_{D^*}^2,\,q^2)}\sqrt{\frac{(q^2-m_\ell^2)\,I_{1c}^\pi+(q^2+m_\ell^2)\,I_{2c}^\pi}{m_\ell^2 (q^2-m_\ell^2)}} .
\nn \eea
Analogously, from the $D \gamma$ mode one has
\bea
A_1(q^2)&=&\frac{1}{2(m_B+m_{D^*})}\left\{ \sqrt{\frac{4 I_{1c}^\gamma}{m_\ell^2+3 q^2}-\frac{ I_{6c}^\gamma}{ q^2}}+\sqrt{\frac{4 I_{1c}^\gamma}{m_\ell^2+3 q^2}+\frac{ I_{6c}^\gamma}{ q^2}} \, \right\} ,
\nn \\
A_2(q^2)&=&\frac{(m_B+m_{D^*})}{2 \lambda(m_B^2,\,m_{D^*}^2,\,q^2)}\Bigg\{(m_B^2-m_{D^*}^2-q^2)\left[ \sqrt{\frac{4 I_{1c}^\gamma}{m_\ell^2+3 q^2}-\frac{ I_{6c}^\gamma}{ q^2}}+\sqrt{\frac{4 I_{1c}^\gamma}{m_\ell^2+3 q^2}+\frac{ I_{6c}^\gamma}{ q^2}} \right]
\nn \\
&-& 4m_{D^*}\sqrt{q^2}\sqrt{-\frac{I_{2s}^\gamma}{q^2-m_\ell^2}}\Bigg\} ,
\nn \\
V(q^2)&=&\frac{(m_B+m_{D^*})}{2 \lambda^{1/2}(m_B^2,\,m_{D^*}^2,\,q^2)}\left\{ \sqrt{\frac{4 I_{1c}^\gamma}{m_\ell^2+3 q^2}-\frac{ I_{6c}^\gamma}{ q^2}} -\sqrt{\frac{4 I_{1c}^\gamma}{m_\ell^2+3 q^2}+\frac{ I_{6c}^\gamma}{ q^2}} \,\, \right\} ,
\label{Vsolgamma}
\\
A_0(q^2)&=&\frac{1}{\sqrt{2}}\frac{\sqrt{q^2}}{ \lambda^{1/2}(m_B^2,\,m_{D^*}^2,\,q^2)}\sqrt{\frac{(q^2-m_\ell^2)\,I_{1s}^\gamma+(q^2+m_\ell^2)\,I_{2s}^\gamma}{m_\ell^2 (q^2-m_\ell^2)}} \,\,\, .
\nn \eea
Such relations require precise signs for the angular coefficient functions and for a few of their combinations.

Considering  the tensor operator  in the effective Hamiltonian (\ref{heff}),
the fully differential decay distribution can still be written as in Eqs.~(\ref{angularpi}),(\ref{angulargamma}), with the coefficients $I_i$ replaced by $I_i+|\epsilon_T|^2I_i^{NP}+2 Re(\epsilon_T) I_i^{INT}$ for $i=1,\dots 6$, and by $I_i+|\epsilon_T|^2I_i^{NP}+2 Im (\epsilon_T) I_i^{INT}$ for $i=7$.
With the  definitions
\bea
H_+^{NP} &=& \frac{1}{2\sqrt{q^2}}\left\{\left[m_B^2-m_{D^*}^2+\lambda^{1/2} (m_B^2,m_{D^*}^2,q^2) \right]({\tilde T}_1+{\tilde T}_2)+q^2({\tilde T}_1-{\tilde T}_2)\right\} \nn \\
H_-^{NP} &=& \frac{1}{2\sqrt{q^2}}\left\{\left[m_B^2-m_{D^*}^2-\lambda^{1/2} (m_B^2,m_{D^*}^2,q^2) \right]({\tilde T}_1+{\tilde T}_2)+q^2({\tilde T}_1-{\tilde T}_2)\right\}
 \\
H_L^{NP}&=&2\left\{
\frac{\lambda (m_B^2,m_{D^*}^2,q^2)}{m_{D^*}(m_B+m_{D^*})^2} \, {\tilde T}_0+2\frac{m_B^2+m_{D^*}^2-q^2}{m_{D^*}}\,{\tilde T}_1
+4m_{D^*}\,{\tilde T}_2 \right\}\,\, \nn
\eea
one has:
\bea
&&
I_{1s}^{NP,\pi}=2[(H_+^{NP})^2+(H_-^{NP})^2](3m_\ell^2+q^2)  , \hskip 1 cm
 I_{1c}^{NP,\pi}=\frac{1}{8}(q^2+m_\ell^2)(H_L^{NP})^2  ,
\nn \\
&&
I_{2s}^{NP,\pi}=2[(H_+^{NP})^2+(H_-^{NP})^2](m_\ell^2-q^2)   ,
 \hskip 1.1 cm 
 I_{2c}^{NP,\pi}=\frac{1}{8}(q^2-m_\ell^2)(H_L^{NP})^2  ,
\nn \\
&&
I_{3}^{NP,\pi}=8H_+^{NP}H_-^{NP}(q^2-m_\ell^2)  ,
 \hskip 2.5 cm 
  I_{4}^{NP,\pi}= \frac{1}{2}(q^2-m_\ell^2)H_L^{NP}[H_+^{NP}+H_-^{NP}]  , \,\,\,\,\,
\nn \\
&&
I_5^{NP,\pi}= -m_\ell^2 H_L^{NP}[H_+^{NP}-H_-^{NP}]  ,
 \\
&&I_{6s}^{NP,\pi}=8m_\ell^2 [(H_+^{NP})^2-(H_-^{NP})^2]  ,
 \hskip 2.5 cm I_{6c}^{NP,\pi}=0  ,
 \nn \\
&&I_7^{NP,\pi}=0  ,
\nn \eea
and
\bea
&&
I_{1s}^{NP,\gamma}=\frac{1}{16}(H_L^{NP})^2(q^2+m_\ell^2)  ,
 \hskip 2 cm
 I_{1c}^{NP,\gamma}=\frac{1}{2}[(H_+^{NP})^2+(H_-^{NP})^2](3m_\ell^2+q^2)  ,
\nn \\
&&
I_{2s}^{NP,\gamma}=\frac{1}{16}(q^2-m_\ell^2)(H_L^{NP})^2  ,
 \hskip 2. cm 
 I_{2c}^{NP,\gamma}=-\frac{1}{2}[(H_+^{NP})^2+(H_-^{NP})^2](q^2-m_\ell^2)  , \hspace*{1.cm}
\nn \\
&&
I_{3}^{NP,\gamma}=-4 H_+^{NP}H_-^{NP}(q^2-m_\ell^2)  ,
 \hskip 2. cm 
  I_{4}^{NP,\gamma}= -\frac{1}{4}(q^2-m_\ell^2)H_L^{NP}[H_+^{NP}+H_-^{NP}] ,
\nn \\
&&
I_5^{NP,\gamma}=\frac{1}{2} m_\ell^2 H_L^{NP}[H_+^{NP}-H_-^{NP}]  ,
 \\
&&I_{6s}^{NP,\gamma}=0  ,
 \hskip 5.5 cm I_{6c}^{NP,\gamma}=
2 m_\ell^2 [(H_+^{NP})^2-(H_-^{NP})^2]  ,
 \nn \\
&&I_7^{NP,\gamma}=0 .
\nn \eea
The interference terms  are given by
\bea
&&
I_{1s}^{INT,\pi}=-4 \sqrt{q^2}\, m_\ell  ( H_+^{NP} H_+ +   H_-^{NP} H_- )  ,
 \hskip 0.8 cm
 I_{1c}^{INT,\pi}=- \sqrt{q^2}\,   m_\ell H_0 H_L^{NP}   ,
\nn \\
&&
I_{2s}^{INT,\pi}=0  ,
 \hskip 6.3 cm 
 I_{2c}^{INT,\pi}=0  ,
\nn \\
&&
I_{3}^{INT,\pi}=0  ,
 \hskip 6.3 cm 
  I_{4}^{INT,\pi}= 0 ,\label{INT-angular}
 \nn \\
&&
I_5^{INT,\pi}= \frac{1}{4}\sqrt{q^2} m_\ell \left[ H_L^{NP} (H_+-H_-)+8H_0(H_+^{NP}-H_-^{NP})+8H_t (H_+^{NP}+H_-^{NP}) \right]  , \hspace*{1.5cm}
 \\
&&I_{6s}^{INT,\pi}=-4 \sqrt{q^2}\, m_\ell    ( H_+^{NP} H_+ -   H_-^{NP} H_- )   ,
 \hskip 0.8 cm I_{6c}^{INT,\pi}= \sqrt{q^2}\,m_\ell H_L^{NP} H_t  ,
 \nn \\
 &&
 I_7^{INT,\pi}= \frac{1}{4}\sqrt{q^2} m_\ell \left[ H_L^{NP} (H_++H_-)-8H_0(H_+^{NP}+H_-^{NP})-8H_t (H_+^{NP}-H_-^{NP}) \right]  , \nn
\nn \eea
and
\bea
&&
I_{1s}^{INT,\gamma}=-\frac{1}{2} \sqrt{q^2}\,   m_\ell H_0 H_L^{NP}   ,
 \hskip 1 cm
 I_{1c}^{INT,\gamma}=- m_\ell  \sqrt{q^2} ( H_+^{NP} H_+ +   H_-^{NP} H_- )  ,
\nn \\
&&
I_{2s}^{INT,\gamma}=0  ,
 \hskip 4.2 cm 
 I_{2c}^{INT,\gamma}=0  ,
\nn \\
&&
I_{3}^{INT,\gamma}=0  ,
 \hskip 4.2 cm 
  I_{4}^{INT,\gamma}= 0   ,\label{INT-angular-gamma} 
 \nn \\
&&
I_5^{INT,\gamma}=\frac{1}{8} \, m_\ell  \sqrt{q^2} \left[ -H_L^{NP}  (H_+-H_-)-8H_0(H_+^{NP}-H_-^{NP})-8H_t (H_+^{NP}+H_-^{NP}) \right]  , \hspace*{1cm}
 \\
&&I_{6s}^{INT,\gamma}=\frac{1}{2}\,  m_\ell \sqrt{q^2}  H_t H_L^{NP}   ,
 \hskip 1.3 cm I_{6c}^{INT,\gamma}=
 - \sqrt{q^2}\,m_\ell   ( H_+^{NP} H_+  -   H_-^{NP} H_- )   , \nn \\
 &&
 I_7^{INT,\gamma}=\frac{1}{8}\sqrt{q^2} m_\ell \left[- H_L^{NP} (H_++H_-)+8H_0(H_+^{NP}+H_-^{NP})+8H_t (H_+^{NP}-H_-^{NP}) \right] . \nn
\eea
The relations (\ref{relationspigamma})  continue to hold.

\section{Standard Model:   scrutinizing CLN vs  BGL parametrization}\label{sec:FF}

\begin{figure}[t]
\begin{center}
\includegraphics[width = \textwidth]{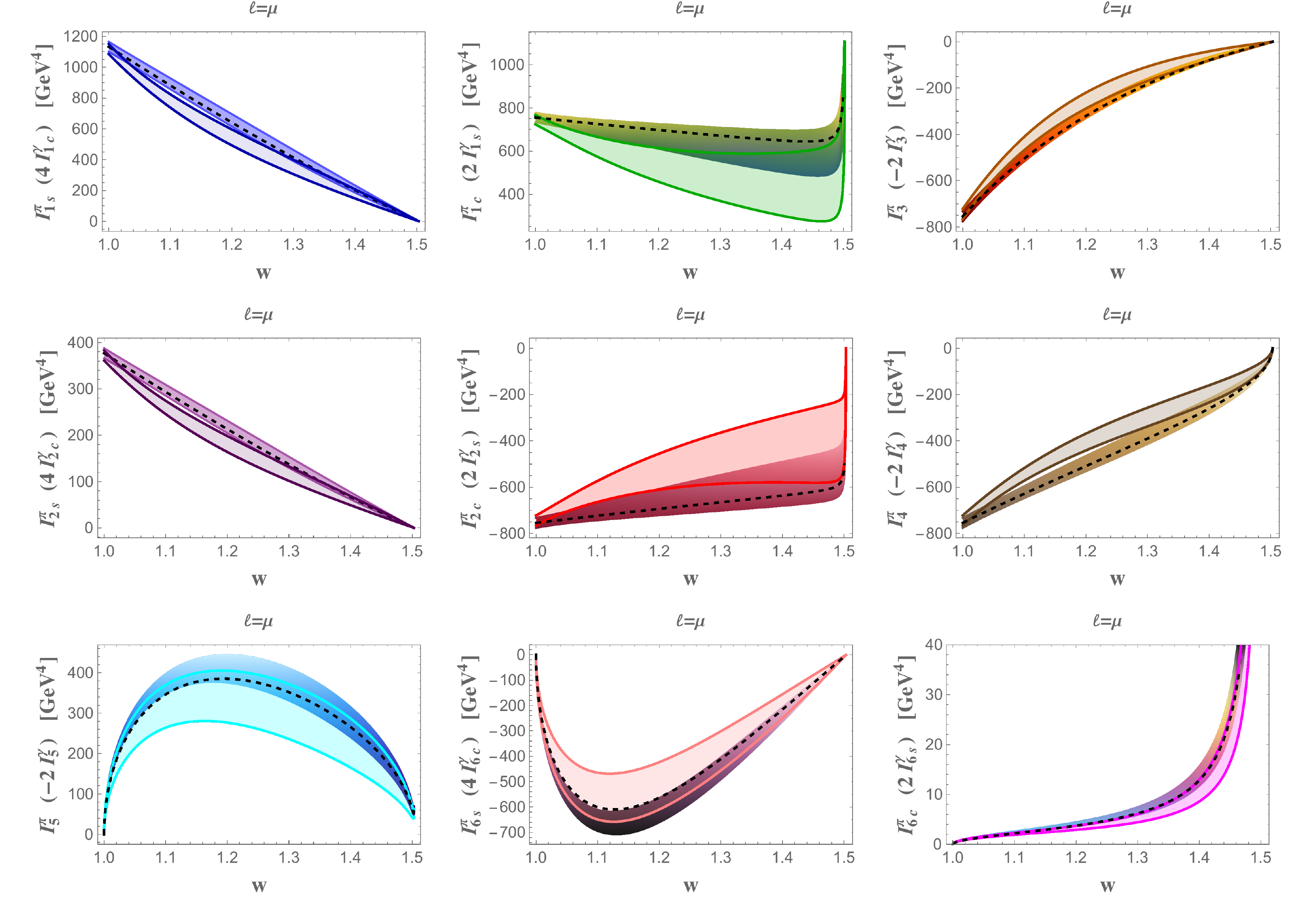}
    \caption{\baselineskip 10pt  Angular coefficients in the fully differential decay distribution Eq.~(\ref{angularpi}) for  $\ell=\mu$ in SM. The  coefficients in (\ref{angulargamma}) are obtained  using the relations (\ref{relationspigamma}).
   The darker regions correspond to the CLN parametrization with parameters in Table \ref{tabCLN}, the lighter regions  to the  BGL parametrization described in appendix \ref{appA}. The dashed  lines are the HQ predictions.}\label{fig:angularpimu}
\end{center}
\end{figure}
\begin{figure}[t]
\begin{center}
\includegraphics[width = \textwidth]{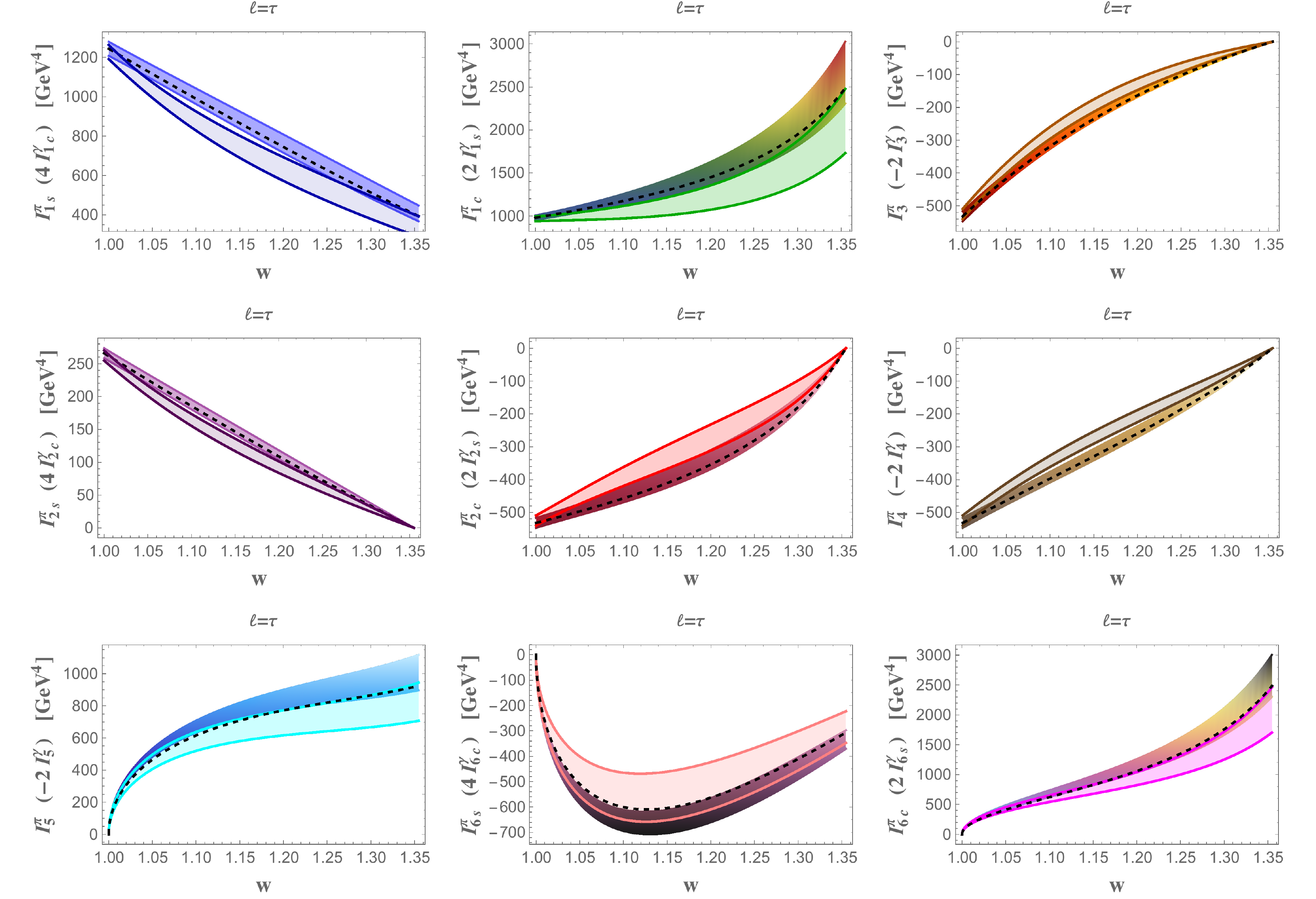}
    \caption{\baselineskip 10pt Angular coefficients in the fully differential decay distribution Eq.~(\ref{angularpi}) for $\ell=\tau$ in  SM. The coefficients in (\ref{angulargamma}) are obtained  using  (\ref{relationspigamma}).    Color code  as in figure \ref{fig:angularpimu}. }\label{fig:angularpitau}
\end{center}
\end{figure}
Understanding the  role of the  form factor parametrization  of the  $B \to D^{*}$ hadronic matrix element  is important before the formulation of  any strategy to disentangle  possible NP effects.
The  angular distributions  can  help identifying  observables  less sensitive to the form factor parametrization, hence  more  suitable to uncover deviations from SM. Observables displaying a  pronounced dependence on such parametrization can help in studying the impact of form factors.

The parametrizations based on the heavy quark limit make use of the relations  among the form factors in HQ, in particular the connection, at the leading order in the $1/m_Q$ expansion, of all the form factors to the single 
 Isgur-Wise function $\xi(w)$, with $w=\frac{m_B^2+m_{D^*}^2-q^2}{2 m_B m_{D^*}}$ the product of $B$ and $D^{(*)}$ four-velocities. $\xi(w)$ is normalized to unity at zero recoil $w=1$.
In the CLN formulation the relations are improved including perturbative $\alpha_s$ and power  $1/m_b$,  $1/m_c$ corrections  \cite{Caprini:1997mu}. 
In terms of  the function $h_{A1}(w)$ defined in  appendix \ref{appA},  which coincides with  $A_1(q^2)$ modulo a $w$-dependent coefficient, one can write
\bea
V(w)&=&\frac{R_1(w)}{R^*}h_{A_1}(w) \nn \\
A_1(w)&=&\frac{w+1}{2}R^*h_{A_1}(w) \nn \\
A_2(w)&=&\frac{R_2(w)}{R^*}h_{A_1}(w) \label{FFVA} \\
A_0(w)&=&\frac{R_0(w)}{R^*}h_{A_1}(w)  \nn \eea
with $R^*=\displaystyle\frac{2\sqrt{m_B m_{D^*}}}{m_B+m_{D^*}}$.
In this approach,  $h_{A_1}(w)$, $R_1(w)$, $R_2(w)$ and $R_0(w)$ are expanded for $w \to 1$, fixing the  series coefficients  using dispersive bounds \cite{Caprini:1997mu}:
\bea
h_{A_1}(w)&=&h_{A_1}(1) \left[ 1-8 \rho^2 z+(53\rho^2-15)z^2-(231\rho^2-91)z^3 \right] \nn
\\
R_1(w)&=&R_1(1) -0.12(w-1)+0.05(w-1)^2 \nn \\
R_2(w)&=&R_2(1) +0.11(w-1)-0.06(w-1)^2 \nn \\
R_0(w)&=&R_0(1) -0.11(w-1)+0.01(w-1)^2 \,\,,\label{clnRi}
\eea
with the conformal variable $z$ defined as $\displaystyle z=\frac{\sqrt{w+1}-\sqrt 2}{\sqrt{w+1}+\sqrt 2}$.
In the HQ limit the predictions 
\be
R_1^{HQ}(1)=1.27 \,\, , \hskip 1.5 cm R_2^{HQ}(1)=0.80 \,\, , \hskip 1.5 cm R_0^{HQ}(1)=1.25\,\,
\label{R12HQET}
\ee
are obtained \cite{Caprini:1997mu,Bernlochner:2017jka}.
However, in the  experimental analyses  making use of this parametrization, not only  the slope $\rho^2$,  but also the ratios $R_1(1)$ and $R_2(1)$ are  fitted parameters, while $h_{A_1}(1)$ is taken  from lattice QCD calculations. $R_0(1)$ is involved in the case of  $\tau$ lepton, and no experimental result is available.
The parameters fitted by Belle Collaboration \cite{Abdesselam:2017kjf}, that we use in our analysis, are   collected in Table \ref{tabCLN}.
We use the last  relation  in (\ref{FFVA}), together with the expressions  (\ref{ha2}) in appendix \ref{appA}, to obtain $R_0(1)$.

In the BGL formulation, recalled in appendix \ref{appA}, the form factors are expressed as functions of the conformal variable $z$. After having included   outer functions \cite{Boyd:1997kz} and subtracted the 
contribution of  $b \bar c$ states, the form factors are expressed as power series of $z$, with the coefficients determined by a fit to the experimental data
 \cite{Boyd:1994tt,Boyd:1995cf,Boyd:1995sq}. The number of parameters for each form factor is larger than in  CLN; on the other hand, no information from the HQ limit is used.
In our analysis we use the parameters in \cite{Bigi:2017njr}, obtained fitting the same data set in  \cite{Abdesselam:2017kjf},  in the case where input from light-cone QCD sum rules is included.
In the absence of results from the fits,  also in this case  we  use the HQ relations to obtain $R_0$, as in   \cite{Bigi:2017jbd}.

A point emphasized in \cite{Grinstein:2017nlq,Bigi:2017njr,Bigi:2017jbd} is that, although  the Belle data in \cite{Abdesselam:2017kjf} can be  well reproduced using both  parametrizations,  the high $q^2$ bins  are better described  by BGL, with a value of  $\vcb$  larger than using CLN  and  closer to the inclusive $\vcb$ determination. 
 Moreover, these Belle data seem to suggest  deviations from HQ symmetry and tensions with preliminary lattice results  for the ratio $R_1$ \cite{Aviles-Casco:2017nge}, as noticed  comparing the data to fits using BGL or various versions of CLN  parametrization  \cite{Bernlochner:2017xyx}.
\begin{table}[b!]
\begin{center}
\begin{tabular}{|c|c|c|c|}
  \hline
  $\vcb \times 10^3 $ & $\rho^2$ & $R_1(1)$ & $R_2(1)$\\
  \hline 
  $37.4 \pm 1.3 $ & $1.03 \pm 0.13 $ & $1.38 \pm 0.07 $& $0.87 \pm 0.10 $
  \\
  \hline                                                                               	
\end{tabular}
\caption{\baselineskip 10pt  CLN parameters fitted by Belle Collaboration \cite{Abdesselam:2017kjf}.}
\label{tabCLN}
\end{center}
\end{table}

In principle,   the angular coefficient functions inferred from the fully differential distribution can be used to reconstruct the form factors. In particular, for the ratios $R_1(w)$ and $R_2(w)$  one has:
\bea
R_1(w) &=& \frac{8 q^2 m_B m_{D^*} (1+w)}{(m_\ell^2+3q^2) \lambda^{1/2}(m_B^2,m_{D^*}^2,q^2)}\frac{1}{I_{6s}^\pi} \Bigg[ \sqrt{(I_{1s}^\pi)^2- \left(\frac{m_\ell^2+3q^2}{q^2} \right)^2 \frac{(I_{6s}^\pi)^2}{16}}-I_{1s}^\pi\Bigg] ,
\label{R1ang}\\
R_2 (w) &=& \frac{2 m_B m_{D^*} (1+w)}{\lambda(m_B^2,m_{D^*}^2,q^2)} \Bigg[ (m_B^2-q^2-m_{D^*}^2) \nn \\
&+&2 \sqrt{2}m_{D^*} q^2\sqrt{-\frac{q^2}{q^2-m_\ell^2}I_{2c}^\pi}\frac{1}{I_{6s}^\pi} \left( \sqrt{\frac{4 I_{1s}^\pi}{m_\ell^2+3 q^2}-\frac{ I_{6s}^\pi}{ q^2}}-\sqrt{\frac{4 I_{1s}^\pi}{m_\ell^2+3 q^2}+\frac{ I_{6s}^\pi}{ q^2}}\right) \label{R2ang} \Bigg] .
\eea
This is interesting, since a difference between the  CLN and  BGL parametrizations emerges in these ratios   \cite{Bigi:2017njr,Bigi:2017jbd,Bernlochner:2017xyx}. 

We now investigate  the angular coefficient functions obtained with CLN and BGL, using their respective set of parameters. The results are collected in figure \ref{fig:angularpimu}.
We use as an input the lattice QCD value $h_{A_1}(1)=0.906 \pm 0.013$ \cite{Bailey:2014tva} times the ew correction factor $\eta_W=1.0066$ \cite{Sirlin:1981ie,Atwood:1989em}.
We  also show the results obtained  in the HQ limit using Eq.~(\ref{R12HQET}).

The functions  $I_{1s}^\pi$, $I_{2s}^\pi$, $I_3^\pi$, $I_4^\pi$, and  $I_{1c}^\gamma$, $I_{2c}^\gamma$,  $I_3^\gamma$,  $I_4^\gamma$ are largely insensitive to  the form factor parametrization.
On the contrary,  $I_{1c}^\pi$, $I_{2c}^\pi$, $I_{6s}^\pi$, and  $I_{1s}^\gamma$, $I_{2s}^\gamma$, $I_{6c}^\gamma$ are more dependent. The coefficients  $I_{6c}^\pi$, $I_{6s}^\gamma$ are proportional to the lepton mass, hence  they are small compared to the others for $\ell=\mu$.
The indication is that the first set of coefficients is more suitable to pin down  deviations from SM. In particular,  $I_7^{\pi(\gamma)}$ vanishes in  SM,  therefore it is able to signal a  NP effect: indeed, in the model with the tensor operator    ${\rm Im}(\epsilon_T^{\ell})$  can be  non-vanishing, as well as  $I_7^{\pi(\gamma)}$.

The second set of angular coefficient functions  can be used to better evaluate the form factor parametrization. The results in BGL display larger uncertainties, and are systematically larger (smaller) than in CLN in 
$I_{2c}^\pi$, $I_{6s}^\pi$  ($I_{1c}^\pi$, $I_5^\pi$). An overlap region spanned by the two parametrizations always exists, and the  HQ result  is  closer to the CLN outcome, sometimes at the limits.
An analogous trend is found for the  $\ell=\tau$ mode, in the angular coefficient functions  displayed in figure \ref{fig:angularpitau}. Comparing  $I_{1c}^\pi$ and  $I_{2c}^\pi$ for $\ell=\mu$ and $\tau$, 
one finds that the uncertainties are smaller in the case of the heavier lepton.
Due to the difficulties discussed in the Introduction, the possibility of accessing the various $I_i$ in the case of $\tau$ is very challenging. In particular, the reconstruction of the angle $\theta$ is not possible when the $\tau$ is reconstructed in decays to final states with multiple neutrinos. However, the reconstruction in visible 3-prong decays opens new interesting perspectives from this point of view, although with the  caveat concerning  the control of the $\pi^+ \pi^+\pi^- \pi^0$ contribution discussed in the Introduction.
Moreover, a set of integrated observables can be considered, with particular attention of those depending on the angle $\theta_V$.

\begin{figure}[t]
\begin{center}
\includegraphics[width = 0.4\textwidth]{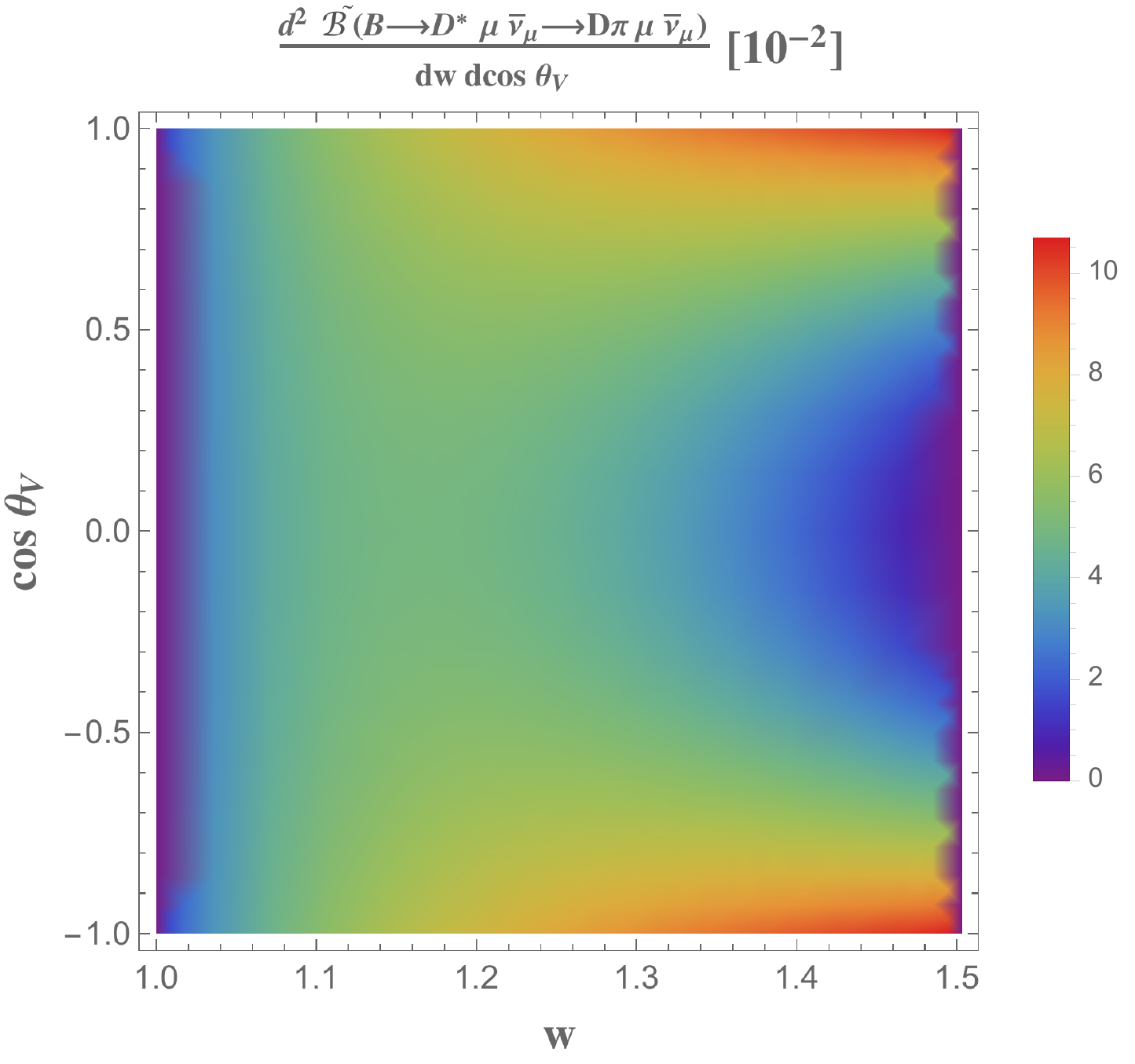}
\includegraphics[width = 0.4\textwidth]{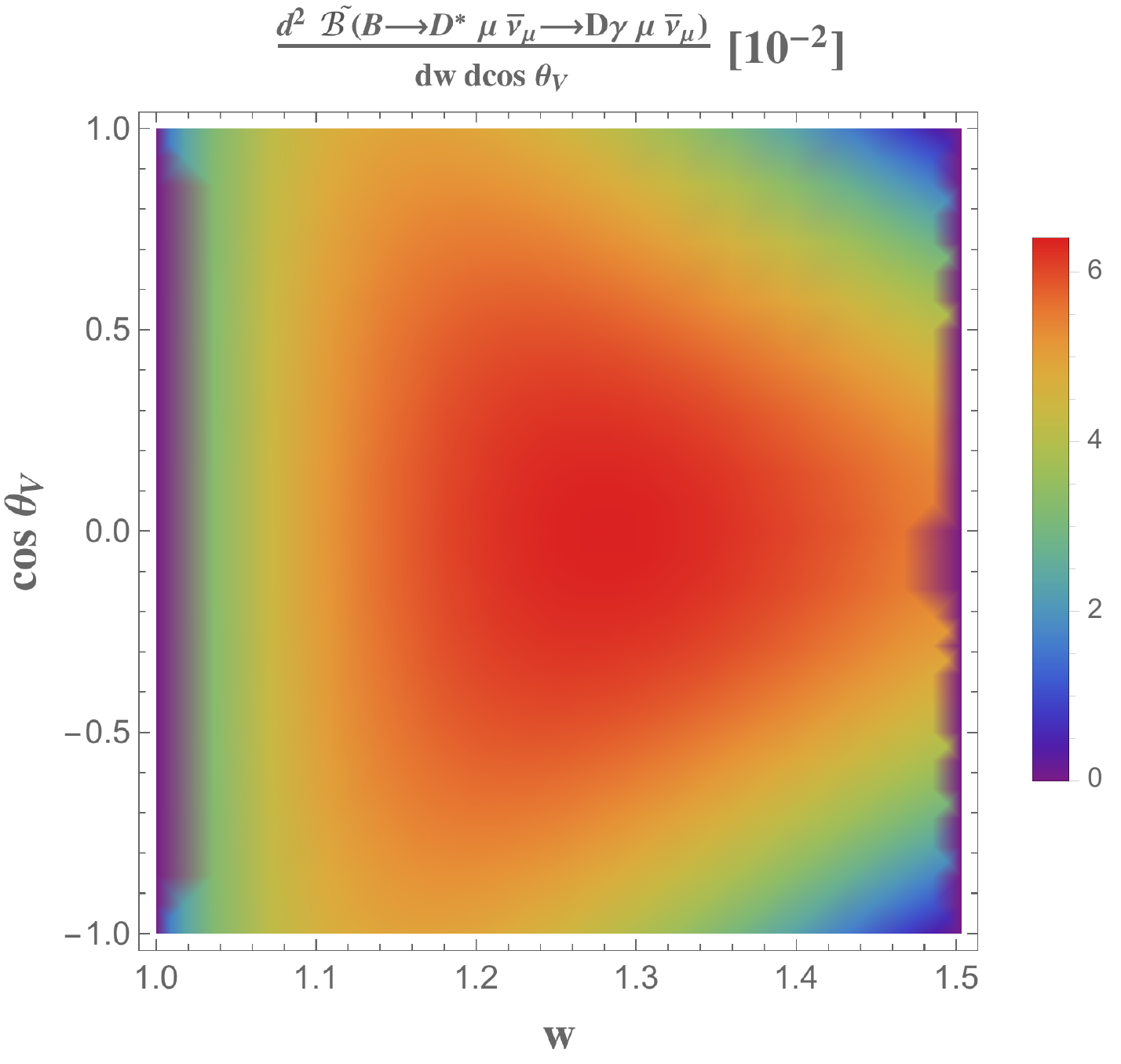}\\
\includegraphics[width = 0.4\textwidth]{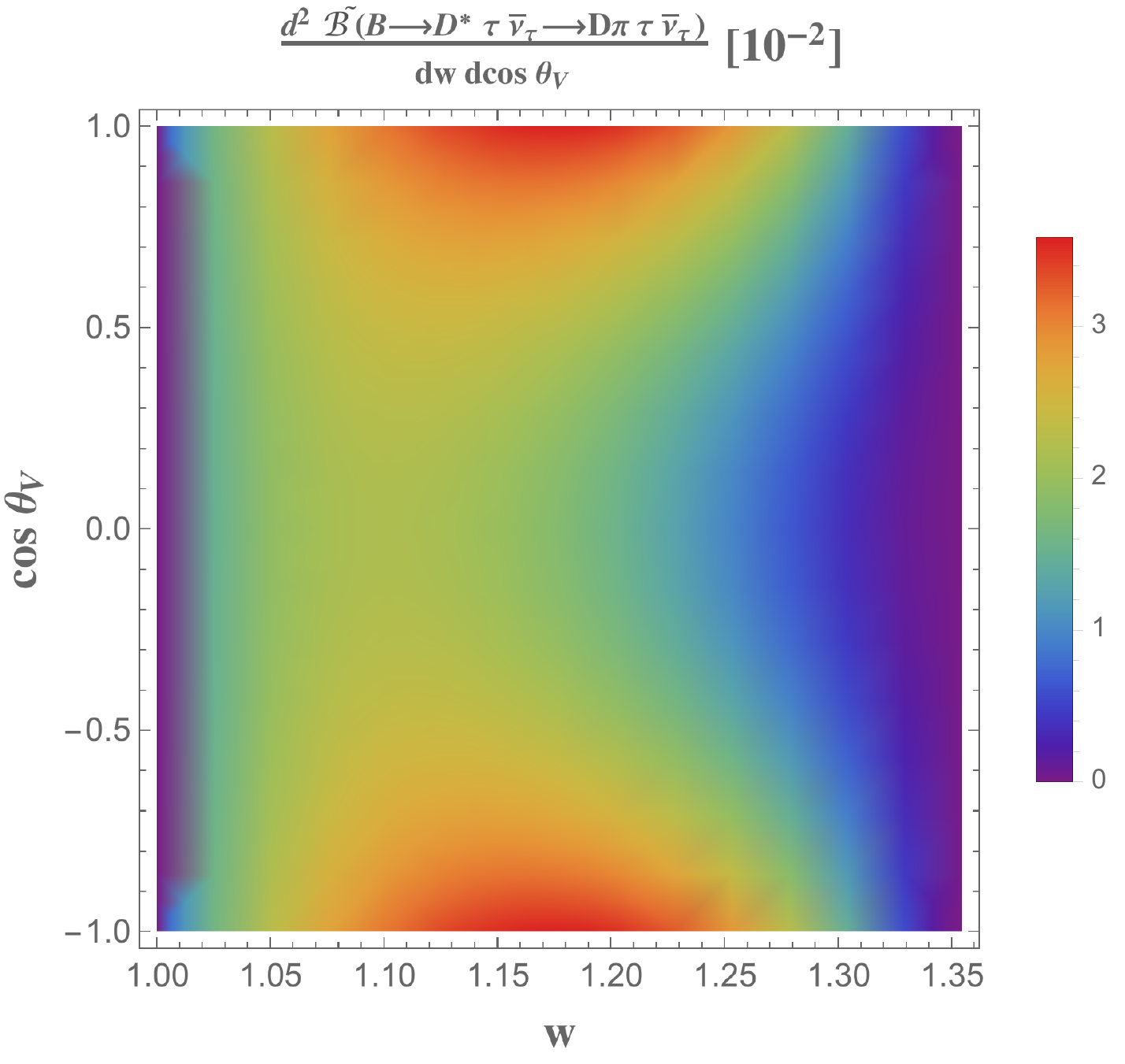}
\includegraphics[width = 0.4\textwidth]{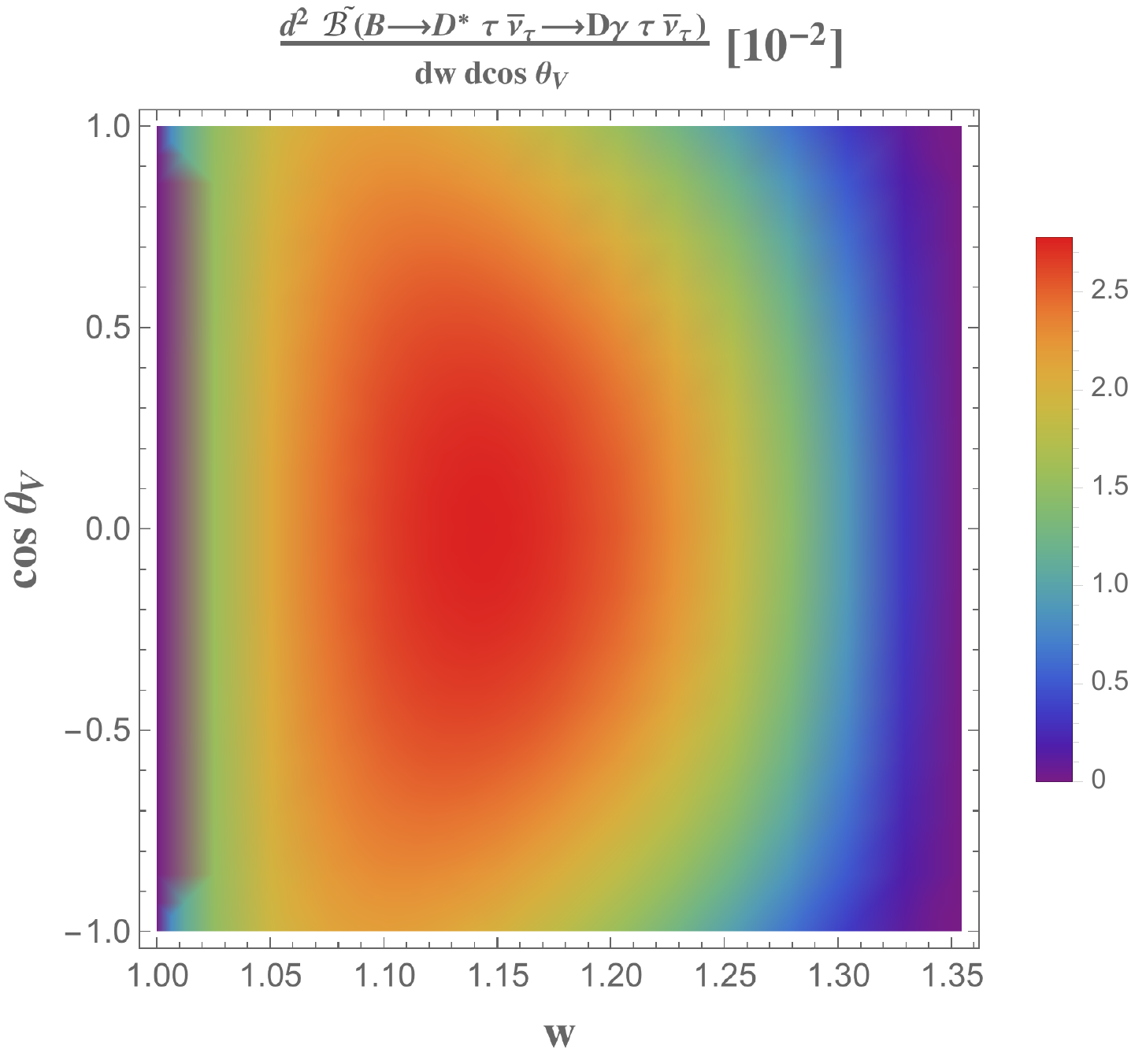}
    \caption{\baselineskip 10pt  SM scatter plots of the double differential  distributions in $w$ and $\cos \theta_V$, with ${\tilde {\cal B}}={\cal B}/{\cal B}(D^* \to D F)$,  using the CLN parametrization.
    The upper and lower plots refer to  $\ell=\mu$  and  $\ell=\tau$ modes, respectively, the left and right  column to  $F=\pi$ and $F=\gamma$.  }\label{densityplot}
\end{center}
\end{figure}

\begin{figure}[t]
\begin{center}
\includegraphics[width = 0.45\textwidth]{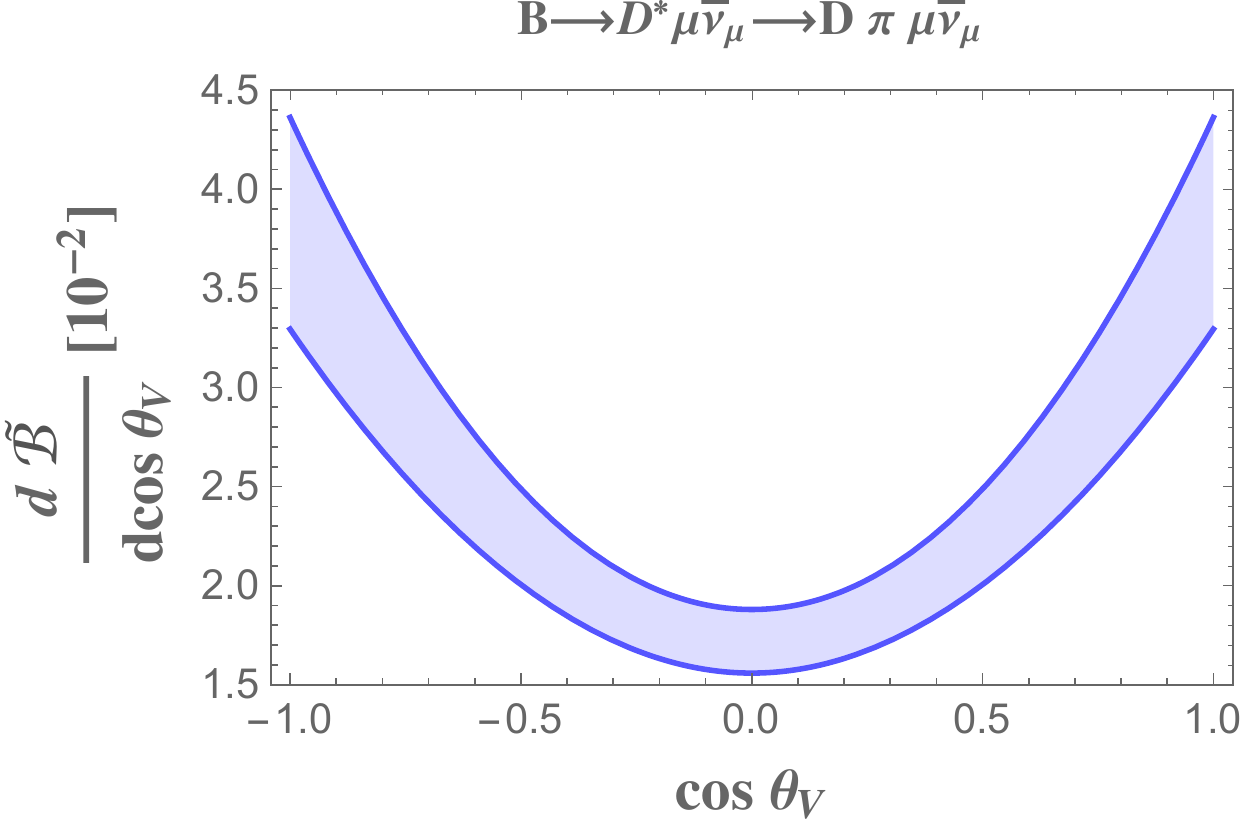} \hskip 0.3cm
\includegraphics[width = 0.45\textwidth]{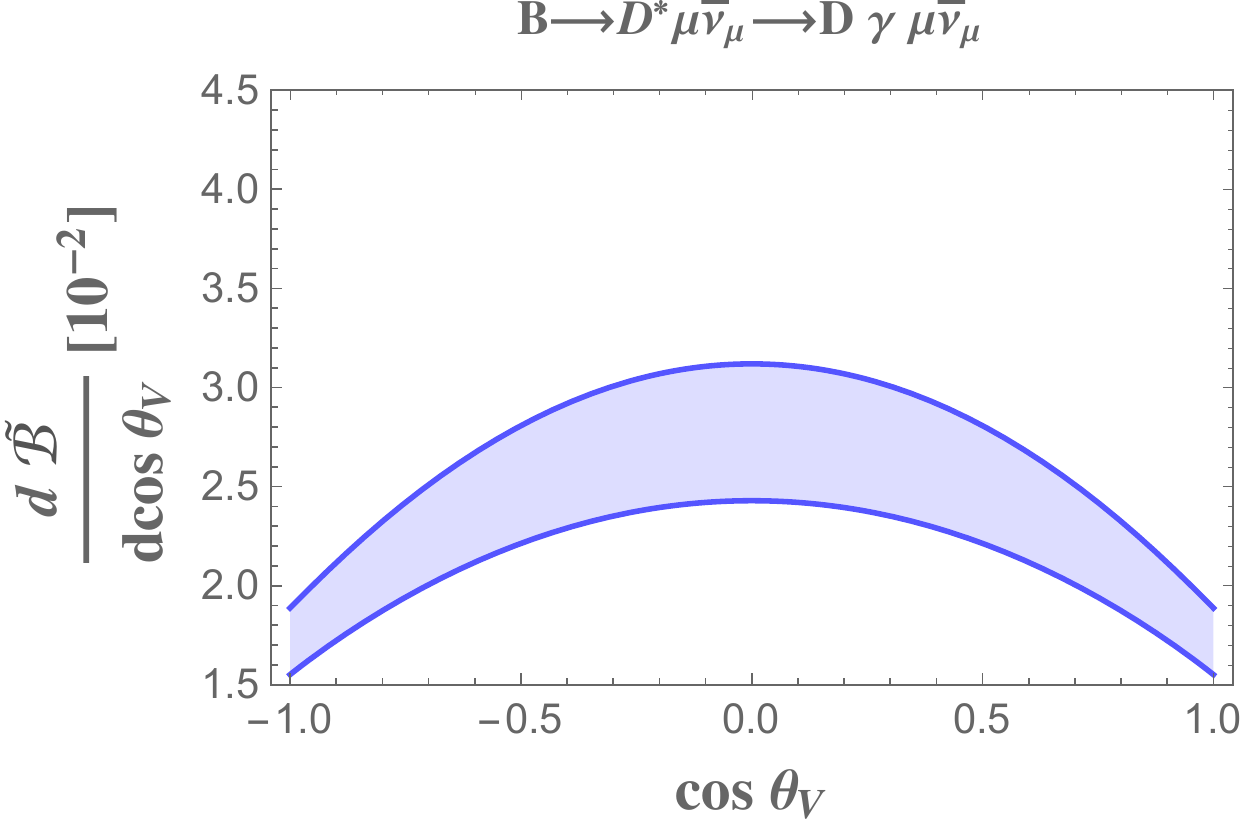}\\
\vskip 0.5 cm
\includegraphics[width = 0.45\textwidth]{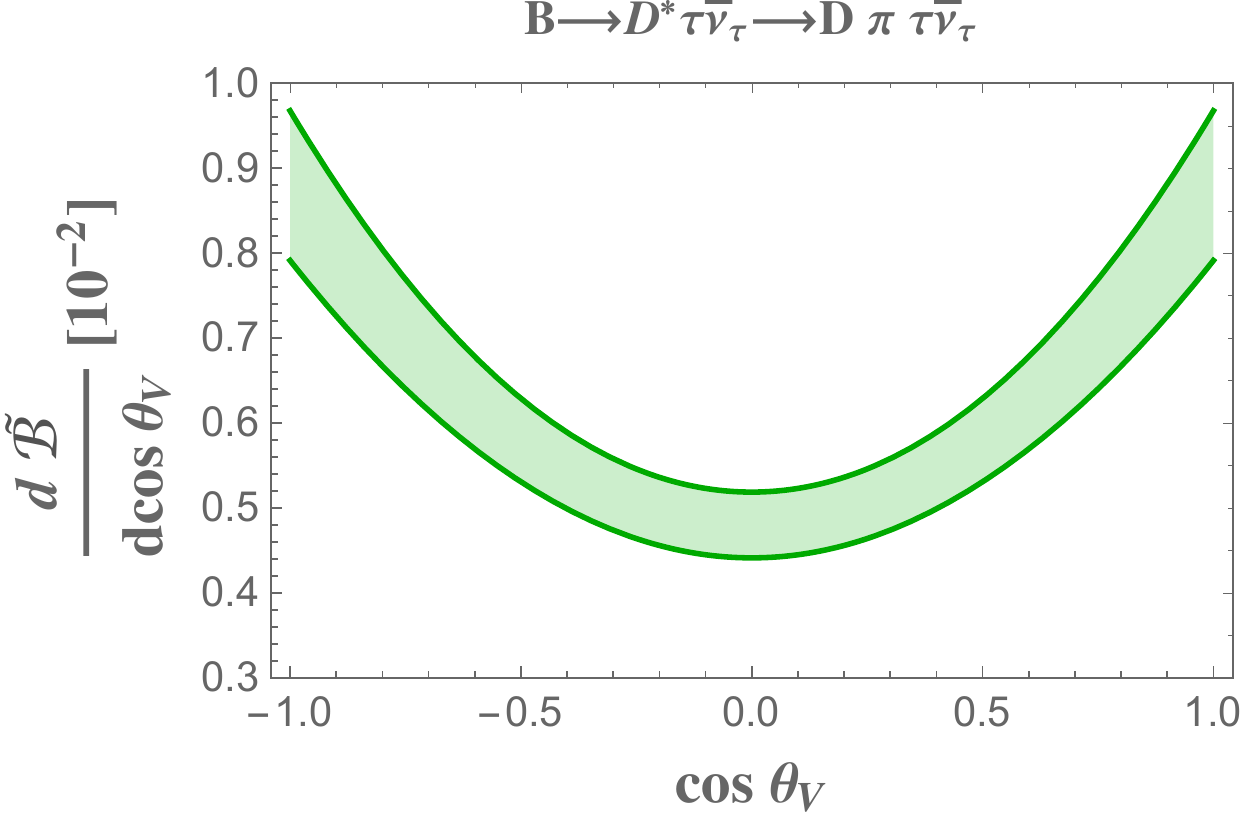}\hskip 0.3cm
\includegraphics[width = 0.45\textwidth]{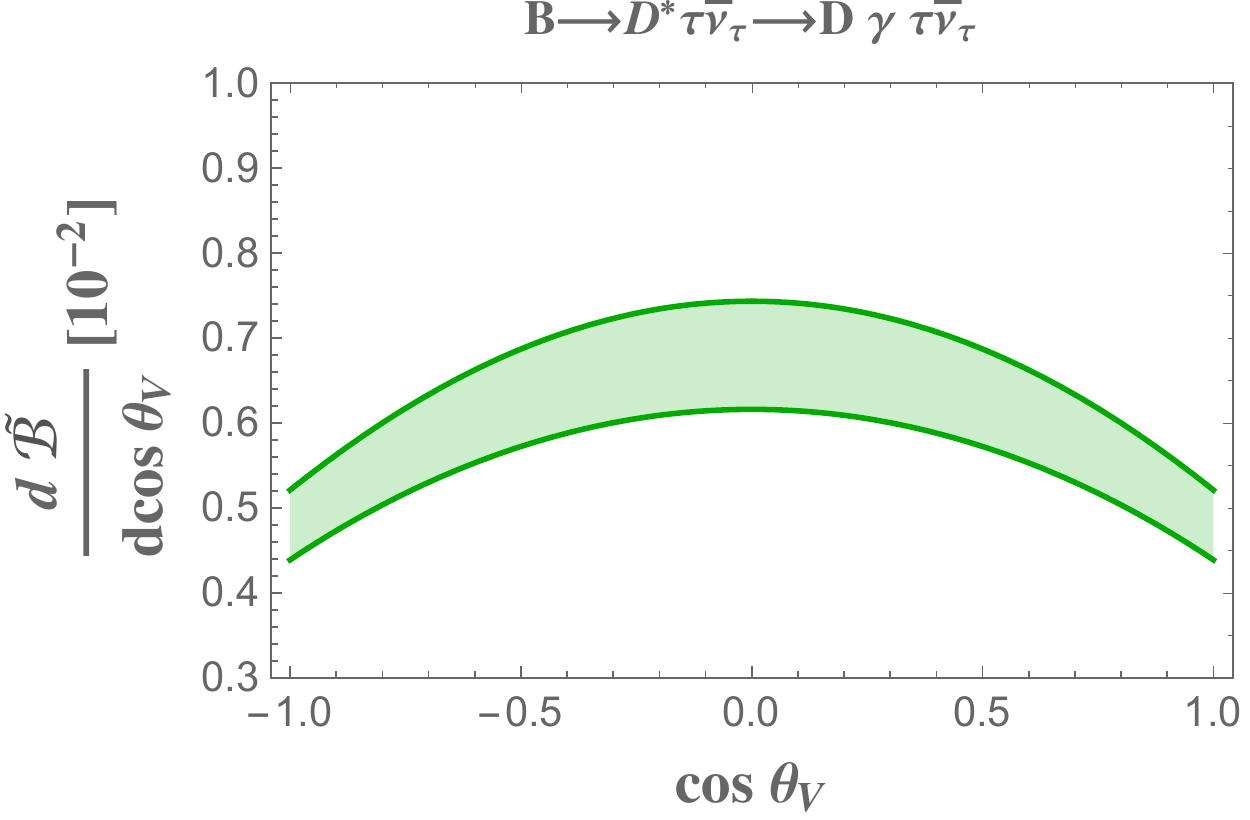}
    \caption{\baselineskip 10pt SM distributions in  $\cos \theta_V$ using CLN, with ${\tilde {\cal B}}= {\cal B}/{ \cal B}(D^* \to D F)$.
    The upper and lower plots refer to  $\ell=\mu$ and  $\ell=\tau$, respectively, the left and right column to  $F=\pi$ and $F=\gamma$. }\label{dGdzV}
\end{center}
\end{figure}

The complementarity of the modes with $D^*$ decaying  to $D \pi$ or to $D \gamma$ emerges from  fig.~\ref{densityplot} and \ref{dGdzV}, obtained   using the CLN parametrization. For $F=\pi$ the events are mainly at the limits of the   $\cos \theta_V$ region, as shown both by the  density plots in figure \ref{densityplot} and by the projections in fig.~\ref{dGdzV}. In the case of the photon,  the most populated region is  for $\cos \theta_V \simeq0$. This should be taken into account in the analysis of $B_s \to D^{*+}_s  \ell^- \bar \nu$, where the final state is dominated by the $D_s \gamma$ mode.

\section{Angular coefficient functions in the NP model}\label{sec:NPcase}
In the case of the effective Hamiltonian  with the tensor operator the angular coefficient functions are modified.
To discuss  the changes with respect to SM  we need to fix a range for the  couplings  $\epsilon_T^\mu$ and $\epsilon_T^\tau$.
In \cite{Biancofiore:2013ki}    $\epsilon_T^\tau$ was constrained by $R(D)$ and $R(D^*)$, assuming $\epsilon_T^\mu=\epsilon_T^e=0$.  In \cite{Colangelo:2016ymy} the latter assumption was relaxed,  $\epsilon_T^\mu \neq0$ and $\epsilon_T^e \neq 0$,  to reproduce  $\bar B \to X_c \ell^- \bar \nu_\ell$ and  $\bar B \to D^{(*)} \ell^- \bar \nu_\ell$ data  in a common range of  $\vcb$. We now consider these constraints, but  since  the ranges for $\epsilon_T^\mu$ and $\epsilon_T^e$ turn out to be almost coincident, we only distinguish  $\epsilon_T^\mu$ and $\epsilon_T^\tau$.
We adopt  the CLN parametrization,  employing  the HQ relations  (including ${\cal O}(\alpha_s)$ and ${\cal O}(1/m_Q)$ corrections \cite{Bernlochner:2017jka}, as reported in Appendix B) to determine  the form factors $T_i$  in (\ref{mat-tensor-Dstar}),   since the BGL parametrization  for such functions has not been developed.

\begin{figure}[t]
\begin{center}
\includegraphics[width = 0.45\textwidth]{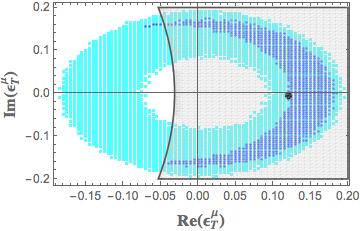} \hskip 0.3cm
\includegraphics[width = 0.45\textwidth]{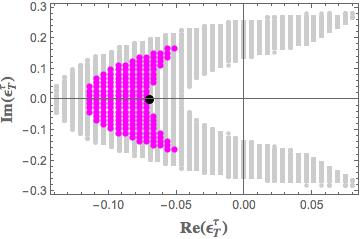}
 \caption{\baselineskip 10pt Parameter space  of $\epsilon_T^\mu$ (left) and $\epsilon_T^\tau$ (right), determined using  $R(D)$ and $R(D^*)$ in (\ref{hfag}) (lighter regions). The darker regions  correspond to  $\chi^2<1.0$.  In  $\epsilon_T^\mu$  the shaded gray region results  using  the Belle measurement of $R_{e\mu} $  \cite{Abdesselam:2017kjf} . The black dots are the values ${\tilde \epsilon}_T^\ell$  defined in the text, used as benchmark points.}\label{epsTmutau}
\end{center}
\end{figure}

We use the range of values of $\epsilon_T^\mu$ selected in  \cite{Colangelo:2016ymy}, restricted to  reproduce  $\vcb$ obtained by the  Belle's fit in Table \ref{tabCLN} (within 2$\sigma$).   In this range  we compute $R(D)$ and $R(D^*)$ using  the averages in Eq.~(\ref{hfag}) at  $1 \sigma$  as constraints.
For  $R(D)$ we use    lattice QCD form factors  \cite{Lattice:2015rga}.  
The  obtained ranges for $\epsilon_T^\mu$ and $\epsilon_T^\tau$ are displayed in figure \ref{epsTmutau}.
The  regions can be restricted imposing  $\chi^2=\left( \frac{R(D)- R(D)^{exp}}{\Delta R(D)^{exp}}\right)^2+\left( \frac{R(D^*)- R(D^*)^{exp}}{\Delta R(D^*)^{exp}}\right)^2 \le 1.0$. 
In this region we select the point  corresponding to the minimum  $|{\tilde \epsilon}_T^\mu|$,
  the black dot  in fig.  \ref{epsTmutau}, together with the corresponding  value for $\epsilon_T^\tau$,
with numerical values  $({\rm Re}({\tilde \epsilon}_T^\mu),\,{\rm Im}({\tilde \epsilon}_T^\mu))=(0.115,\,-0.005)$ and $({\rm Re}({\tilde \epsilon}_T^\tau),\,{\rm Im}({\tilde \epsilon}_T^\tau))=(-0.067,\, 0)$. 
This is a benchmark point used to describe the sensitivity of the angular observables and the  pattern of correlated deviations from SM in this scenario. In correspondence to this value,  the fraction of longitudinally polarized  $D^*$ measured  by   Belle in \cite{Dungel:2010uk} is reproduced, considering the various uncertainties, while the fraction of tranversely polarized $D^*$ in the maximum recoil region turns out to deviate by more than $2\sigma$ in the last two bins of $w$.  Indeed,  this distribution is found to be SM-like:  in SM for massless leptons the $D^*$ is fully longitudinally polarized at $q^2=0$. Compatibility with data in this kinematical region would be obtained for  ${\rm Re}({\epsilon}_T^\mu)\leq 0.05$, in agreement with  the findings in \cite{Jung:2018lfu}. However, the purpose of our analysis is not to obtain the best fit of the NP coupling, a task deferred to different studies based on the full knowledge of the data sets with their systematics, but to provide the overview on how the various observables coherently deviate from SM  in this scenario.
It should be remarked that in the selected parameter region, for $\epsilon_T^\mu=\epsilon_T^e$,  the ratio $R_{e\mu}=\displaystyle{\frac{{\cal B}({\bar B}^0 \to D^{*+}e^- {\bar \nu}_e)}{{\cal B}({\bar B}^0 \to D^{*+}\mu^- {\bar \nu}_\mu)}}=1.04 \pm 0.05 \pm 0.01$   \cite{Abdesselam:2017kjf}  is reproduced: in figure \ref{epsTmutau},  the shaded gray region is constrained by $R_{e\mu}$. 

\begin{figure}[t]
\begin{center}
\includegraphics[width = \textwidth]{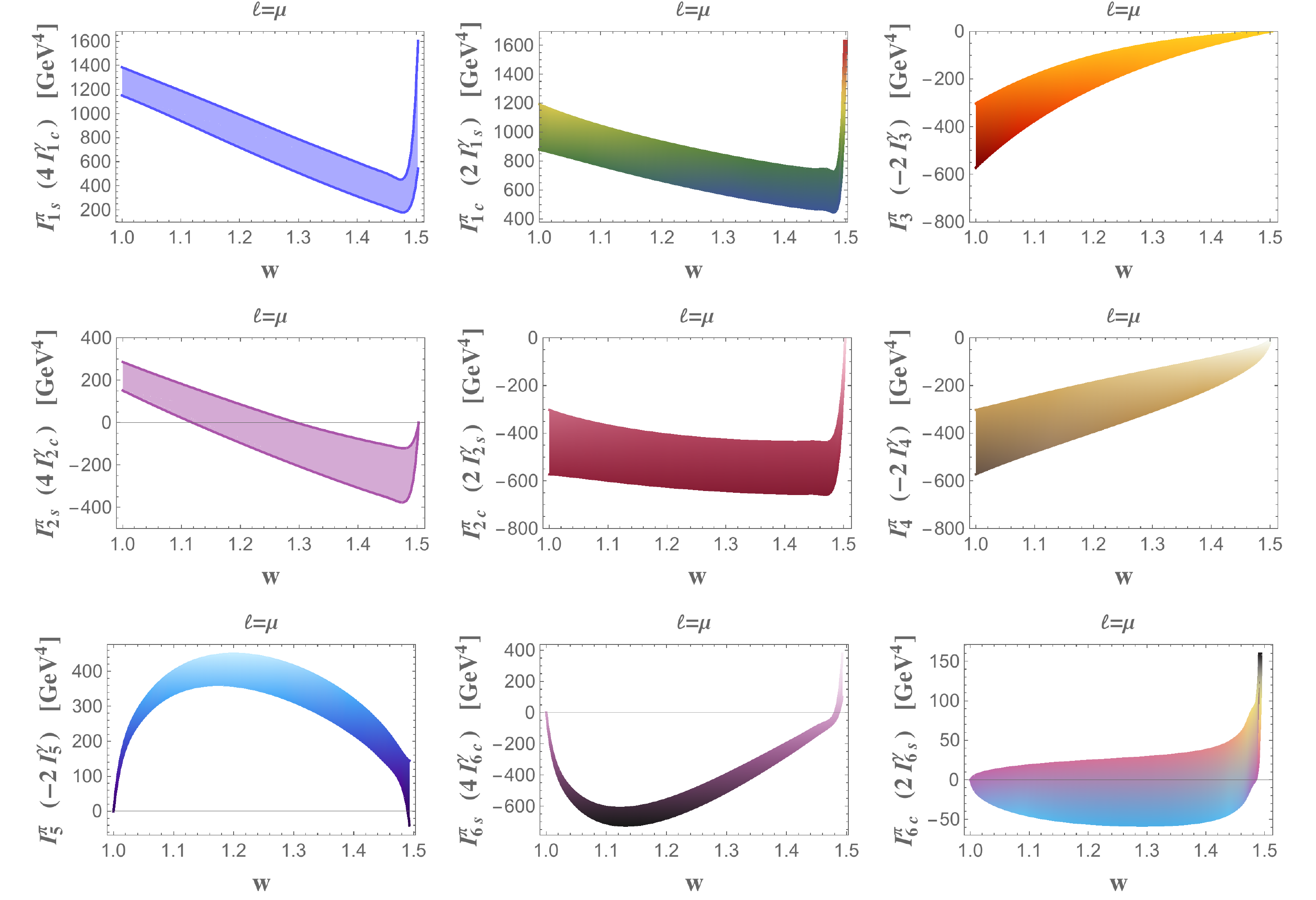}
    \caption{\baselineskip 10pt Angular coefficients in the fully differential decay distribution Eq.~(\ref{angularpi}) for $\ell=\mu$, with the  tensor operator  in the effective Hamiltonian and  coupling $\epsilon_T^\mu$
in the low $\chi^2$ region displayed in figure \ref{epsTmutau}. }\label{fig:angularpimuNP}
\end{center}
\end{figure}

\begin{figure}[t]
\begin{center}
\includegraphics[width = \textwidth]{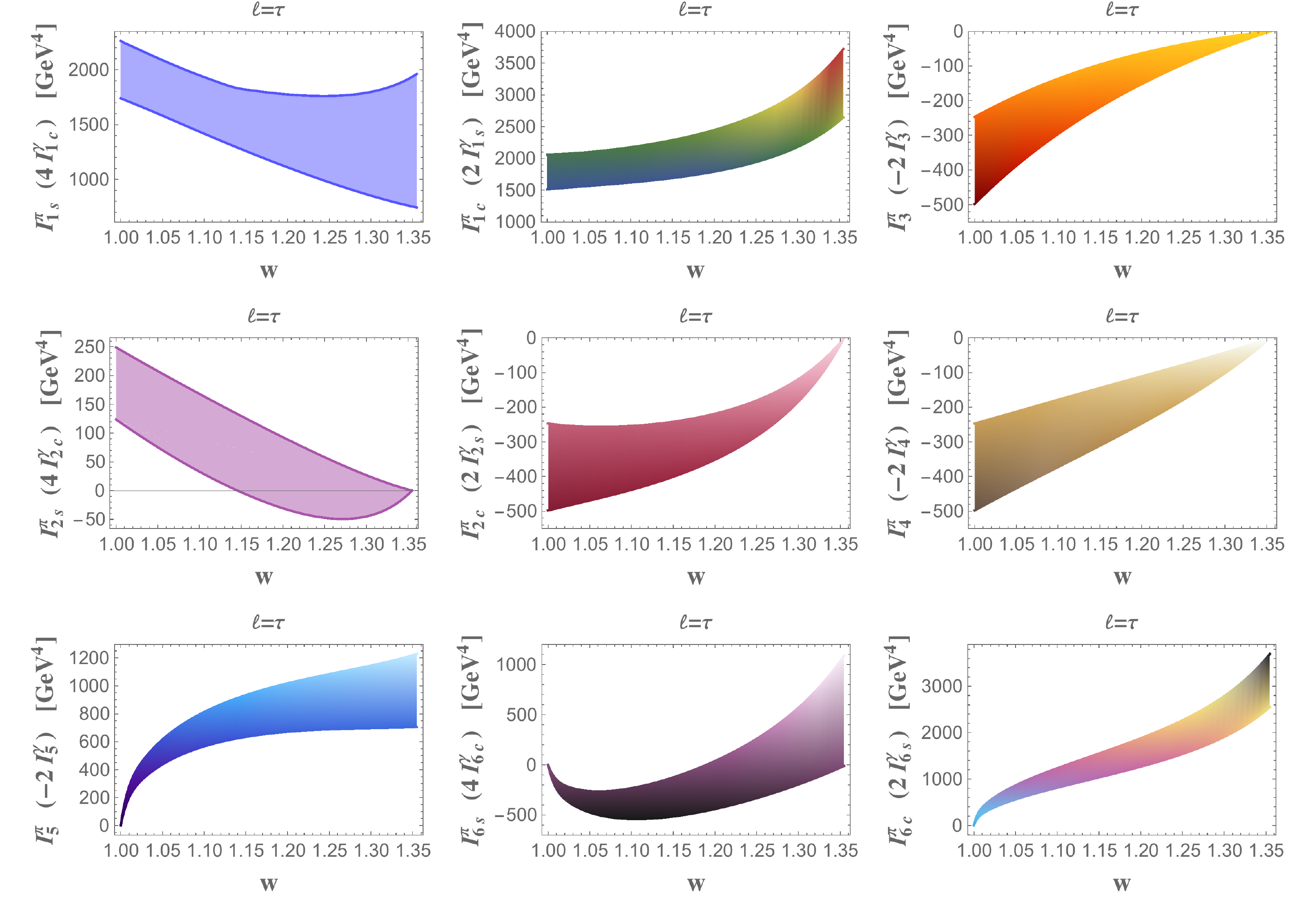}
    \caption{\baselineskip 10pt  Angular coefficients in the fully differential decay distribution Eq.~(\ref{angularpi}) for $\ell=\tau$, with the  tensor operator  in the effective Hamiltonian and coupling $\epsilon_T^\tau$ in the low $\chi^2$ region displayed in figure \ref{epsTmutau}.   }\label{fig:angularpitauNP}
\end{center}
\end{figure}

\begin{figure}[t]
\begin{center}
\includegraphics[width =0.45 \textwidth]{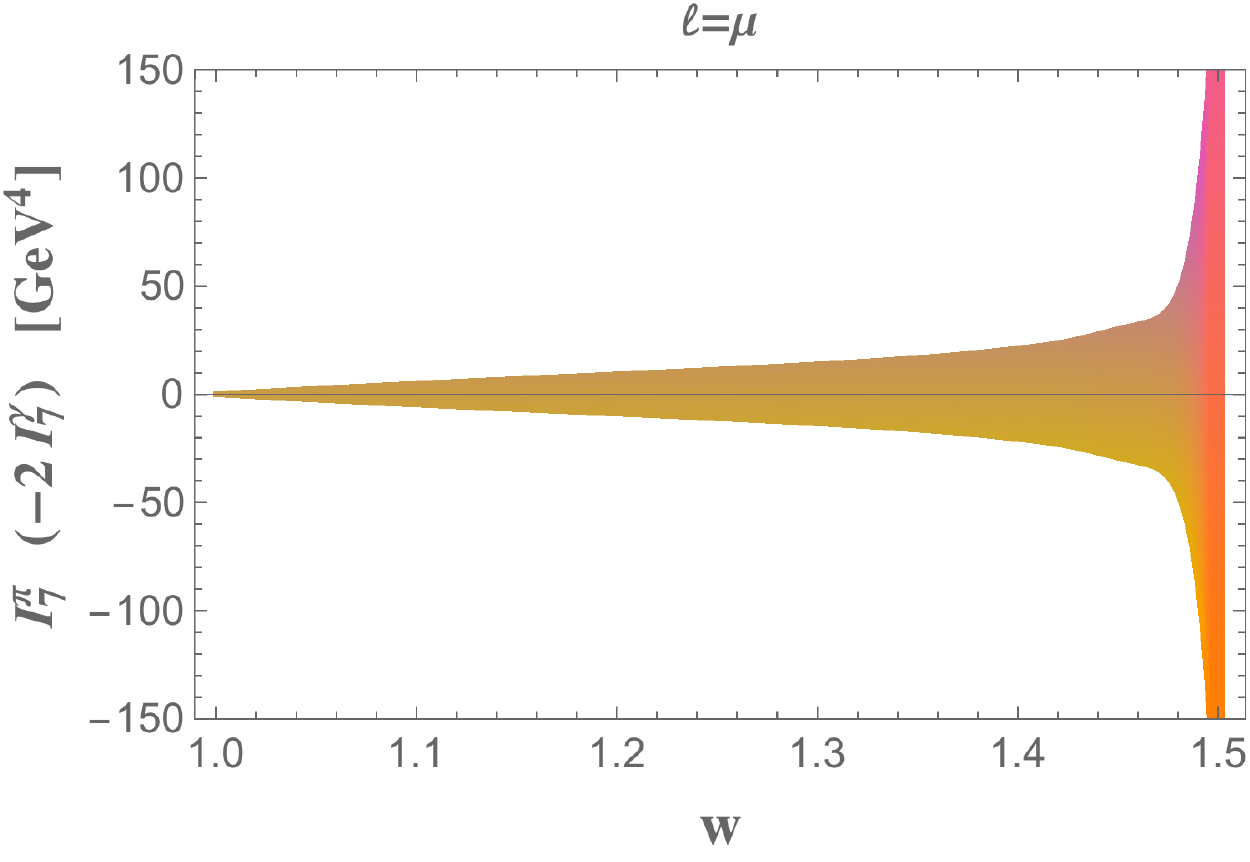}
\includegraphics[width = 0.45\textwidth]{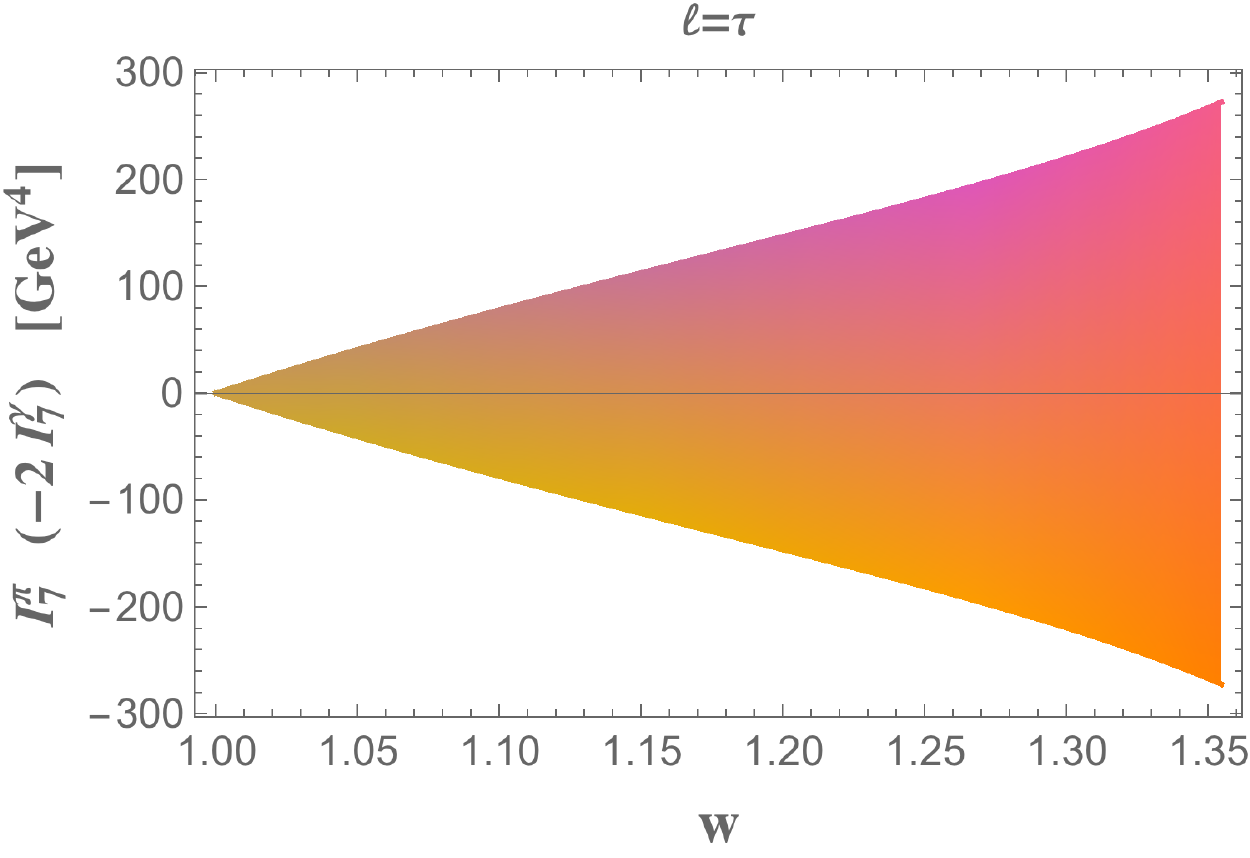}
    \caption{\baselineskip 10pt  Coefficient $I_7$ in  Eq.~~(\ref{angularpi}) for $\ell=\mu$ and    (left), and  $\ell=\tau$ (right).  The coefficients   $ \epsilon_T^\mu$ and $ \epsilon_T^\tau$  are varied in the low $\chi^2$ region displayed in figure \ref{epsTmutau}.
$I_7$  does not vanish  when the  tensor operator is included in the effective Hamiltonian.}\label{I7fig}
\end{center}
\end{figure}

We compute the angular coefficients $I_i$ using the parameters  $\epsilon_T^\mu$,\, $\epsilon_T^\tau$ in the low $\chi^2$ region  in figure \ref{epsTmutau}, with the results shown in figs.~ \ref{fig:angularpimuNP} and \ref{fig:angularpitauNP} for $\ell=\mu$ and $\ell=\tau$.   
Comparing the results in SM, the impact of NP is  to modify  the size of the coefficients, in several cases  mainly near the maximum recoil point $w \to w_{max}$,  as noticed in \cite{Jung:2018lfu}.   $I_{2s}^\pi \, (I_{2c}^\gamma)$, always positive in  SM, has a zero in  NP. 
$I_7$ is displayed in figure \ref{I7fig}; it is proportional to the lepton mass, hence it is  small in the muon case, but it can be different from zero if  $\epsilon_T^\ell$ has  non-zero imaginary part.

\section{Scrutinizing deviations from SM}\label{sec:obs}
\begin{figure}[t]
\begin{center}
\includegraphics[width =0.32 \textwidth]{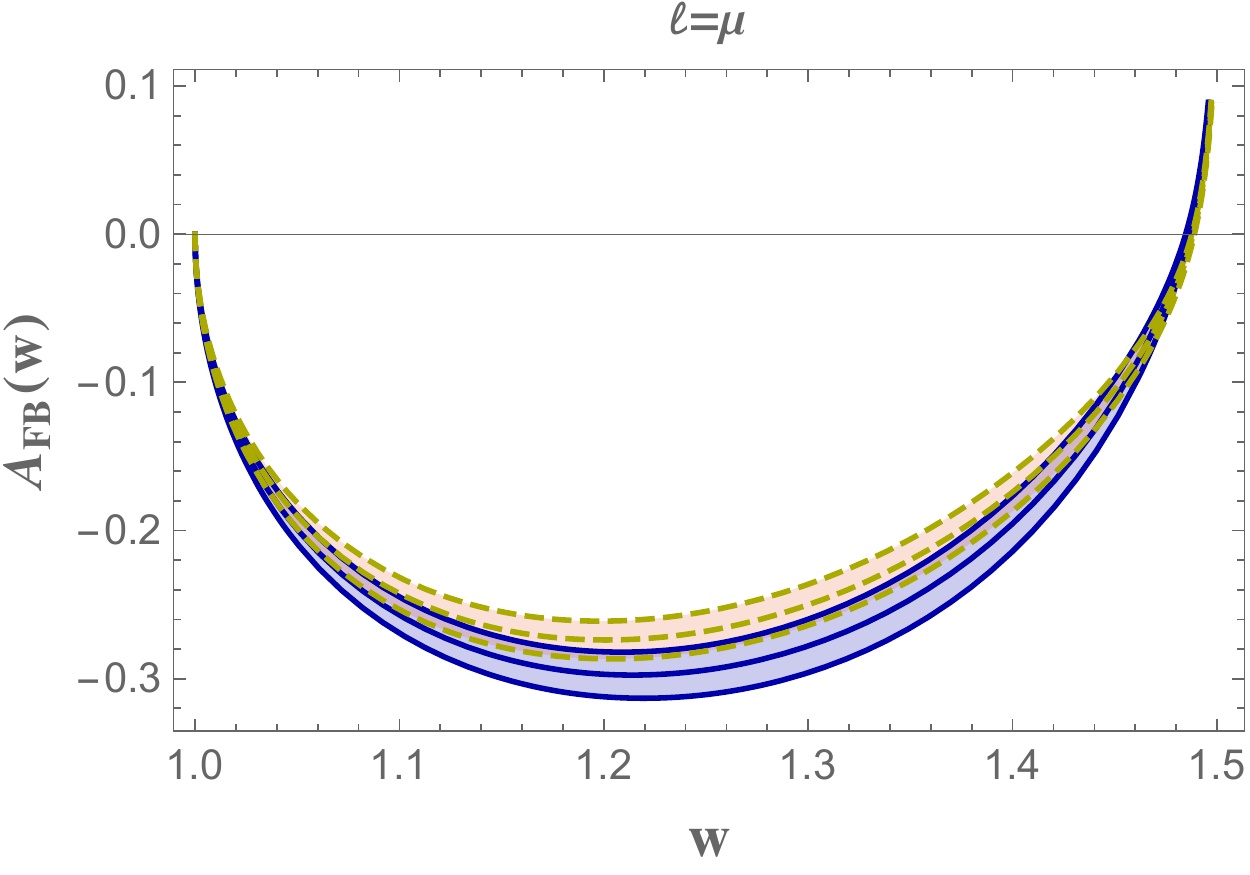}
\includegraphics[width = 0.32\textwidth]{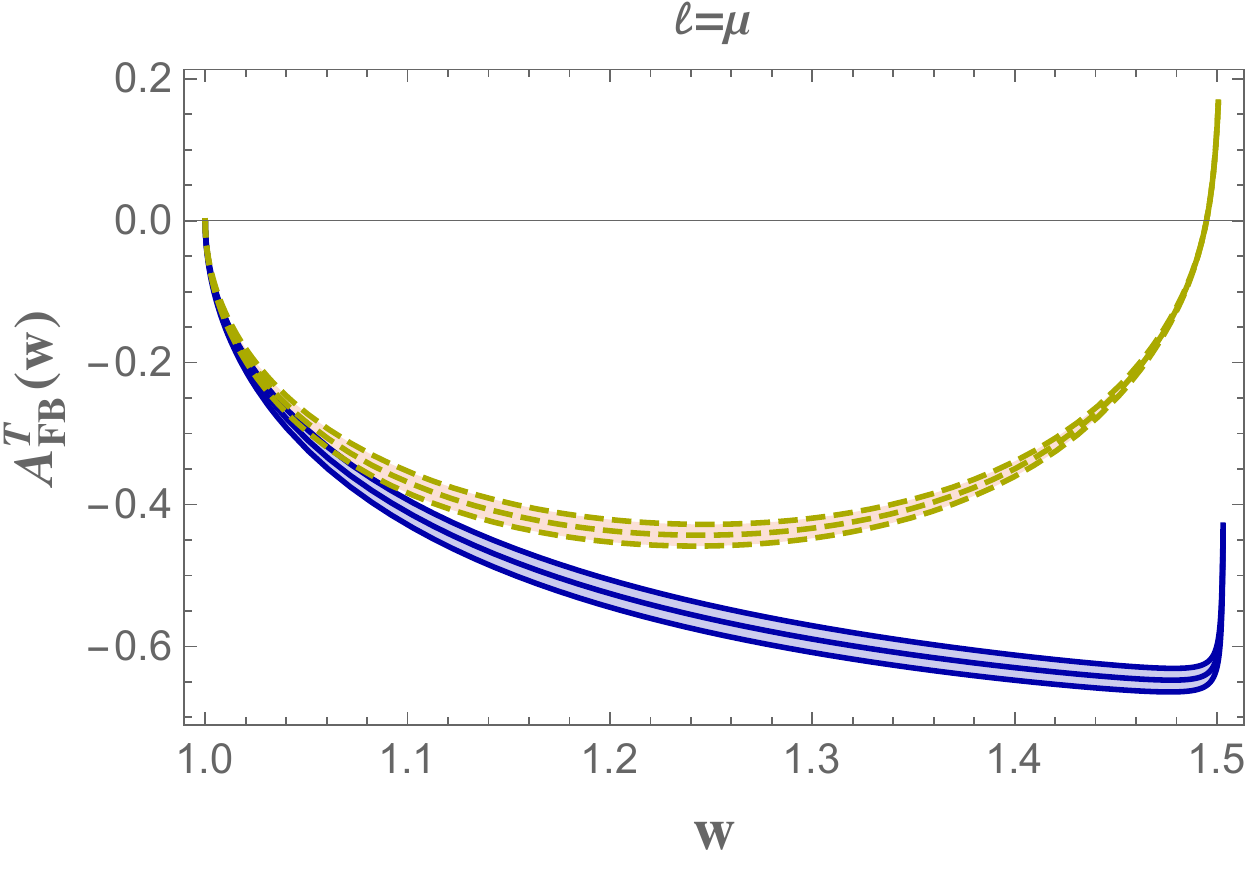}
\includegraphics[width = 0.32\textwidth]{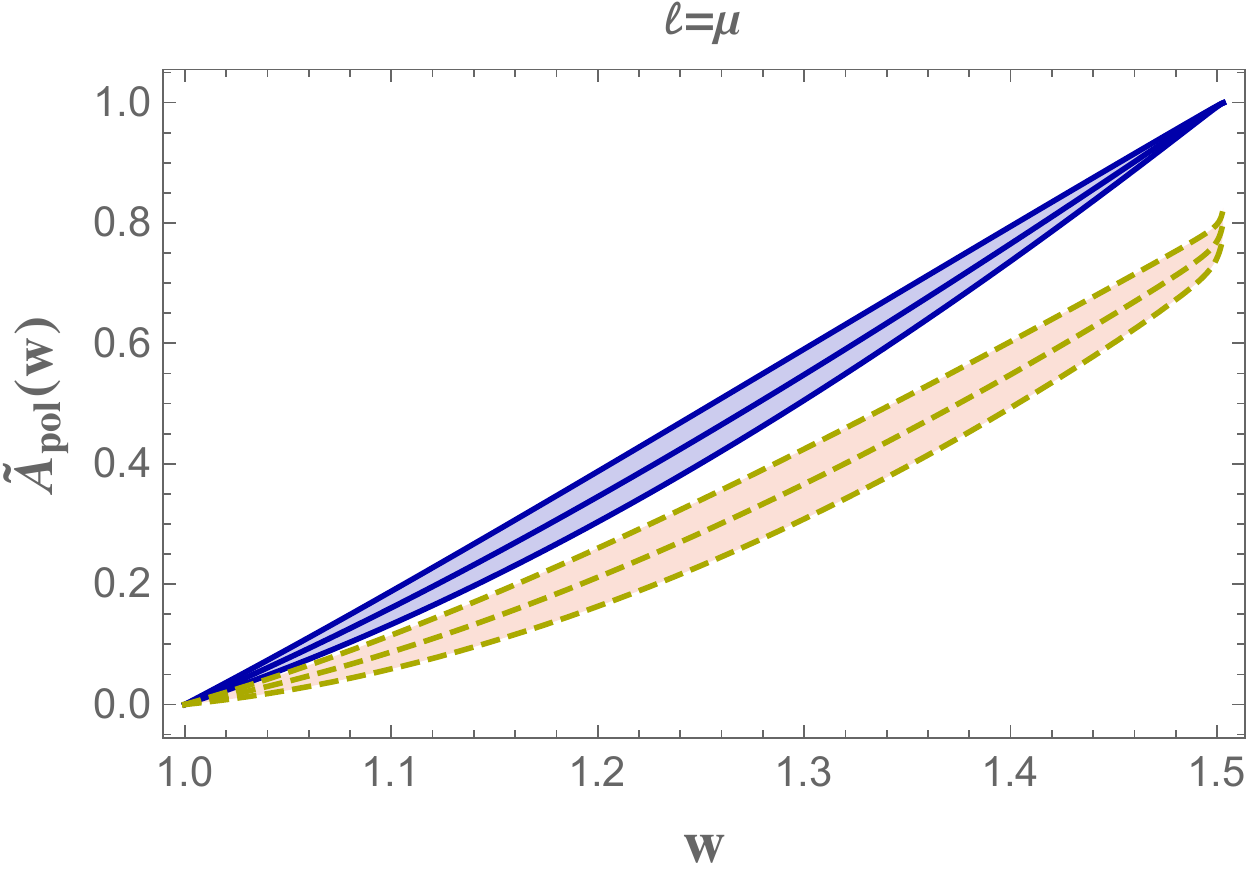}\\
\includegraphics[width =0.32 \textwidth]{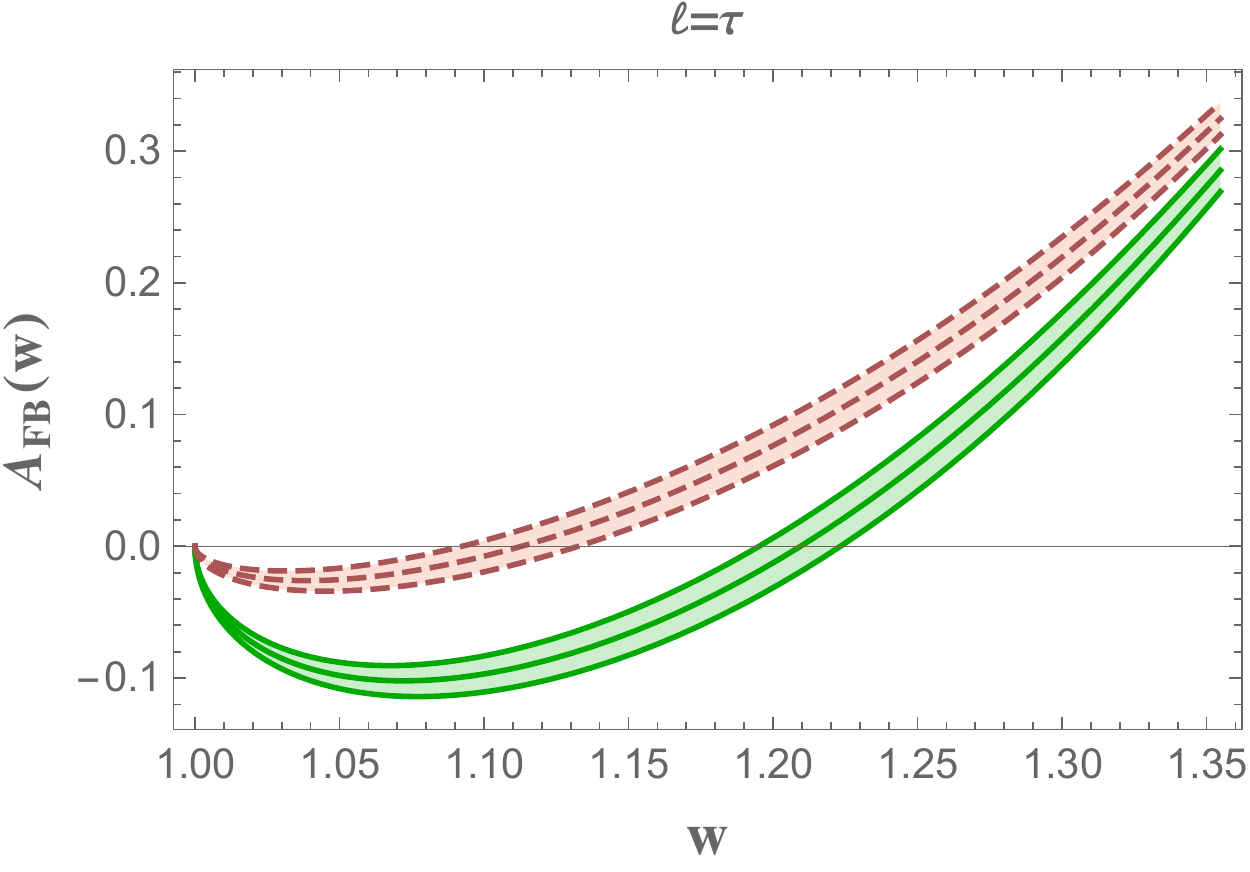}
\includegraphics[width = 0.32\textwidth]{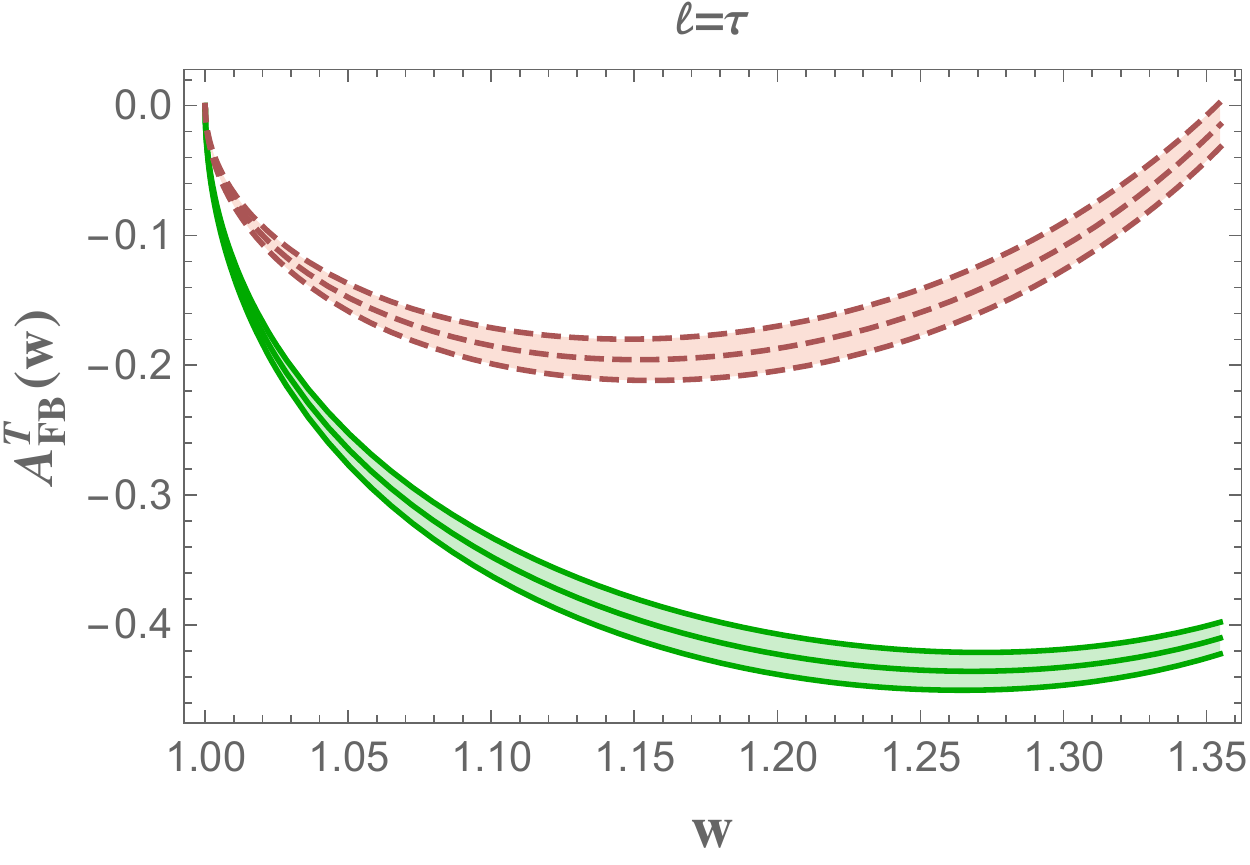}
\includegraphics[width = 0.32\textwidth]{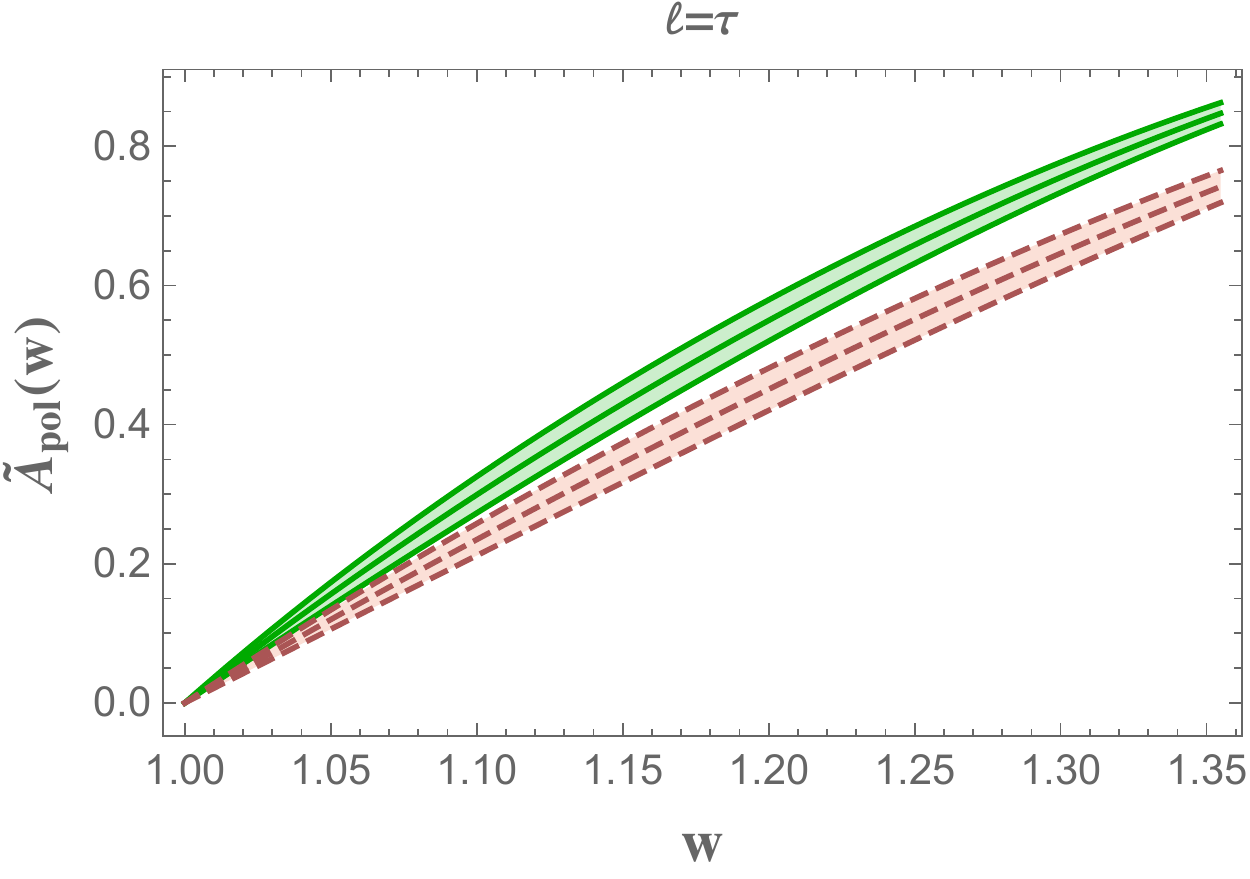}
    \caption{\baselineskip 10pt  Observables defined in Eq.~(\ref{AFB}) (left column),  (\ref{AFBT}) (middle) and   (\ref{Apol}) (right).
    The upper and lower plots refer to $\ell=\mu$ and $\ell=\tau$, respectively.  The solid curves correspond to  SM, the dashed ones to NP at the benchmark point   ${\tilde \epsilon}_T^\ell$.}\label{obs}
\end{center}
\end{figure}

 Starting from  the set of angular coefficient functions, several observables can be constructed to scrutinize  SM and possible anomalies.
A few observables  are independent of the $D^*$ decay mode.
\begin{itemize}
\item The {\bf  $q^2$-dependent forward-backward (FB) lepton asymmetry} is defined as\\
\be
A_{FB}(q^2)=\left[\int_0^1 \, dcos \, \theta \, \displaystyle{\frac{d^2 \Gamma}{dq^2 dcos \, \theta}} -\int_{-1}^0 \, dcos \, \theta \, \displaystyle{\frac{d^2 \Gamma}{dq^2 dcos \, \theta}} \right]\big/{\displaystyle{\frac{d \Gamma}{dq^2}}}  \,\,\, . \label{AFB}
\ee
It can be expressed in terms of the coefficient functions:
\be
A_{FB}(q^2)=\frac{3(I_{6c}^\pi+2I_{6s}^\pi)}{6I_{1c}^\pi+12 I_{1s}^\pi-2I_{2c}^\pi-4I_{2s}^\pi}
=
\frac{3(I_{6s}^\gamma+4I_{6c}^\gamma)}{6I_{1s}^\gamma+24 I_{1c}^\gamma-2I_{2s}^\gamma-8I_{2c}^\gamma} \,\,\, , \label{AFBangular}
\ee
and in SM   in terms of the helicity amplitudes  
\be
A_{FB}(q^2)|_{SM}=\frac{3q^2 \left( H_+^2-H_-^2\right)-6m_\ell^2 H_0 H_t}{ 2 m_\ell^2 \left(H_0^2+3H_t^2+H_+^2+H_-^2 \right) +4 q^2 \left(H_0^2+H_+^2+H_-^2 \right)} \,\,\, .
\ee
\item The { \bf transverse forward-backward (TFB) asymmetry} 
is the  FB asymmetry  for transversely polarized $D^*$.
In  SM it is expressed in terms of the helicity amplitudes 
\be
A_{FB}^T(q^2)|_{SM}=\frac{3q^2\left(H_+^2-H_-^2 \right)}{2(m_\ell^2+2q^2)\left(H_+^2+H_-^2 \right)}\,\, .\label{AFBT}
\ee
$A_{FB}^T$ only depends  on the form factor ratio $R_1$, hence it is useful to check the HQ prediction for such a quantity \cite{Neubert:1993mb}.
\item{ \bf $D^*$ polarization  asymmetry.}
Defining the distributions ${d \Gamma_{L(T)}}/{dq^2}$ for  longitudinally (L) and transversely (T) polarized $D^*$, a 
 polarization asymmetry can be defined:
\be
\frac{dA_{pol}^{D^*}(q^2)}{dq^2}=2 \displaystyle\frac{d \Gamma_{L}}{dq^2} \big/\displaystyle\frac{d \Gamma_{T}}{dq^2}-1 \,\, .
\ee
A combination regular at $w \to w_{max}$ is
\be
{\tilde A}_{pol}^{D^*}(q^2)=\frac{\displaystyle{\frac{dA_{pol}^{D^*}(q^2)}{dq^2}}}{1+\displaystyle{\frac{dA_{pol}^{D^*}(q^2)}{dq^2}}}\,\,. \label{Apol}
\ee
In  SM this quantity  is expressed in terms of the helicity amplitudes:
\be
{\tilde A}_{pol}^{D^*}(q^2)|_{SM}=1-\frac{(m_\ell^2+2q^2)\left(H_+^2+H_-^2 \right)}{6m_\ell^2 H_t^2+2(m_\ell^2+2q^2)H_0^2} \,\,\, .
\ee
\end{itemize}
In figure \ref{obs} we depict the SM results for these observables and the  NP ones obtained at the benchmark point   ${\tilde \epsilon}_T^\mu$ and ${\tilde \epsilon}_T^\tau$. 
The  SM results are systematically modified in  NP; in particular,  a zero appears in $A_{FB}(w)$    when $\ell=\tau$  \cite{Biancofiore:2013ki}.
  ${\tilde A}_{pol}^{D^*}$ shows a high sensitivity to the tensor structure for $w \to w_{max}$, in particular in the case $\ell=\mu$, as noticed in \cite{Jung:2018lfu}, since in  SM, for $m_\ell \to 0$, $D^*$ is  fully longitudinally polarized at this kinematical point.

\begin{figure}[t]
\begin{center}
\includegraphics[width =0.45 \textwidth]{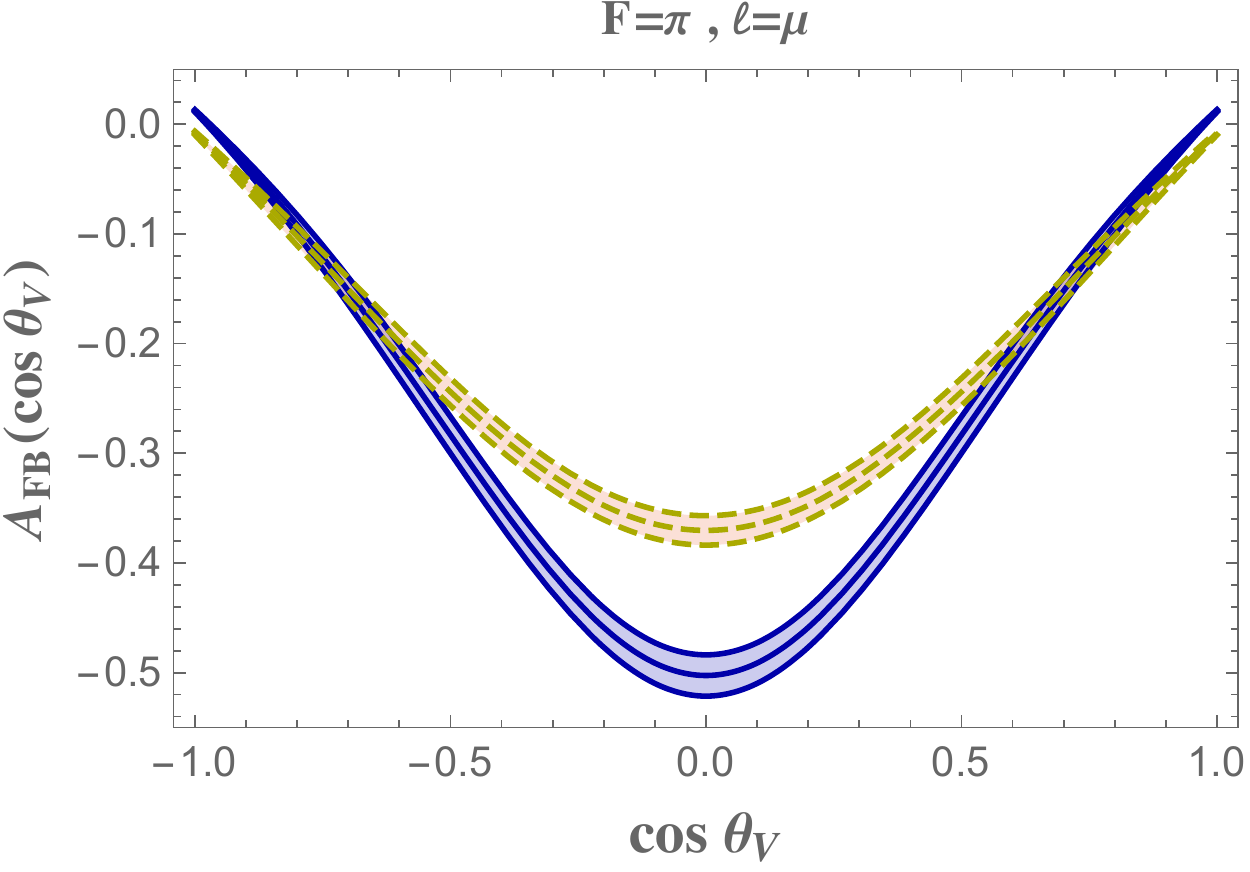}
\includegraphics[width = 0.45\textwidth]{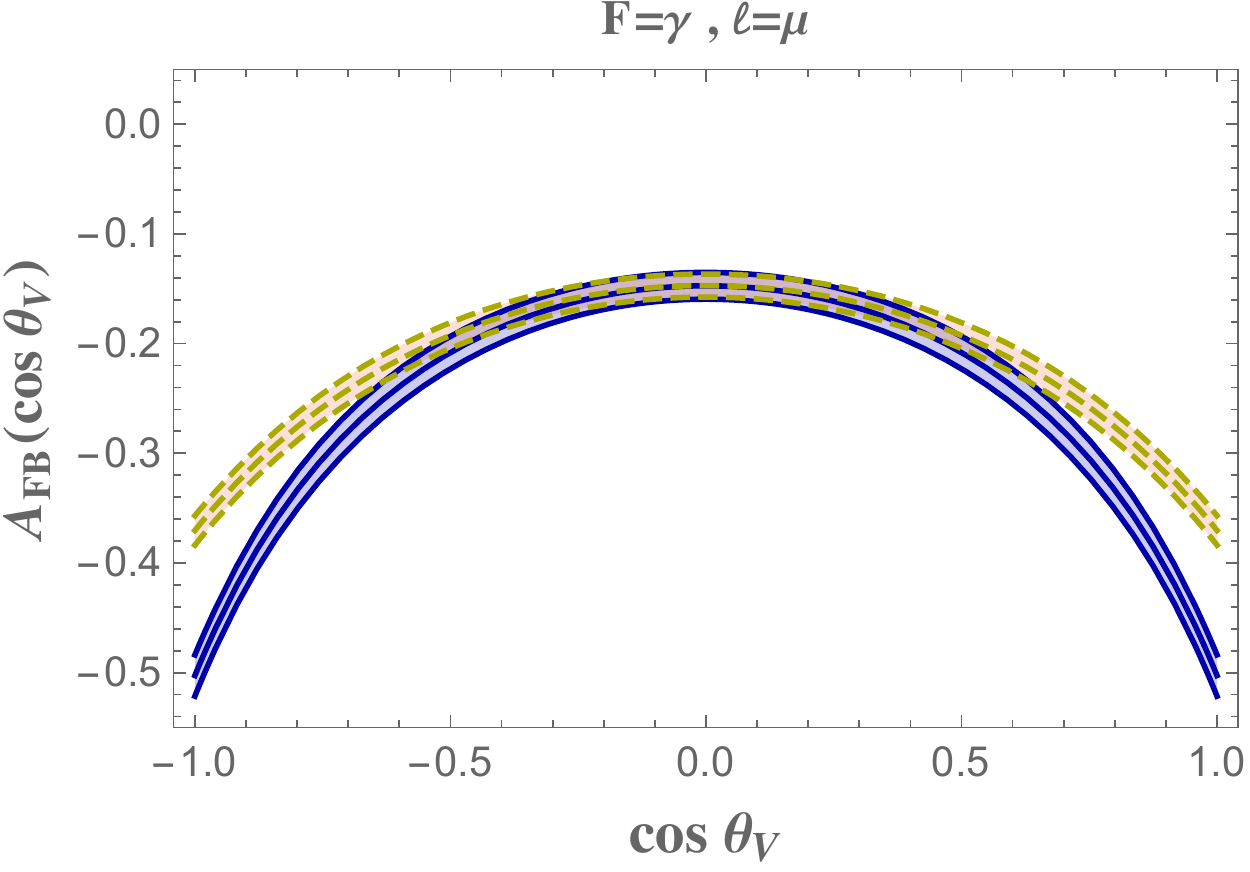}\\
\vskip0.2cm
\includegraphics[width =0.45 \textwidth]{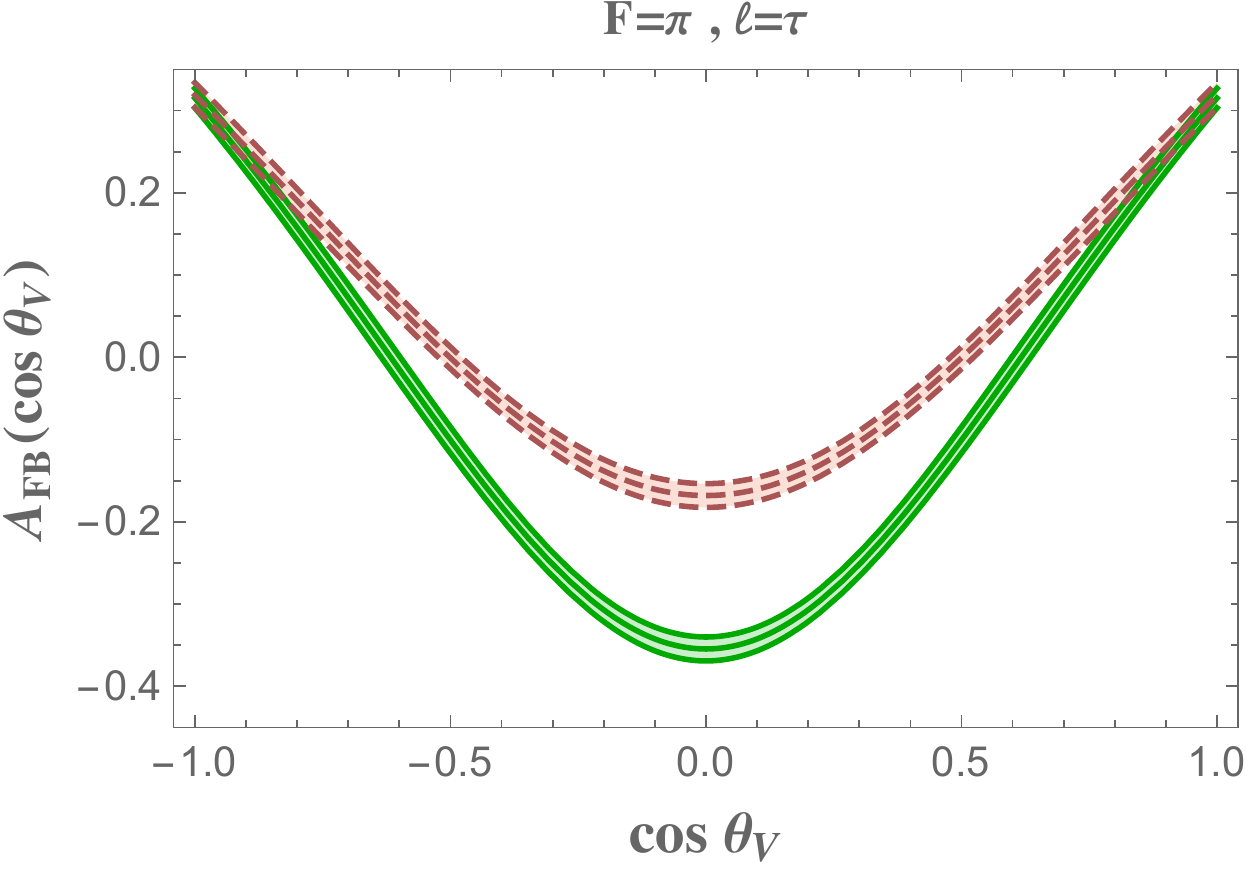}
\includegraphics[width = 0.45\textwidth]{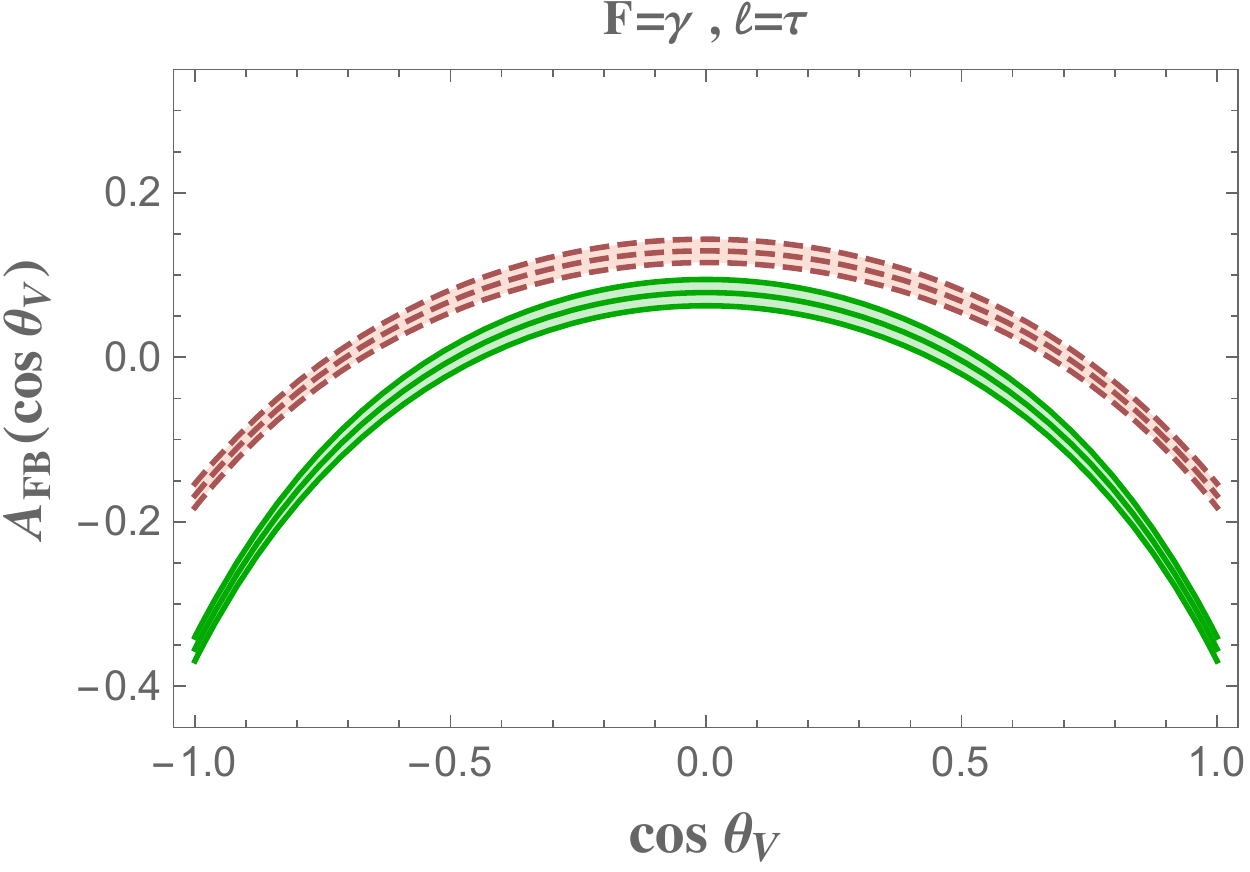}
    \caption{\baselineskip 10pt $\cos \theta_V$-dependent forward-backward asymmetry  defined in (\ref{AFBzV}).
    The upper and lower plots refer to $\ell=\mu$ and  $\ell=\tau$, respectively,  the  left and right column to $F=\pi$ and $F=\gamma$. The solid curves correspond to  SM, the dashed ones to NP at the benchmark point   ${\tilde \epsilon}_T^\ell$.
     }\label{zVobs}
\end{center}
\end{figure}

An observable  different when the final state involves a pion $F=\pi$ or a photon $F=\gamma$ is the
\begin{itemize}
\item { \bf $\cos \theta_V$-dependent forward-backward asymmetry}, defined as 
\be
A_{FB}(cos \theta_V)=\frac{\left[\int_0^1 \, dcos \, \theta \, \displaystyle{\frac{d^2 \Gamma}{dcos \theta_V dcos \, \theta}} -\int_{-1}^0 \, dcos \, \theta \, \displaystyle{\frac{d^2 \Gamma}{dcos \theta_V dcos \, \theta}}\right] }{\displaystyle{\frac{d \Gamma}{dcos \theta_V}}} \,\,. \label{AFBzV}
\ee
\end{itemize}
Figure \ref{zVobs} shows the result in  SM compared to   NP  for  ${\tilde \epsilon}_T^\ell$. 
The deviation from SM is largest for $\cos \theta_V \simeq 0$ in the case of pion,  and for $\cos \theta_V \simeq \pm 1$ when $F=\gamma$.

\begin{figure}[t]
\begin{center}
\includegraphics[width = 0.45\textwidth]{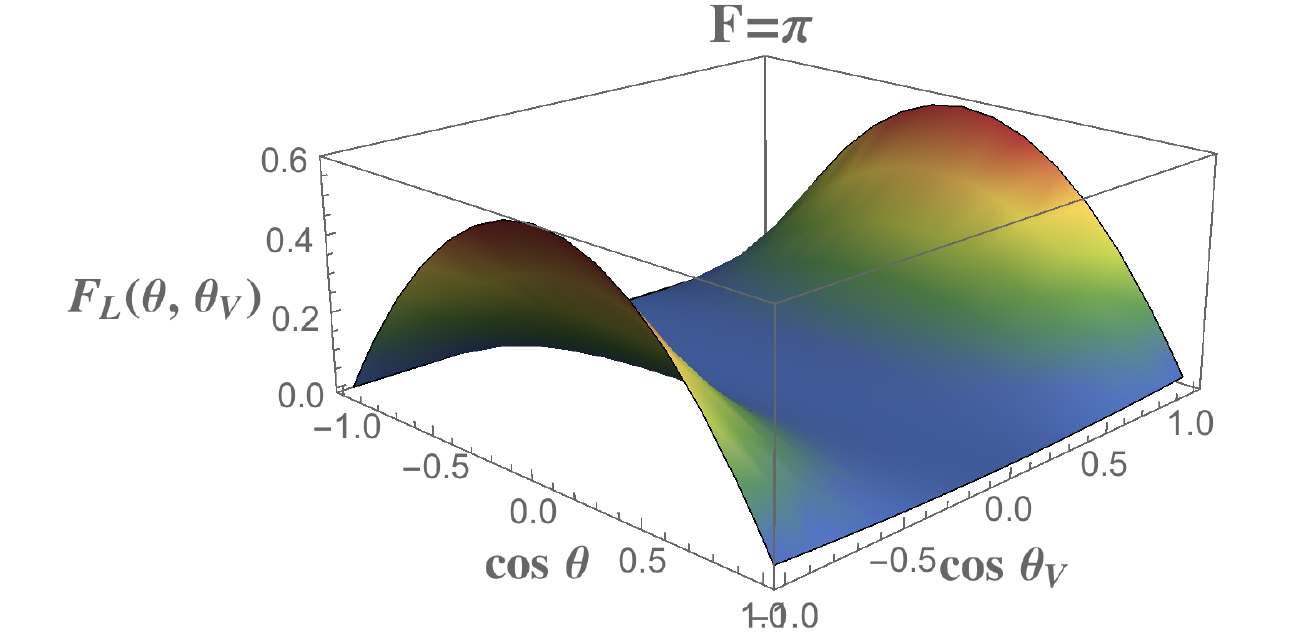}
\includegraphics[width = 0.45\textwidth]{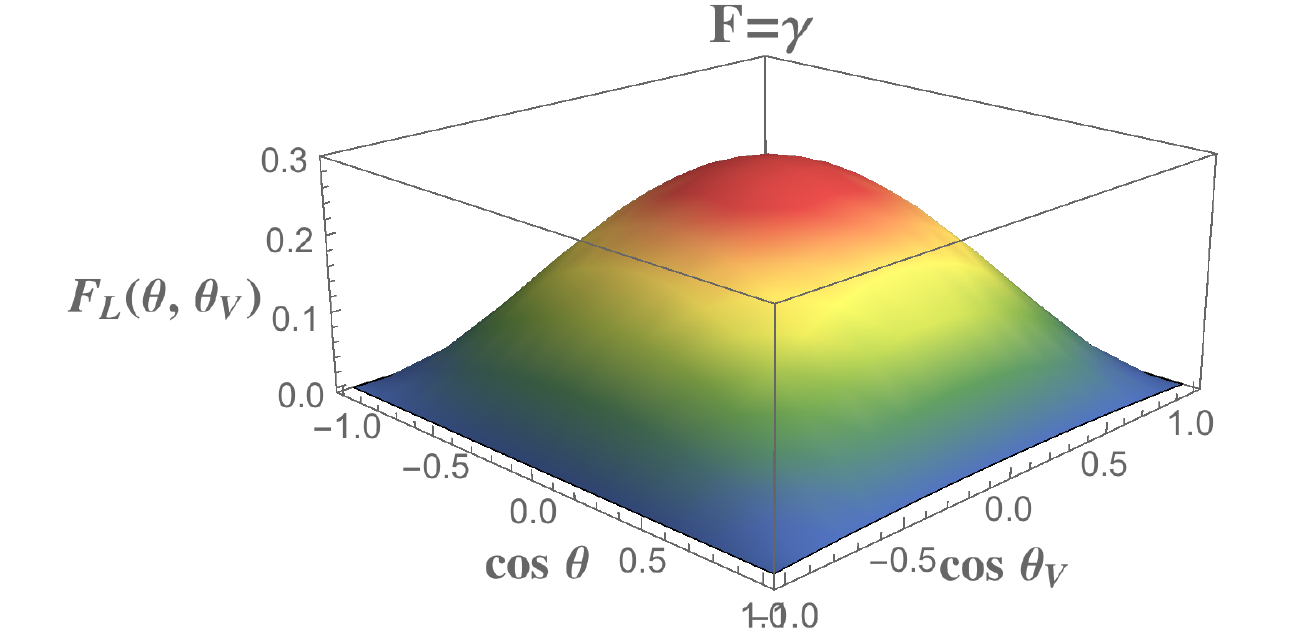}\\
\vskip 0.3cm
\includegraphics[width = 0.45\textwidth]{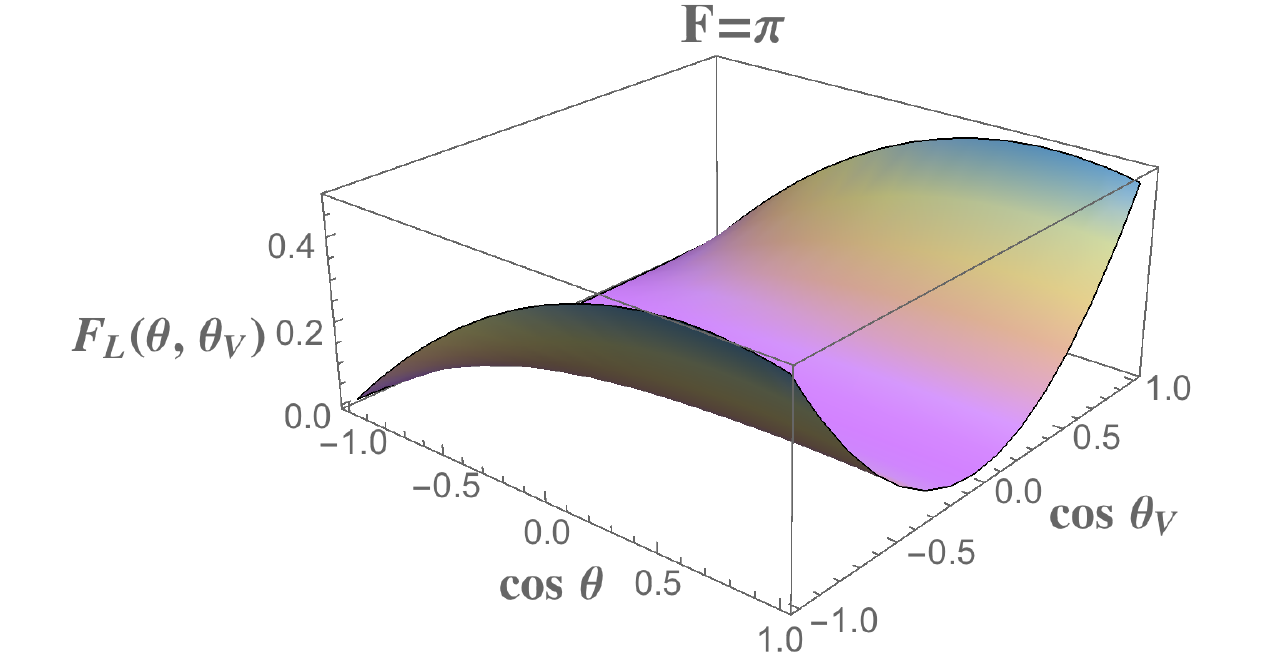}
\includegraphics[width = 0.45\textwidth]{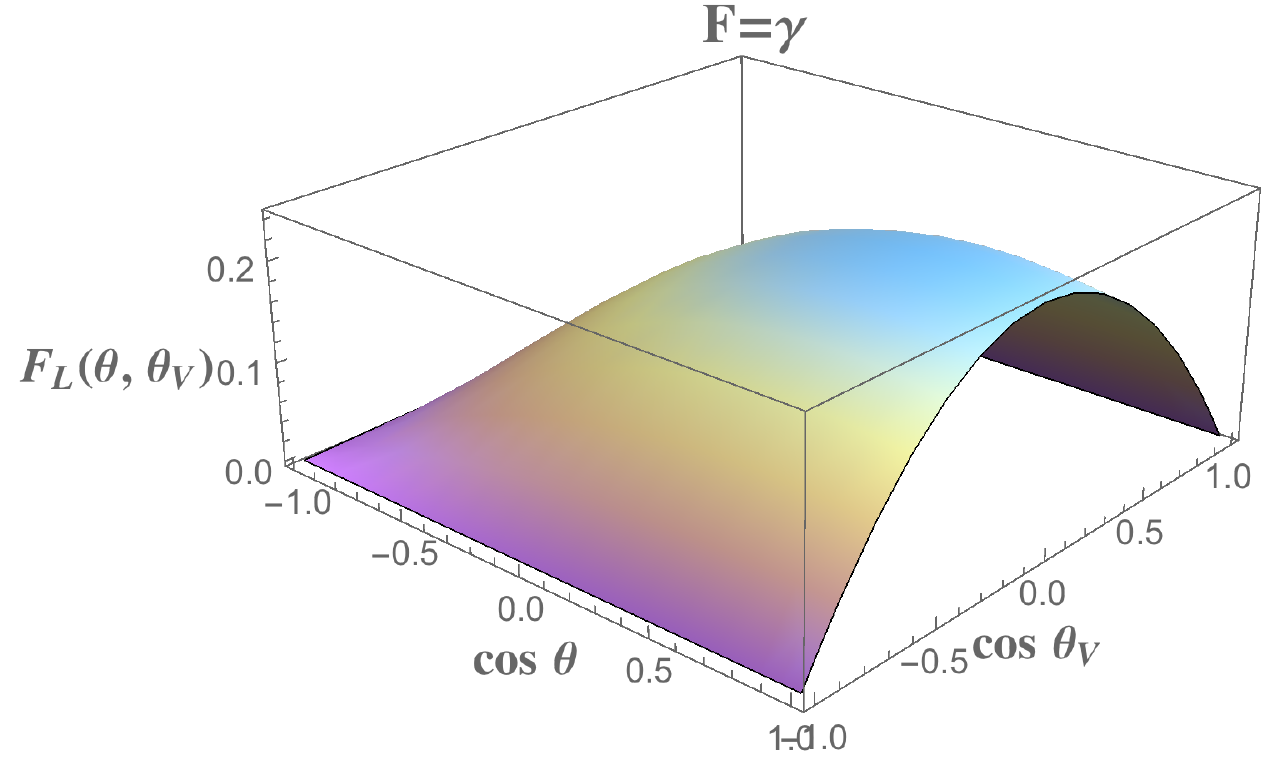}
\caption{\baselineskip 10pt SM distributions $F_L(\theta,\theta_V)$ defined in Eq.~(\ref{FL}). Upper and lower plots refer to $\ell=\mu$ and   $\ell=\tau$, respectively, the left and right column  to  $F=\pi$ and  $F=\gamma$.
}\label{3DFLSM}
\end{center}
\end{figure}

\begin{figure}[t]
\begin{center}
\includegraphics[width = 0.45\textwidth]{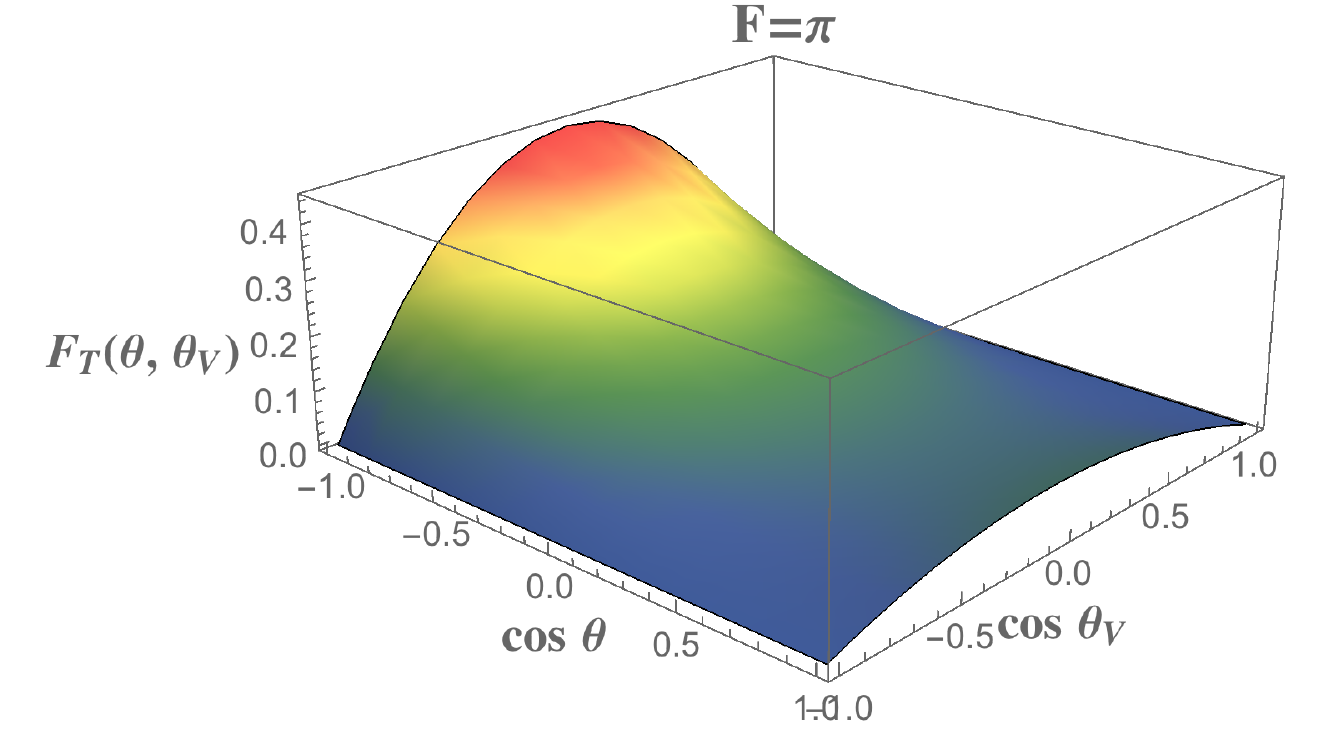}
\includegraphics[width = 0.45\textwidth]{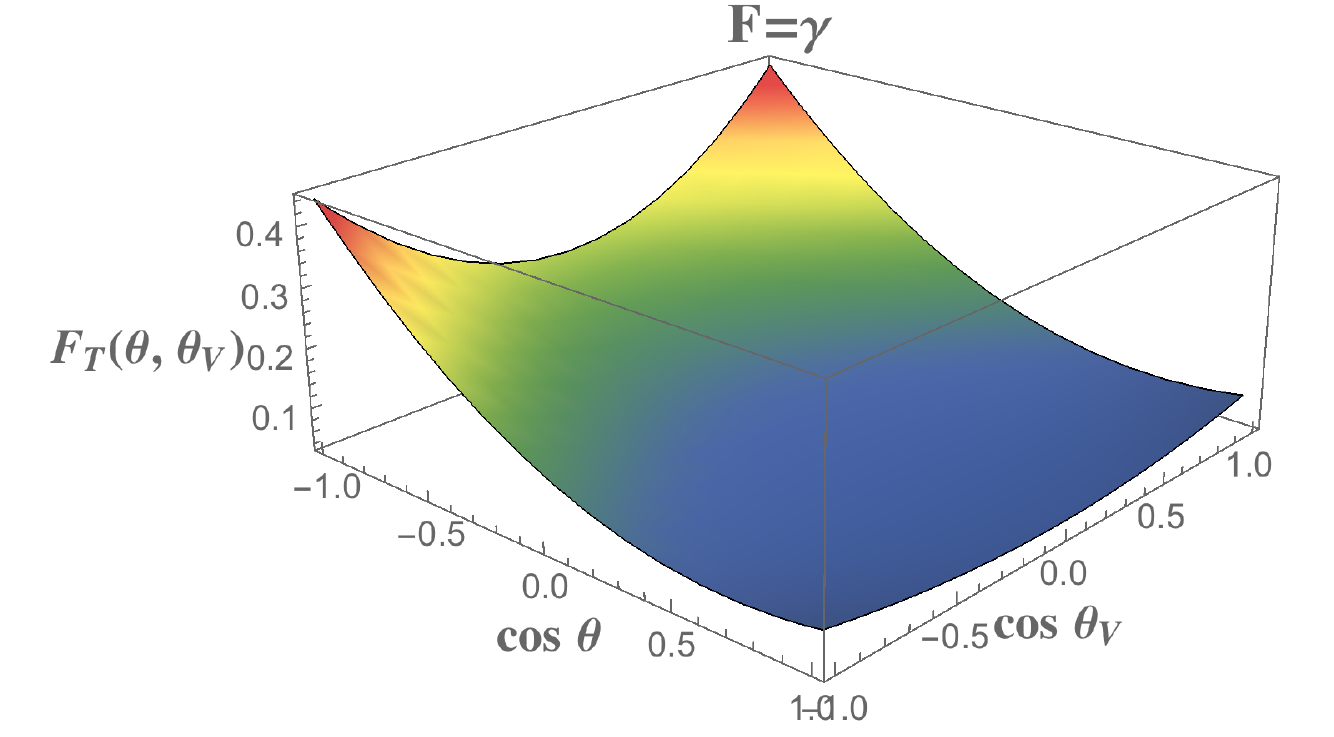}
\\
\vskip 0.3cm
\includegraphics[width = 0.45\textwidth]{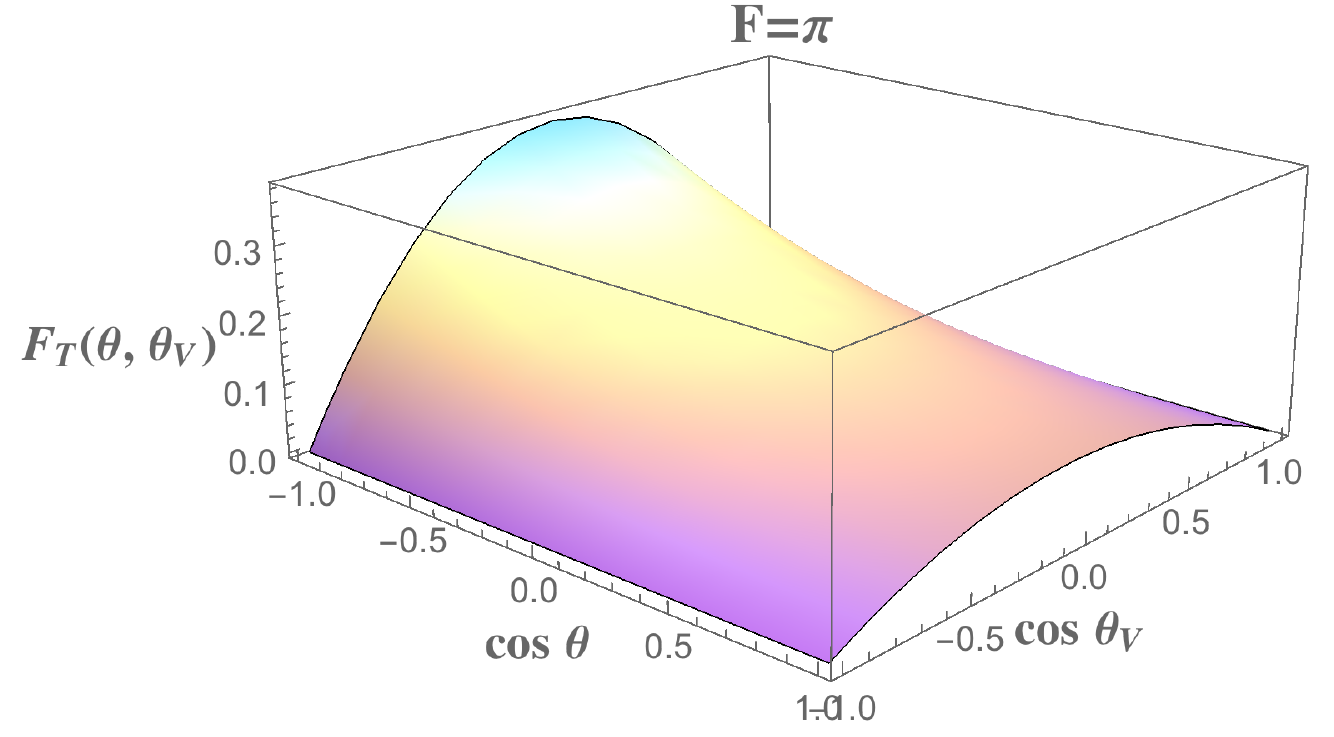}
\includegraphics[width = 0.45\textwidth]{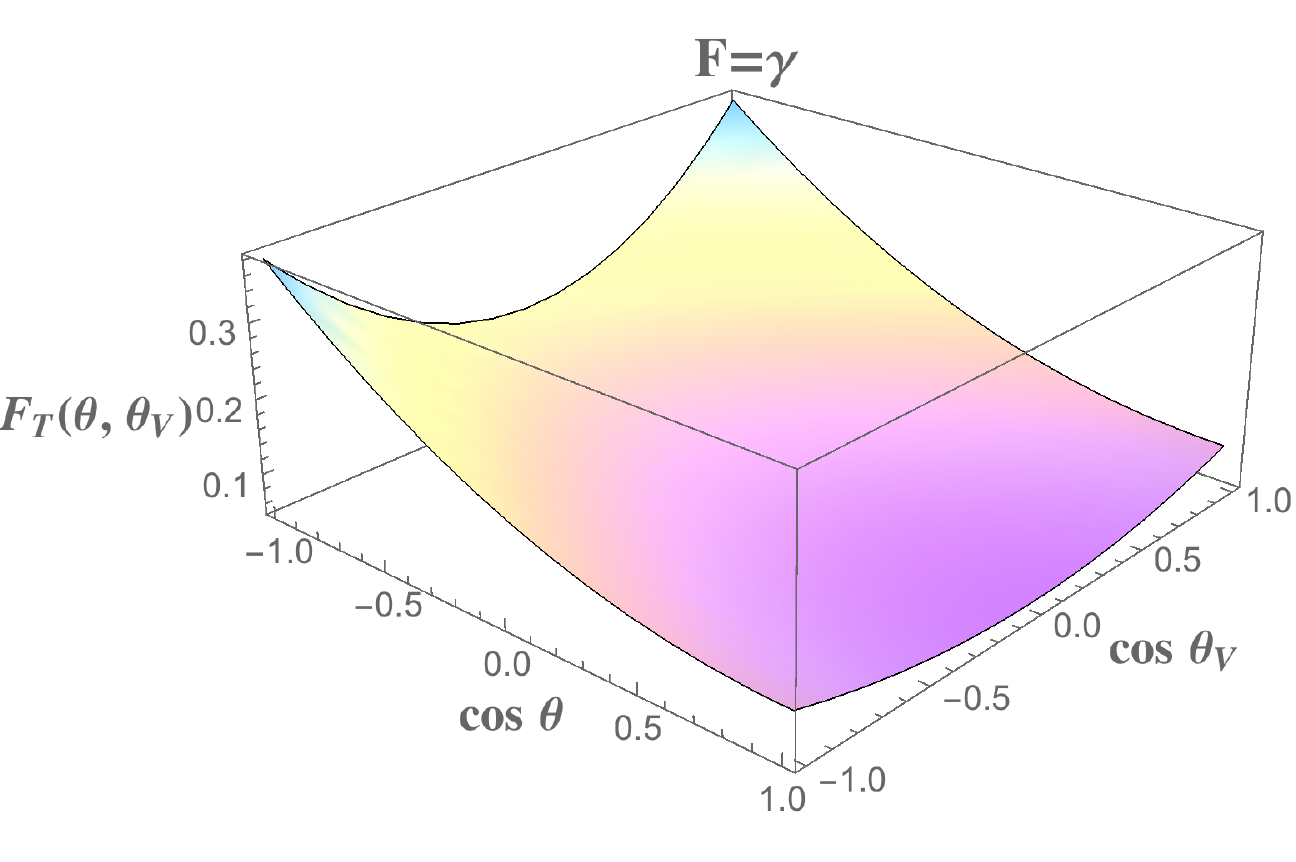}
\caption{\baselineskip 10pt SM distributions $F_T(\theta,\theta_V)$  defined in Eq.~(\ref{FT}). Upper and lower plots refer to  $\ell=\mu$ and  $\ell=\tau$, respectively, the left and right column  to  $F=\pi$ and  $F=\gamma$. }\label{3DFTSM}
\end{center}
\end{figure}

The sensitivity of the angular distributions to the $D^*$ polarization can be studied
considering  the triple differential distributions obtained from (\ref{angularpi}) and (\ref{angulargamma}) after integration in the angle $\phi$.
When $D^*$ is longitudinally polarized one has 
\bea
\frac{d^3 \Gamma_L}{dq^2d \cos \theta_V d \cos \theta}\Big|_{F=\pi}&=&{\cal N}_\pi |{\vec p}_{D^*}|\left(1-  \frac{ m_\ell^2}{q^2}\right)^2 \, 2 \pi \left[I_{1c}^\pi  +I_{2c}^\pi  \cos 2 \theta +I_{6c}^\pi \cos \theta \right] \,\cos \theta_V^2
\label{longpi} \\
\frac{d^3 \Gamma_L}{dq^2d \cos \theta_V d\cos \theta}\Big|_{F=\gamma}&=&{\cal N}_\gamma |{\vec p}_{D^*}|\left(1-  \frac{ m_\ell^2}{q^2}\right)^2 \, 2 \pi \left[I_{1s}^\gamma  +I_{2s}^\gamma  \cos 2 \theta +I_{6s}^\gamma \cos \theta \right] \,\sin \theta_V^2 , \label{longgamma}
\eea
and for transversely polarized $D^*$ (summing over  the two transverse polarizations)
\bea
\frac{d^3 \Gamma_T}{dq^2d \cos \theta_V d\cos \theta}\Big|_{F=\pi}&=&{\cal N}_\pi |{\vec p}_{D^*}|\left(1-  \frac{ m_\ell^2}{q^2}\right)^2 \, 2 \pi \left[I_{1s}^\pi  +I_{2s}^\pi  \cos 2 \theta +I_{6s}^\pi \cos \theta \right] \,\sin \theta_V^2 
\label{transpi} \\
\frac{d^3 \Gamma_T}{dq^2d \cos \theta_V d\cos \theta}\Big|_{F=\gamma}&=&{\cal N}_\gamma |{\vec p}_{D^*}|\left(1-  \frac{ m_\ell^2}{q^2}\right)^2 \, 2 \pi \left[I_{1c}^\gamma  +I_{2c}^\gamma  \cos 2 \theta +I_{6c}^\gamma \cos \theta \right] \,(3+ \cos  2 \theta_V). \,\,\,\,\,\,\,\,\,\,\, \label{transgamma}
\eea
Double  differential $D^*$ polarization fractions can be defined:
\bea
F_L(\theta,\,  \theta_V)&=&\frac{1}{\Gamma (\bar B \to D^* (D F) \ell ^- {\bar \nu}_\ell)}\int_{q^2_{min}}^{q^2_{max}}\, dq^2
\frac{d^3 \Gamma_L}{dq^2d \cos \theta_V d\cos \theta} \label{FL} \\
F_T(\theta,\, \theta_V)&=&\frac{1}{\Gamma (\bar B \to D^* (D F) \ell ^- {\bar \nu}_\ell)}\int_{q^2_{min}}^{q^2_{max}}\, dq^2
\frac{d^3 \Gamma_T}{dq^2d \cos \theta_V d\cos \theta}\,\,. \label{FT} \eea
These quantities keep the same angular dependence as in (\ref{longpi})-(\ref{transgamma}). In particular, they are simmetric under $\cos \theta_V \to -\cos \theta_V$, but they have no definite behavior when $\cos \theta \to -\cos \theta$, since the first two terms are  invariant under this transformation, while the last one changes sign. 
In  $F_L$  this term involves the angular coefficient $I_{6c}^\pi \, (I_{6s}^\gamma)$ proportional to the lepton mass: therefore,  the  distribution is expected  to be nearly symmetric when $\cos \theta \to - \cos \theta$ in the muon case, not for $\tau $. For SM this is shown in fig.~\ref{3DFLSM}. The analogous plots for $F_T$ are shown in figure ~\ref{3DFTSM}.

\begin{figure}[t]
\begin{center}
\includegraphics[width = 0.45\textwidth]{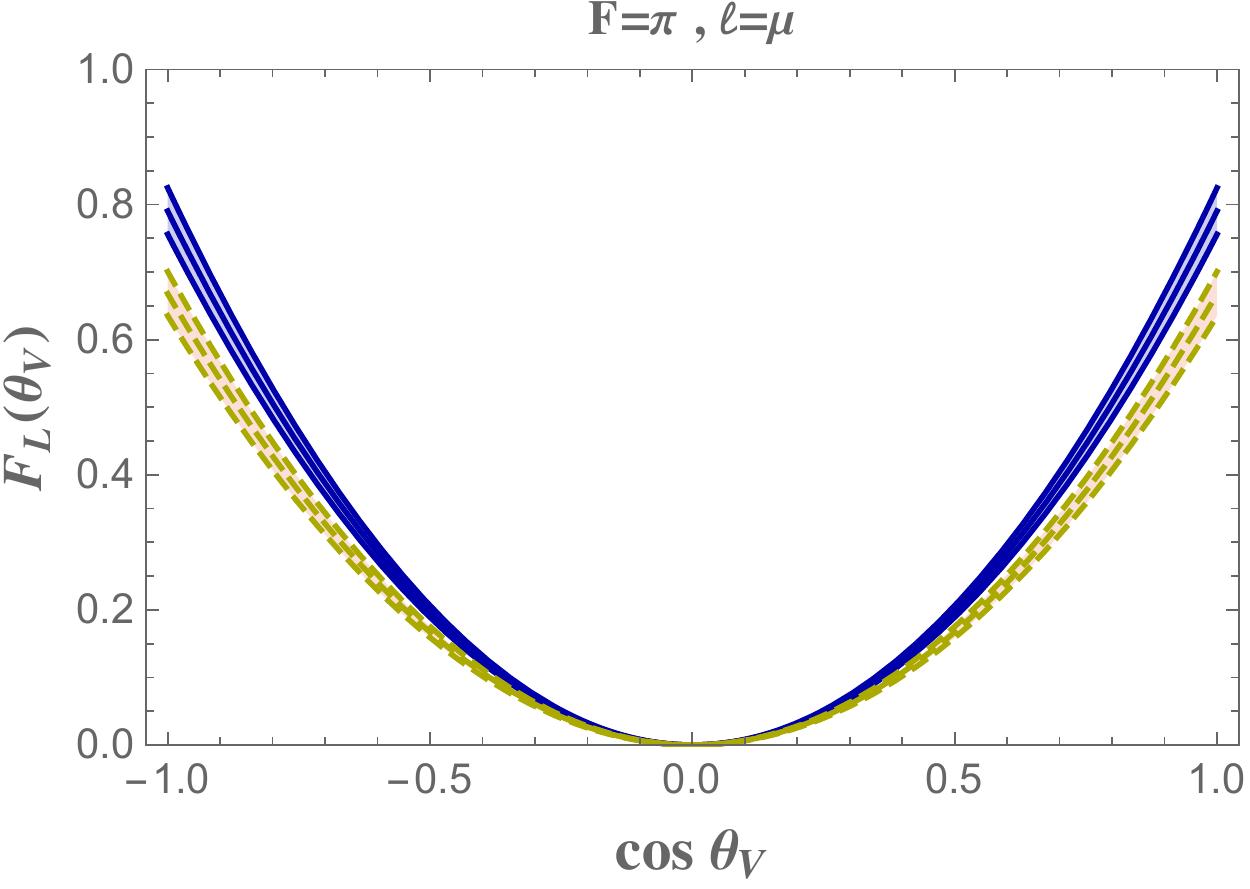}
\includegraphics[width = 0.45\textwidth]{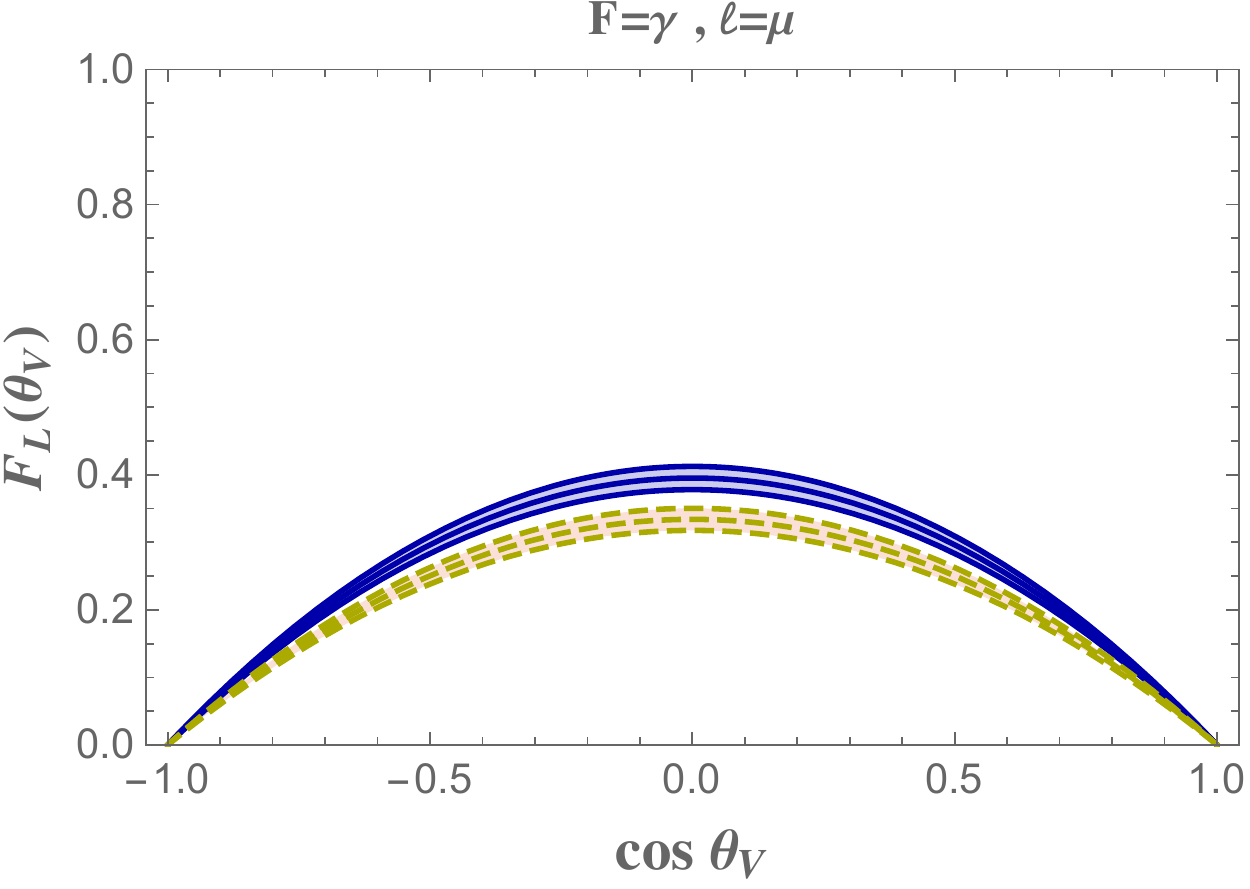}
\\
\vskip 0.3cm
\includegraphics[width = 0.45\textwidth]{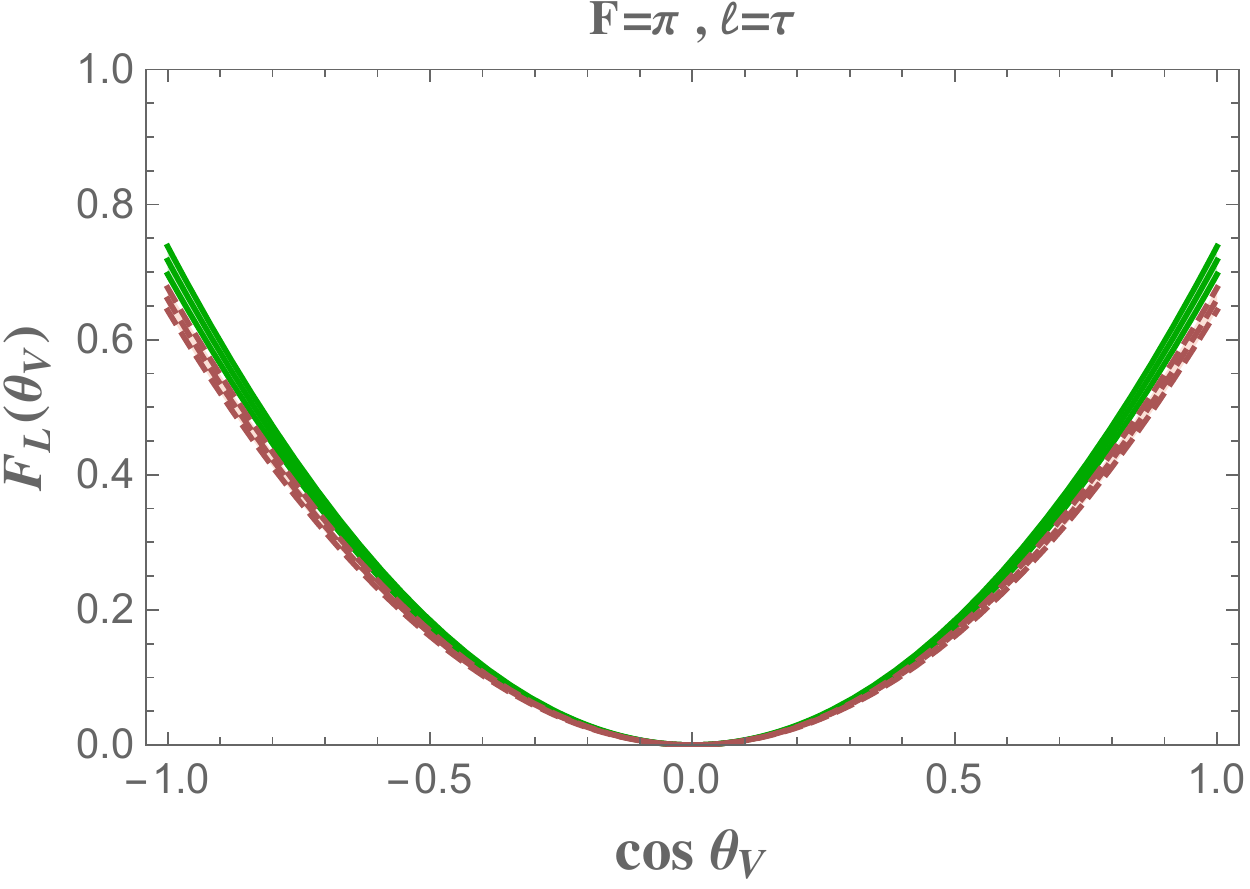}
\includegraphics[width = 0.45\textwidth]{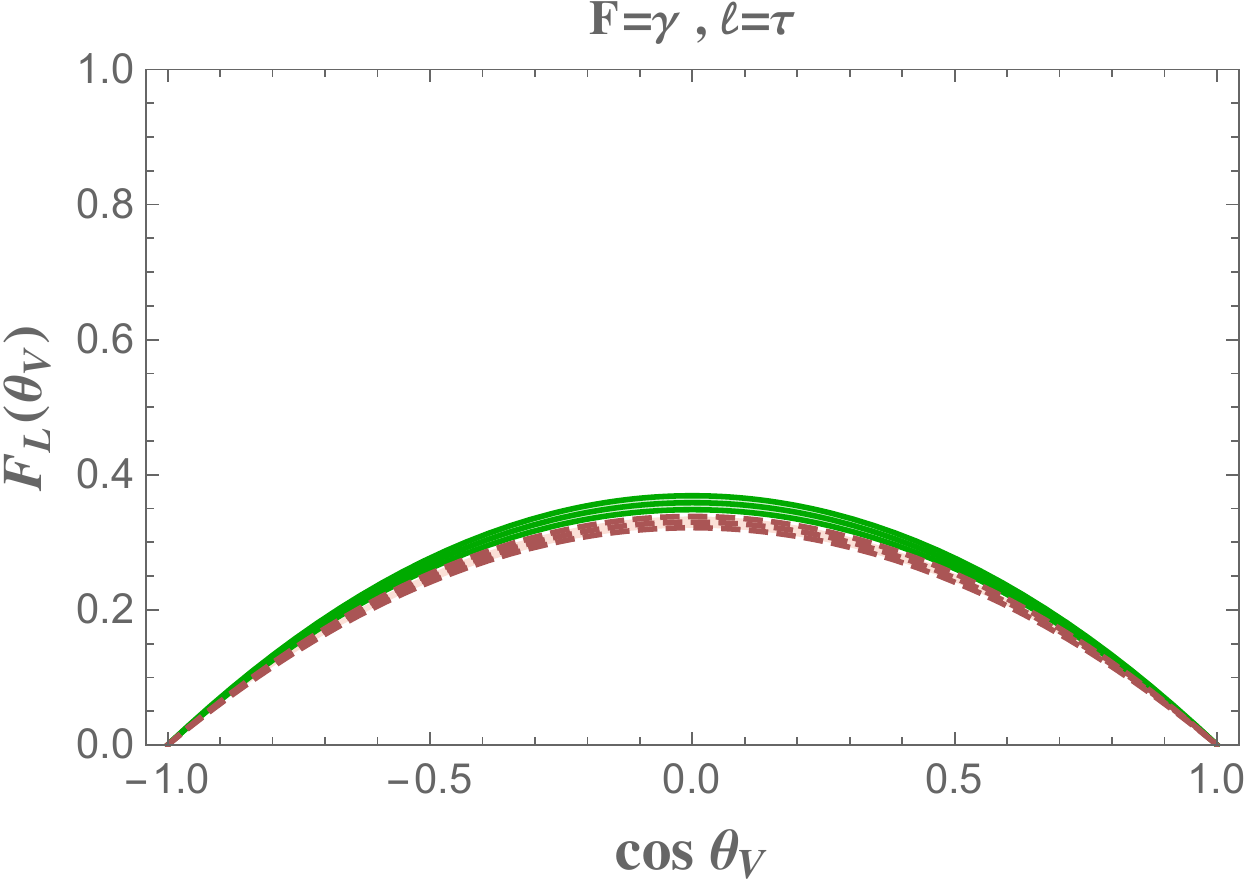}
    \caption{\baselineskip 10pt Distribution $F_L(\theta_V)$. The upper and lower plots refer to  $\ell=\mu$ and  $\ell=\tau$, the left and right column  to  $F=\pi$ and   $F=\gamma$. The continuous lines show the SM result, the dashed lines the NP result at the benchmark point $\tilde \epsilon_T^\ell$. }\label{FLthetaV}
\end{center}
\end{figure}

\begin{figure}[t]
\begin{center}
\includegraphics[width = 0.45\textwidth]{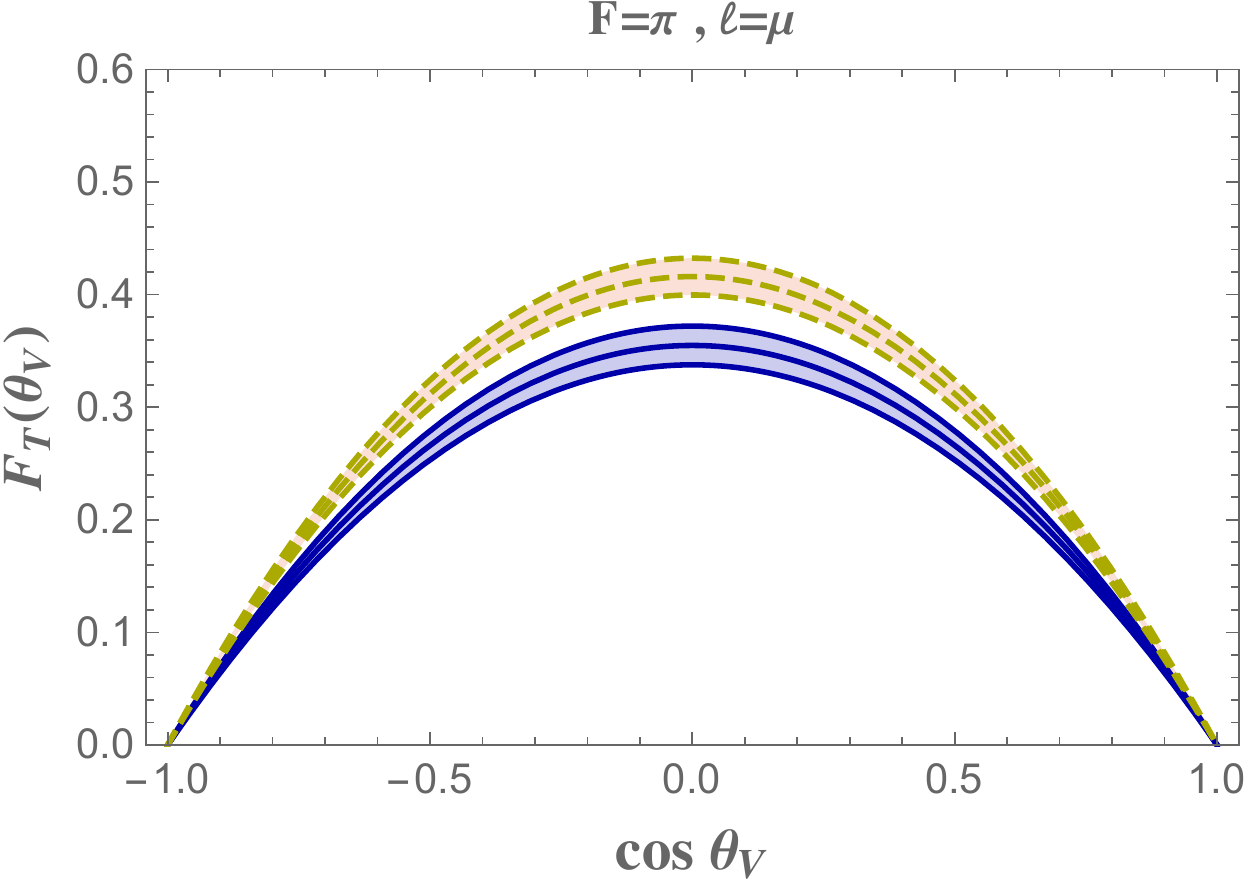}
\includegraphics[width = 0.45\textwidth]{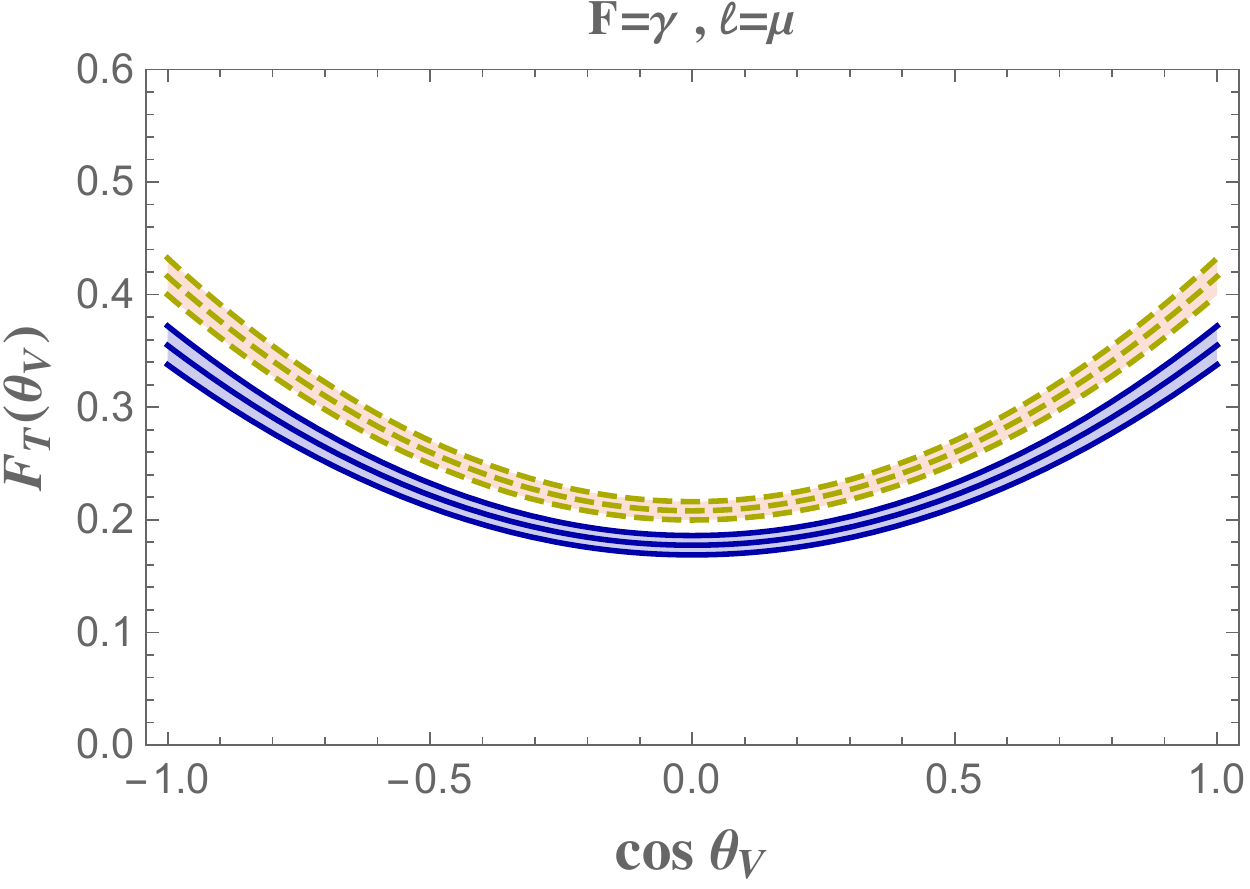}
\\
\vskip 0.3cm
\includegraphics[width = 0.45\textwidth]{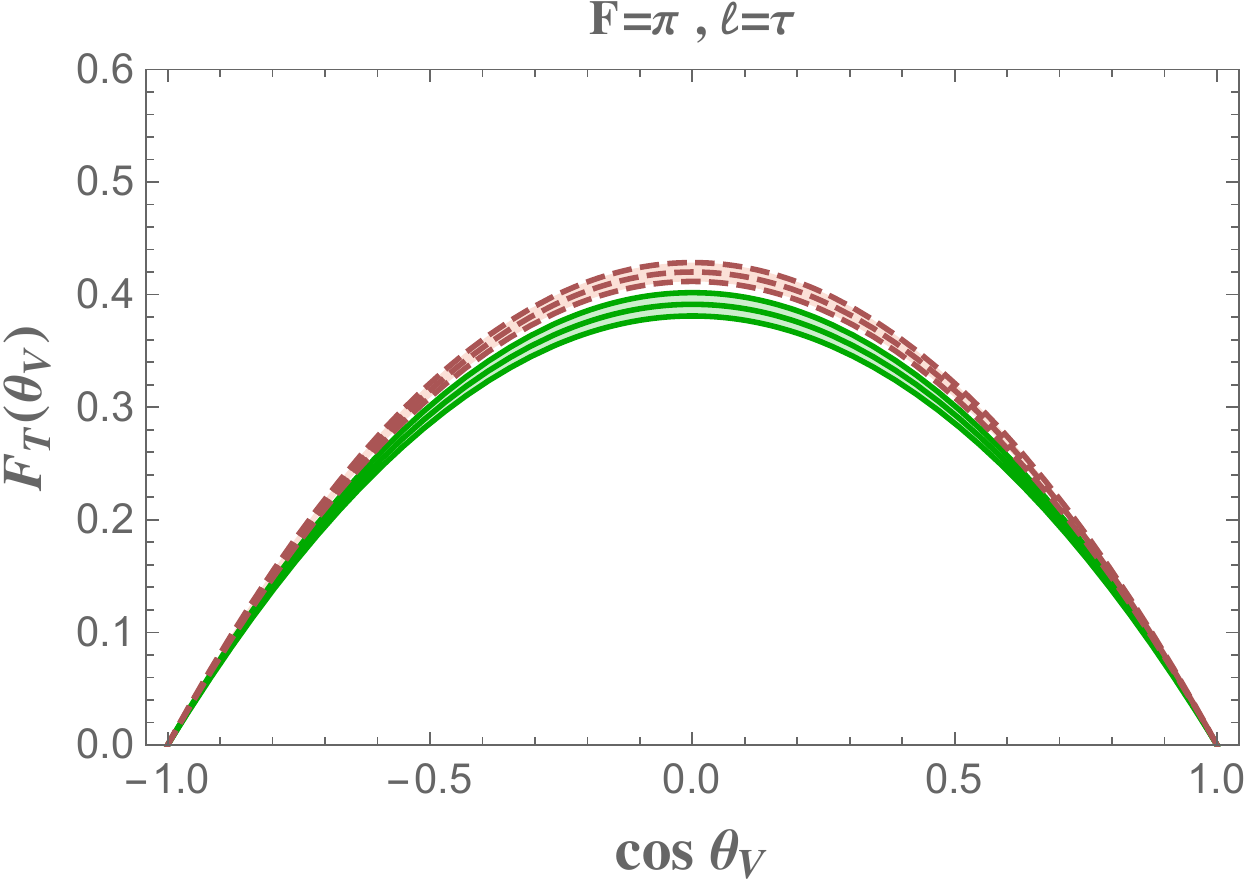}
\includegraphics[width = 0.45\textwidth]{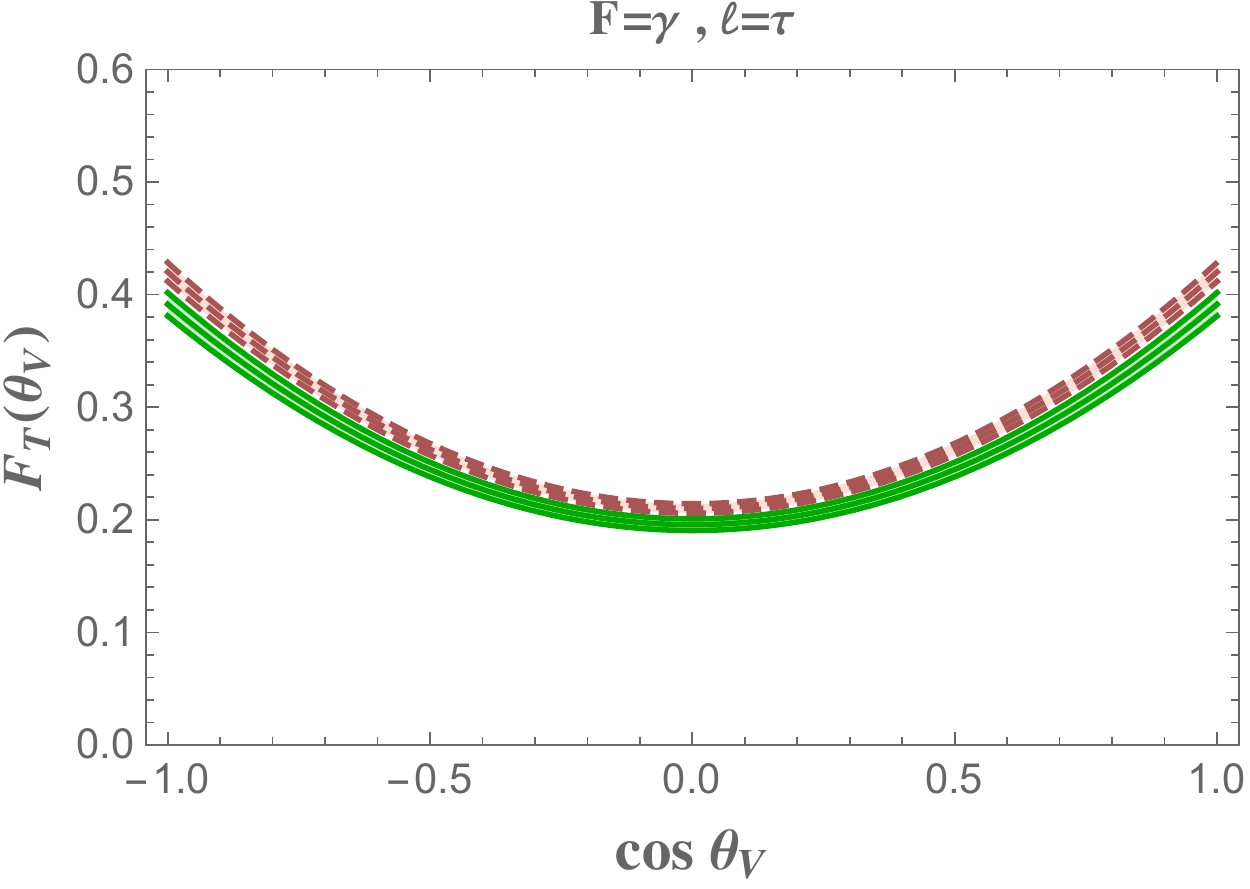}
    \caption{\baselineskip 10pt Distribution $F_T(\theta_V)$. The upper and lower plots refer to the case $\ell=\mu$ and $\ell=\tau$, the left and right column  to  $F=\pi$ and $F=\gamma$. Color codes as in fig.~\ref{FLthetaV}.
    }\label{FTthetaV}
\end{center}
\end{figure}

\begin{figure}[t]
\begin{center}
\includegraphics[width = 0.45\textwidth]{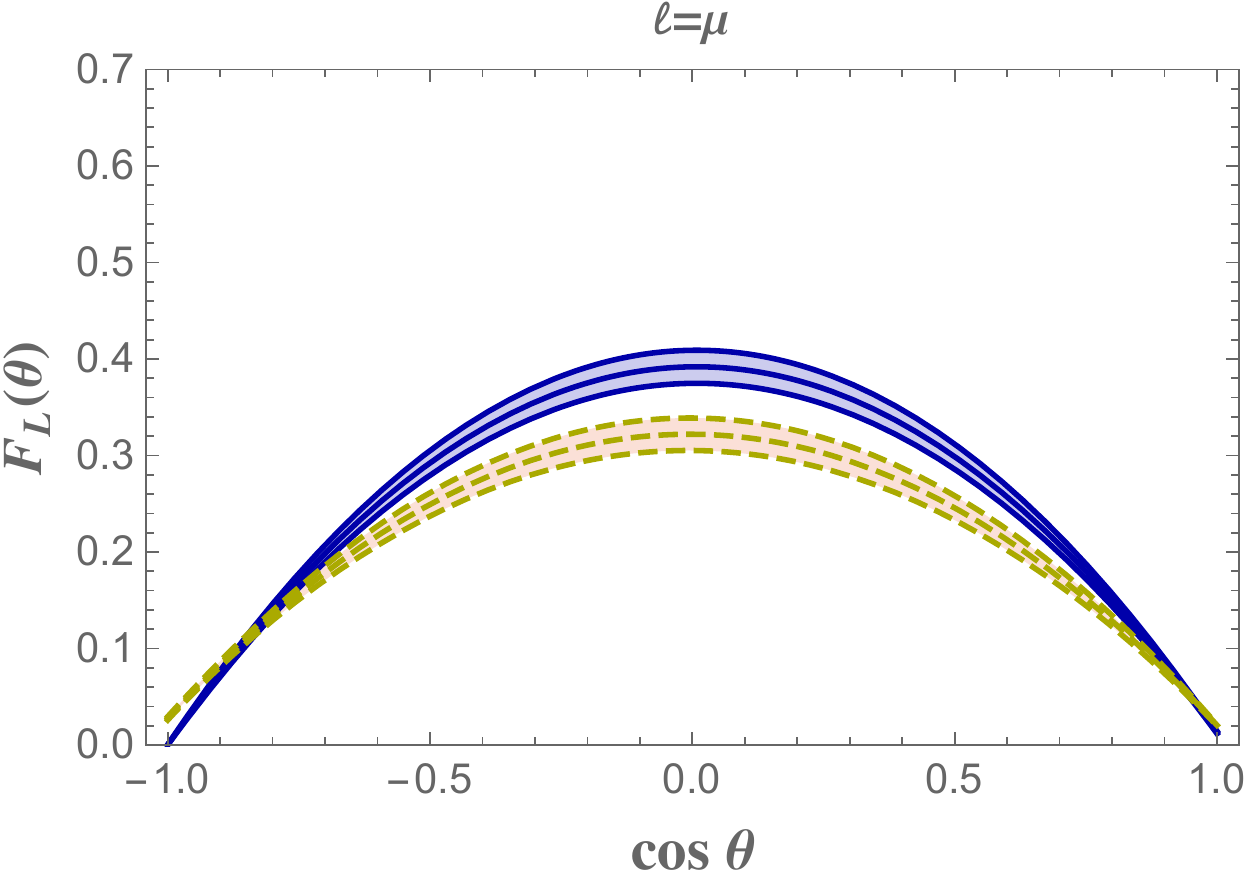}
\includegraphics[width = 0.45\textwidth]{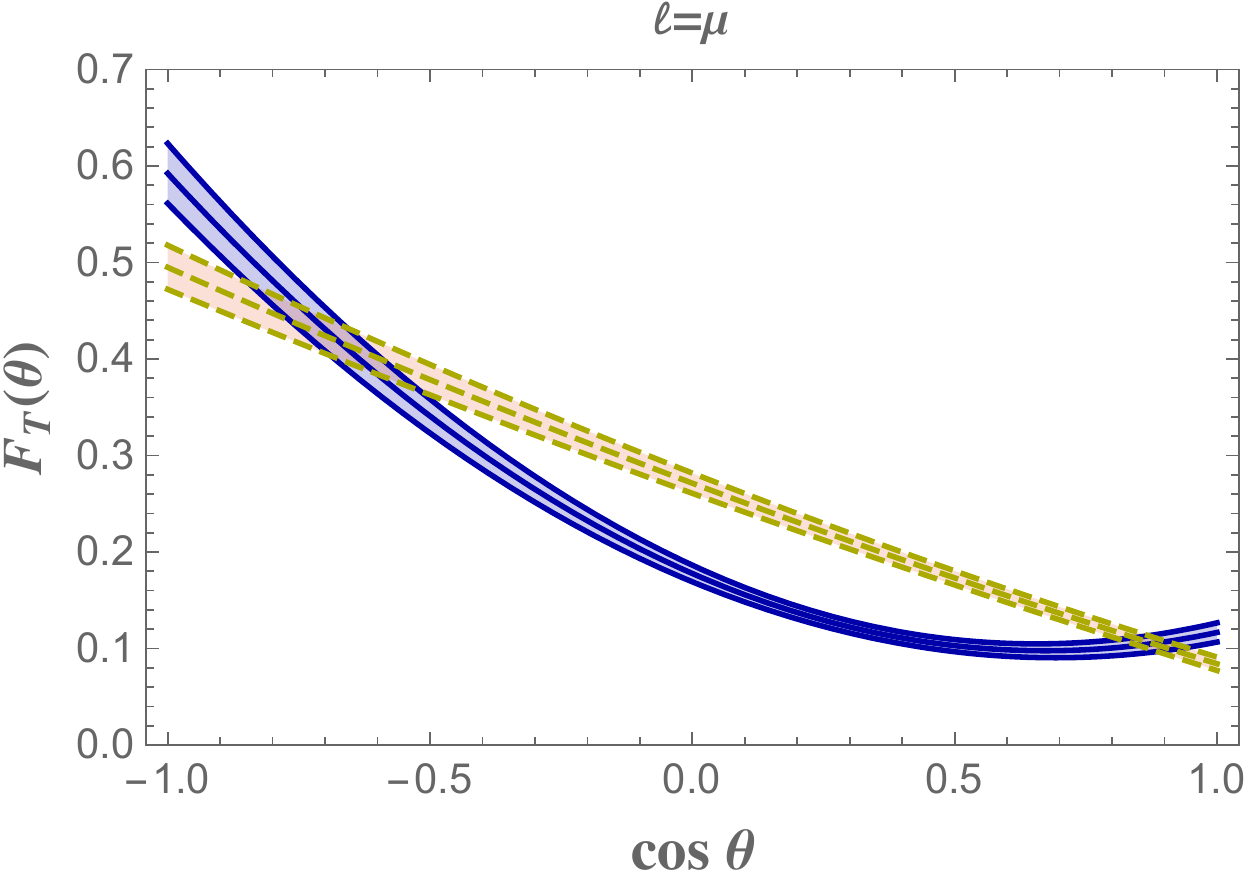}
\\
\vskip 0.3cm
\includegraphics[width = 0.45\textwidth]{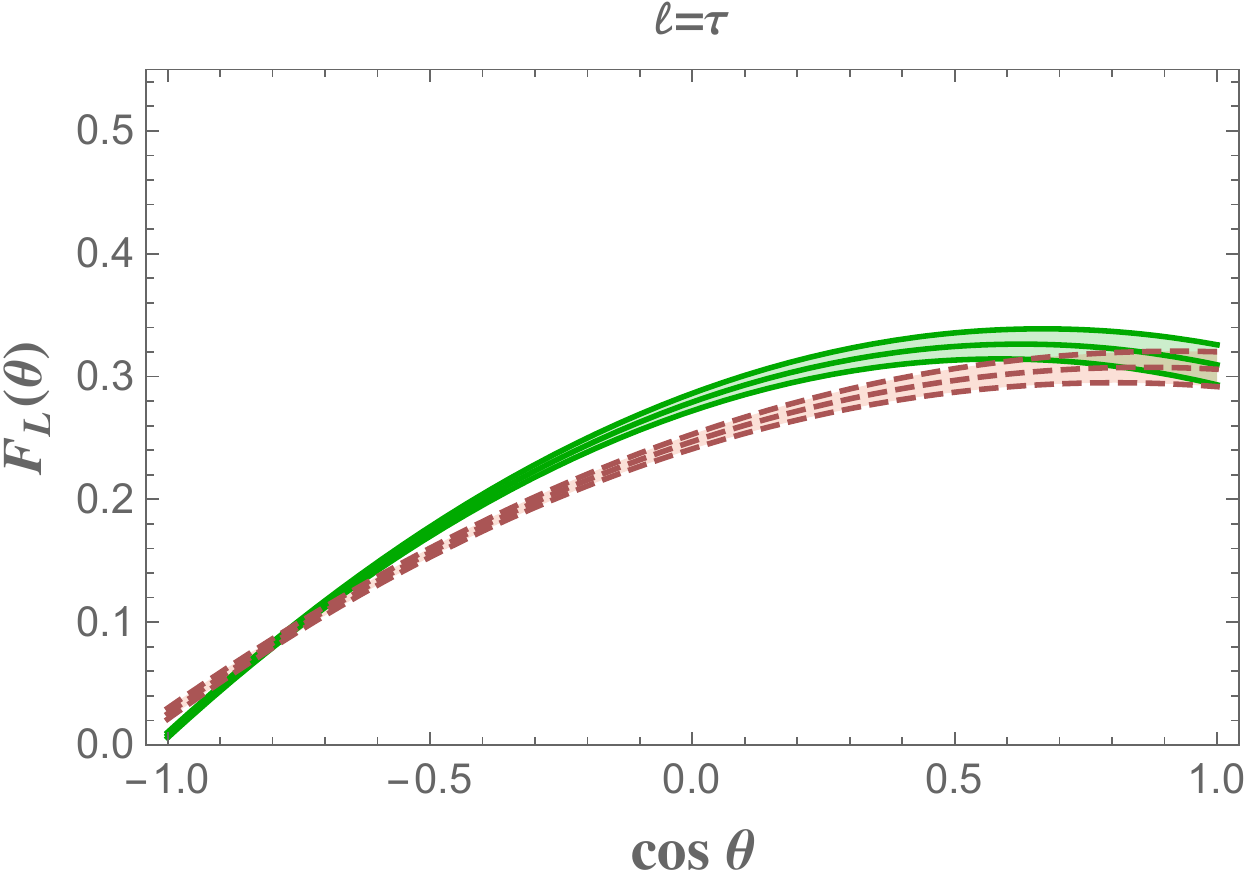}
\includegraphics[width = 0.45\textwidth]{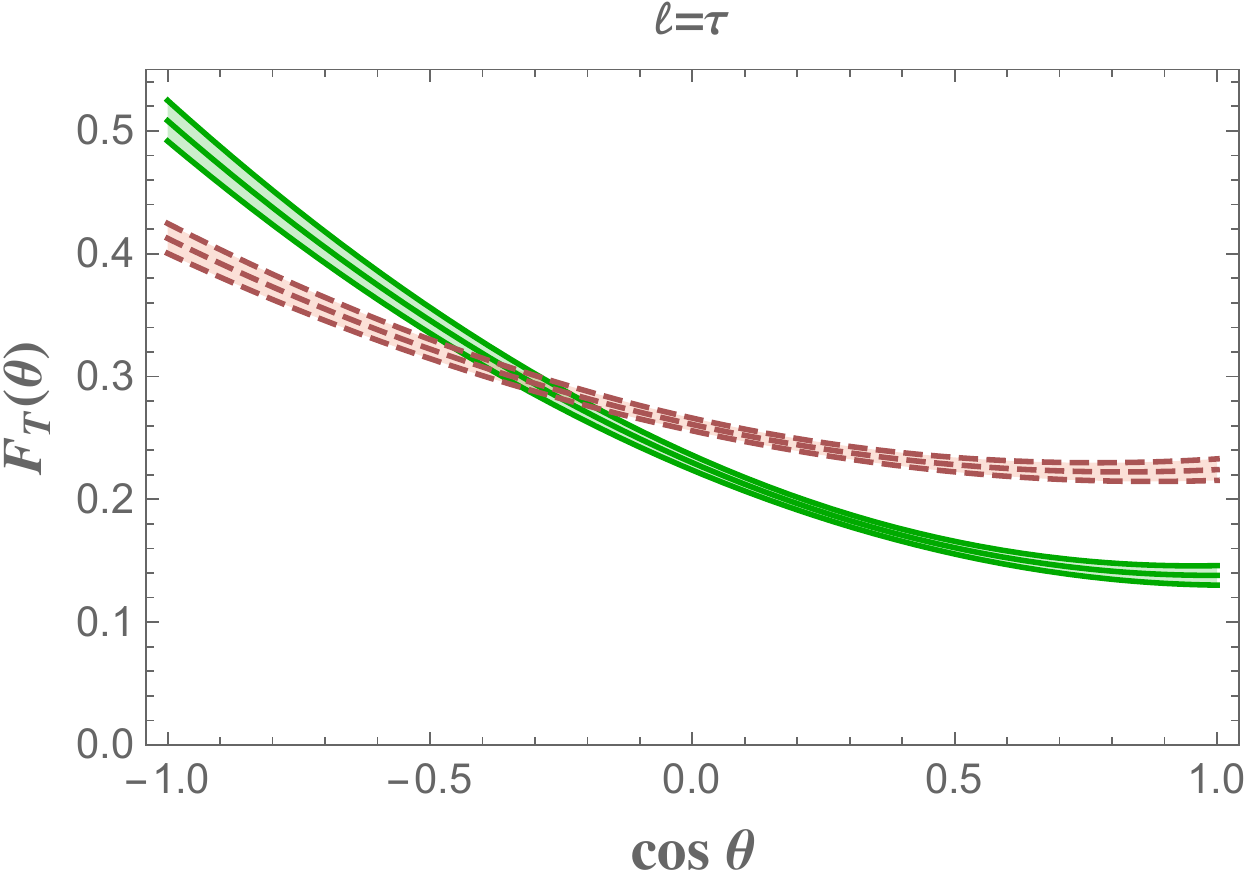}
    \caption{\baselineskip 10pt  Distributions $F_L(\theta)$ (left) and $F_T(\theta)$ (right). Upper and lower plots refer to  $\ell=\mu$ and  $\ell=\tau$, respectively.  Color codes as in fig.~\ref{FLthetaV}.}\label{FLTtheta}
\end{center}
\end{figure}

When $F=\pi$, the
 direction $\cos \theta_V=0,\, \cos \theta=-1$  selects the transverse $D^*$ polarization,  while for $\cos \theta_V=\pm 1,\, \cos \theta=0$ $D^*$ is longitudinally polarized.
For $F=\gamma$, $F_L$ has a  maximum at $\cos \theta_V=0,\, \cos \theta=0$, while $F_T$ is largest at $\cos \theta_V=\pm 1,\, \cos \theta=-1$. 
The sensitivity to NP can be visualized integrating  the double differential distributions  in $\cos \theta$ or in $\cos \theta_V$. At the benchmark point,  
integrating over $\cos \theta$ we obtain  $F_{L,T}(\theta_V)=\int_{-1}^1\, d \cos \theta \, F_{L,T}(\theta,\,\theta_V)$   in 
figs.~\ref{FLthetaV} and \ref{FTthetaV}.
Integrating in $\cos \, \theta_V$, the distributions $F_{L,T}(\theta)=\int_{-1}^1\, d \cos \theta_V \, F_{L,T}(\theta,\,\theta_V)$ coincide for $F=\pi$ and $ F=\gamma$: they are  shown in fig.~\ref{FLTtheta} in SM and  NP case. 

The observables for  $\ell=\mu$ are more sensitive to NP: in the case of  $F_L(\theta_V)$ the deviation is larger for $\cos \theta_V \simeq \pm 1$ for $F=\pi$, and for $\cos \theta_V \simeq 0$ for $F=\gamma$.
Highest sensitivity to NP is in the function $F_T(\theta)$,  which can probe 
the sign of the angular coefficient $I_{2s}^\pi \, (I_{2c}^\gamma)$ through its concavity.
Indeed,  the sign of the second derivative of $F_T(\theta) $ with respect to $\cos \, \theta$ depends on the sign of this coefficient, that is positive in  SM but could have a different sign in other scenarios. Indeed, comparing figs.~\ref{fig:angularpimuNP} and  \ref{fig:angularpimu}  one sees that NP can produce a sign reversal for this coefficient.

\vspace*{1.cm} 
\noindent{\bf Tests of LFU}\\

The angular coefficient functions in the fully differential  distribution provide  LFU tests.
This is interesting, considering that  after  integration over the angles only four coefficients contribute to the decay rate, therefore only those are probed by ratios of branching fractions.

Information from the fully differential decay rate can be exploited defining \\
${\tilde I}_i=\left(1-\frac{m_\ell^2}{q^2} \right)^2 |{\vec p}_{D^*}|_{BRF} \, I_i$, and  the ratios
\be
R_i^{\ell_1 \, \ell_2}=\frac{\int_{w=1}^{w_{max}(\ell_1)} ( {\tilde I}_i^\pi(w))_{\ell_1} dw }{\int_{w=1}^{w_{max}(\ell_2)} ( {\tilde I}_i^\pi(w) )_{\ell_2} dw} \label{ri}
\ee
for $\ell_1 \ell_2=\tau \, \mu, \tau \, e, \mu \, e$.
The SM predictions for these ratios, using CLN,  are collected in Table \ref{tab:RiSM}. The errors reflect the form factor uncertainties.
\begin{table}[b!]
\begin{center}
\begin{tabular}{|c|c|c|c|}
  \hline
 &  $\ell_1=\tau \,\,, \, \ell_2=\mu$ & $\ell_1=\tau \,\,, \, \ell_2=e$ &   $\ell_1=\mu \,\,, \, \ell_2=e$ \\[1mm]  
 \hline\hline 
 $R_{1s}^\pi$ &	 $0.263 \pm 0.006$ & $0.262 \pm 0.005$ & $0.9957 \pm 0.0001$ 
 \\
  $R_{1c}^\pi$ & $0.28\pm 0.02$ &$0.28 \pm 0.02$ & $1.008\pm 0.004$ 
\\
 $R_{2s}^\pi$ & $0.134 \pm 0.003$ &$0.133 \pm 0.003$ &$0.9923 \pm 0.0002$ 
 \\
  $R_{2c}^\pi$ &$0.079 \pm 0.005 $ &  $0.077 \pm 0.005 $ &$0.975 \pm 0.002 $ 
 \\ 
 $R_{3}^\pi$ &$0.153 \pm 0.004$ &$0.152 \pm 0.004$ &$0.9932 \pm 0.0002$ 
 \\
 $R_{4}^\pi$ & $0.112 \pm 0.004$ &$0.111 \pm 0.004$ &$0.9891 \pm 0.0004$ 
 \\
 $R_{5}^\pi$ &$0.30 \pm 0.02$ & $0.30 \pm 0.02$  &  $0.999 \pm 0.001$
 \\
 $R_{6s}^\pi$ &$0.197 \pm 0.004$ & $0.196 \pm 0.004$ &$0.9943 \pm 0.0001$
 \\
$R_{6c}^\pi$ & $5.90 \pm 0.45$ &$76000 \pm 7000$ &$12900 \pm 200$
  \\[1mm]
  \hline                                                                               	
\end{tabular}
\caption{\baselineskip 10pt  SM predictions for the ratios in Eq.(\ref{ri}) using CLN. }
\label{tab:RiSM}
\end{center}
\end{table}
\begin{table}[b!]
\begin{center}
\begin{tabular}{|c|c|c|c|}
  \hline
 &  $\ell_1=\tau \,\,, \, \ell_2=\mu$ & $\ell_1=\tau \,\, ,\, \ell_2=e$ &   $\ell_1=\mu \,\, ,\, \ell_2=e$ \\[1mm]  
 \hline\hline 
 $R_{1s}^\pi$ &	 $0.32 \pm 0.01$ & $0.304 \pm 0.008$ & $0.957 \pm 0.002$ 
 \\
  $R_{1c}^\pi$ & $0.36 \pm 0.03$ &$0.34\pm 0.02$ & $0.956\pm 0.003$ 
\\
 $R_{2s}^\pi$ & $0.37 \pm 0.02$ &$0.38 \pm 0.02$ &$1.04 \pm 0.01$ 
 \\
  $R_{2c}^\pi$ &$0.082 \pm 0.006 $ &  $0.080 \pm 0.006 $ &$0.973 \pm 0.002 $ 
 \\ 
 $R_{3}^\pi$ &$0.183 \pm 0.005$ &$0.182 \pm 0.005$ &$0.9932 \pm 0.0002$ 
 \\
 $R_{4}^\pi$ & $0.131 \pm 0.005$ &$0.130 \pm 0.005$ &$0.9890 \pm 0.0004$ 
 \\
 $R_{5}^\pi$ &$0.35 \pm 0.03$ & $0.33 \pm 0.03$  &  $0.96 \pm 0.01$
 \\
 $R_{6s}^\pi$ &$0.150 \pm 0.006$ & $0.152 \pm 0.006$ &$1.012 \pm 0.003$
 \\
$R_{6c}^\pi$ & $-11.6 \pm 1.5$ &$-944 \pm 40$ &$81.2 \pm 9.1$
\\
$R_{7}^\pi$ & $0$ &$0$ &$184 \pm 2$
  \\[1mm]
  \hline                                                                               	
\end{tabular}
\caption{\baselineskip 10pt  Ratios (\ref{ri})  in the NP scenario with the tensor operator, using CLN and at the benchmark point   ${\tilde \epsilon}_T^\ell$.}
\label{tab:RiNP}
\end{center}
\end{table}
Since $I_{6c}^\pi$ is proportional to the lepton mass squared, the ratios $R_{6c}^\pi$ are much larger than  the others. Analogous ratios in the case of  photon  can be defined  using  (\ref{relationspigamma}).
The same quantities  predicted in the NP scenario are collected
in Table \ref{tab:RiNP}. In the case of the ratios  $R_i^{\mu \, e}$,  assuming $\epsilon_T^\mu=\epsilon_T^e$, a deviation with respect to the SM result would  signal NP but not LFU violation.

Although the measurement of these ratios is challenging,  the high statistics foreseen, e.g.,  at Belle II   is promising   \cite{Guido:2017tpe}. For ratios involving the $\tau$ lepton,    the  use  of the   $\tau$ reconstruction  through the three-prong decays, as done at LHCb, can result in improved signal-to-background ratio and in a higher statistical significance \cite{Aaij:2017deq}.

\section{Conclusions}
To  understand the  experimental results on semileptonic $B$ decays,  the $R(D^{(*)})$ anomaly   and the tension in the exclusive vs inclusive $\vcb$ determinations,  it is mandatory  to control  the uncertainties  in the SM predictions and to explore all possible ways  in which  deviations can be observed. 
Considering the angular coefficient functions in the fully differential decay distribution in  $\bar B \to D^* \ell^- {\bar \nu}_\ell$, with $D^*$ decaying either as $D^* \to D \pi$ or as $D^* \to D \gamma$, we  have studied several observables able  to discern   effects of the form factor parametrization and to identify the
cases with minimal  sensitivity to hadronic uncertainties,  useful  to pin down  deviations.  As a testing example, we have considered a NP model with a  tensor operator.
 
Comparing the results obtained using the CLN and the BGL parametrization, we have identified the  angular coefficients  less sensitive to  the parametrization.
We have worked out   relations  allowing to extract the form factors from   measured angular coefficients.
Moreover,   the relations between the angular coefficients for $D^*$ decaying to $\pi$ and to $\gamma$ can be used as tests, exploiting the complementary of the two modes. 

Considering the SM extension,  we have shown that some angular coefficients, absent in the SM,  can be found  in NP.
A number of  observables display peculiar features in the NP model, e.g. the $q^2$-dependent forward-backward asymmetry for $\tau$ , and   the $\theta_V$-dependent forward-backward asymmetry  both for $\ell=\mu$ and for $\ell=\tau$.  The $D^*$ transverse polarization fraction $F_T(\theta)$ for $\ell=\mu$ is sensitive to the sign of one of the angular coefficients,  different in SM and NP.
Finally,  ratios to probe LFU and show possible violations have been constructed.
Although  the measurement of several observables is challenging, in particular in the  $\tau$ mode,  the forthcoming analyses at LHCb and Belle II are surely encouraging and  provide  exciting perspectives for  SM tests and NP searches. 

\vspace*{1cm}
\noindent {\bf Acknowledgements.}
We thank C. Bozzi and M. Rotondo for  discussions,  and M. Jung and D. Straub for  comments. This study has been  carried out within the INFN project (Iniziativa Specifica) QFT-HEP.

\appendix
\section{Four-body phase-space}
\label{kinematics}
We remind that the  four-body phase-space integration can be carried out using the identities
\bea
d \Pi_4=\frac{1}{2m_B}&& [dk_1][dk_2][dp_D][dp_F] (2\pi)^4 \delta^4 (p_B-p_D-p_F-k_1-k_2)\nn \\
=\frac{ (2\pi)^4 }{2m_B} 
&&
\left\{d^4 q d^4 p_{D^*} \delta^4(p_B-q-p_{D^*}) \right\}\times \nn \\
&&\left\{ [dk_1][dk_2]\delta^4 (q-k_1-k_2)\right\}\,\times\left\{[dp_D][dp_F] \delta^4 (p_{D^*}-p_D-p_F)\right\}\nn
\\=\frac{ (2\pi)^4 }{2m_B}  &&d\Pi_2^{(q,p_{D^*})} \times d\Pi_2^{(k_1,k_2)} \times d\Pi_2^{(p_D,p_F)}\,\,, \label{fourbody}
 \eea
using the  notation $[dp]=\displaystyle\frac{d^3p}{(2 \pi)^3 2p^0}$.
$d\Pi_2^{(k_1,k_2)}$ and $ d\Pi_2^{(p_D,p_F)}$ are the  two-body  phase-spaces
\bea
d\Pi_2^{(k_1,k_2)}&=&\frac{1}{ (2\pi)^6 }\frac{1}{4 \sqrt{q^2}}|{\vec k_1}|_{LRF} \,d\Omega_L
\label{pik1k2} \\
d\Pi_2^{(p_D,p_F)}&=&\frac{1}{ (2\pi)^6 }\frac{1}{4 \sqrt{p_{D^*}^2}}|{\vec p_D}|_{D^*RF} \, d\Omega_D \,\,.
\label{pipdpf}
\eea
In (\ref{pik1k2}), $|{\vec k_1}|_{LRF}=\displaystyle\frac{q^2-m_\ell^2}{2 \sqrt{q^2}}$ is the lepton three-momentum in the lepton-pair rest-frame, and   $d \Omega_L=d \cos \theta d \phi$.  In (\ref{pipdpf}), $|{\vec p_D}|_{D^*RF}$ is the $D$ three-momentum in the $D^*$ rest-frame, and  $d\Omega_D=(2 \pi) d \cos \theta_V $, with  $\theta_V$  the angle between the $D$ momentum in the $D^*$ rest-frame  and the $z$ axis;  the integration over the azimuthal angle in this frame is trivial.
 $d\Pi_2^{(q,p_{D^*})}$ can be evaluated exploiting   the narrow width approximation (\ref{atotfin}):
\be
d\Pi_2^{(q,p_{D^*})}\delta(p_{D^*}^2-m_{D^*}^2)=\frac{ \pi}{m_B}|{\vec p}_{D^*}|_{BRF} \, dq^2 \,\,\,\, , \label{pipdstarq}
\ee
where $|{\vec p}_{D^*}|_{BRF}$ is the $D^*$ three-momentum  in the $B$ rest-frame. 

\section{Hadronic matrix element   parametrizations}\label{appA}
In the CLN parametrization \cite{Caprini:1997mu}  the  $\bar B \to D^*$ matrix elements are written as
\bea
\langle D^*(v^\prime,\epsilon)|{\bar c} \gamma_\mu b| {\bar B}(v) \rangle 
&=& \sqrt{m_B m_{D^*}} i \, h_V(w)\epsilon_{\mu \nu \alpha \beta} \epsilon^{*\nu} v^{\prime \alpha} v^\beta \,\,\, ,
\label{hai} \\
\langle D^*(v^\prime,\epsilon)|{\bar c} \gamma_\mu \gamma_5 b| {\bar B}(v) \rangle
&=& \sqrt{m_B m_{D^*}} \Big[h_{A_1}(w)(w+1)\epsilon^*_\mu-\left[h_{A_2}(w) v_\mu+h_{A_3}(w) v^\prime_\mu \right](\epsilon^* \cdot v) \Big] \,\,\ , 
\nn\\
\langle D^*(v^\prime,\epsilon)|{\bar c} \sigma_{\mu \nu} b| {\bar B}(v) \rangle
&=& -\sqrt{m_B m_{D^*}} \epsilon_{\mu \nu \alpha \beta}
\Big[h_{T_1}(w)\epsilon^{*\alpha}(v+v^\prime)^\beta+h_{T_2}(w) \epsilon^{*\alpha}(v-v^\prime)^\beta\nn \\
&&\hskip 3 cm +h_{T_3}(w) v^\alpha v^{\prime \beta} (\epsilon^* \cdot v) \Big] \,\,\ , 
\nn
\eea
with $v$ and $v^\prime$  the $B$ and $D^*$ four-velocities and $w=v \cdot v^\prime$. The  factor $\sqrt{m_B m_{D^*}}$ accounts for the mass-dependent normalization of the states  (in  \cite{Caprini:1997mu} the mass-independent normalization is adopted).
This parametrization is  related to  the one in (\ref{FF-D*-mio})-(\ref{mat-tensor-Dstar}) through
 \bea
 V(q^2)&=& {m_B+m_{D^*} \over 2 \sqrt{m_B m_{D^*}}} h_V(w)   \nn \\
 A_1(q^2) &=& \sqrt{m_B m_{D^*}}{w+1 \over m_B+m_{D^*}} h_{A_1}(w)  \nn \\
 A_2(q^2) &=& { m_B+m_{D^*} \over 2 \sqrt{m_B m_{D^*}}} \left[h_{A_3}(w)+{m_{D^*} \over m_B} h_{A_2}(w)\right]   \label{noi-cln} \\
A_0(q^2) &=& { 1 \over 2 \sqrt{m_B m_{D^*}}} \left[ m_B (w+1) h_{A_1}(w) -(m_B-m_{D^*} w)h_{A_2}(w)-(m_Bw-m_{D^*})h_{A_3}(w) \right] , \nn
\eea
and
\bea
T_0(q^2)&=&-\frac{(m_B+m_{D^*})^2}{m_B m_{D^*}}\sqrt{\frac{m_{D^*}}{m_B}}h_{T_3}(w) \nn \\
T_1(q^2)&=& \sqrt{\frac{m_{D^*}}{m_B}}\left(h_{T_1}(w) +h_{T_2}(w) \right)\,\, \label{hTi}
\\
T_2(q^2)&=& \sqrt{\frac{m_B}{m_{D^*}}} \left(h_{T_1}(w) -h_{T_2}(w) \right) \,\,\, , \nn
\eea
with $q^2=m_B^2+m_{D^*}^2 -2  m_B m_{D^*} w$.
The  form factors $T_3,\,T_4,\,T_5$ in (\ref{mat-tensor-Dstar}) are related to $T_0,\,T_1,\,T_2$ by the identity:
$\sigma_{\mu \nu}\, \gamma_5=\displaystyle{\frac{i}{2}}\epsilon_{\mu \nu \alpha \beta} \sigma^{\alpha \beta}$.
The  relations of the form factors in (\ref{hai}) to the Isgur-Wise function, $h_V(w)=h_{A_1}(w)=h_{A_3}(w)=h_{T_1}(w)=\xi (w)$ and $h_{A_2}=h_{T_2}=h_{T_3}=0$
hold in the HQ limit. 
Such relations  can be improved
including radiative $\alpha_s$  and  power $\displaystyle{\frac{1}{m_{b}}}$,  $\displaystyle{\frac{1}{m_{c}}}$ corrections. 
In the case of the functions in (\ref{noi-cln}) they   have been worked out in \cite{Neubert:1993mb,Caprini:1997mu}:
\bea
h_V(w) &=& \left[ C_1 +\epsilon_c (L_2 -L_5) +\epsilon_b (L_1 -L_4) \right] \, \xi(w) 
\nn \\
h_{A_1}(w) &=& \left[ C_1^5 +\epsilon_c \left(L_2-{w-1 \over w+1}L_5 \right) +\epsilon_b \left(L_1 -{w-1 \over w+1}L_4 \right)  \right] \,  \xi(w)
\nn \\
h_{A_2}(w) &=& \left[ C_2^5 +\epsilon_c (L_3+L_6) \right] \, \xi(w) 
\label{ha2}
\\
h_{A_3}(w) &=& \left[ C_1^5+C_3^5 +\epsilon_c (L_2 -L_3 -L_5+L_6) + \epsilon_b (L_1 -L_4) \right] \, \xi(w) \,\,\, . \nn
\eea
The coefficients $C_i$  incorporate the radiative corrections.   $L_i$  account for  ${\cal O}(1/m_Q)$ corrections in the HQ expansion, and  their numerical values have been  obtained  using
QCD sum rule determinations of the subleading  form factors \cite{Neubert:1993mb}. 
Their expressions can be found in the original papers \cite{Neubert:1993mb,Caprini:1997mu}, and are collected in the appendix of \cite{Biancofiore:2013ki}.
The analogous relations for the form factors in (\ref{hTi}) have been worked out in \cite{Bernlochner:2017jka}:
\bea
h_{T_1}(w) &=&  \left[ {\tilde C}_1+\epsilon_c L_2+\epsilon_b L_1  \right] \, \xi(w)  \nn \\
h_{T_2}(w) &=&  \left[ {\tilde C}_2+\epsilon_c L_5-\epsilon_b L_4   \right] \, \xi(w) \label{hTNLO} \\
h_{T_3}(w) &=&  \left[ {\tilde C}_3+\epsilon_c ( L_6 - L_3)  \right] \, \xi(w) \nn  \eea
where ${\tilde C}_i$  incorporate the radiative corrections. 
Among the $C_i$ and ${\tilde C}_i$,  the set $C_2^5$, ${\tilde C}_2$ and ${\tilde C}_3$ starts at ${\cal O}(\alpha_s)$. We refer to \cite{Bernlochner:2017jka} for the expressions of the parameters in (\ref{hTNLO}).

The BGL  parametrization uses the form factors $g$, $f$, $a_+$ and $a_-$:
\bea
\langle D^*(p^\prime,\epsilon)|{\bar c} \gamma_\mu b| {\bar B}(p) \rangle 
&=& i \,  \epsilon_{\mu \nu \alpha \beta} \epsilon^{* \nu} p^{\prime \alpha} p^\beta g \,\, ,
\nn \\
\langle D^*(v^\prime,\epsilon)|{\bar c} \gamma_\mu \gamma_5 b| {\bar B}(v) \rangle
&=& \epsilon^*_\mu \, f + (\epsilon^* \cdot p) \left[(p+p^\prime)_\mu a_+ + (p-p^\prime_\mu) a_-\right] \,\,,
\label{matBGL}
\eea
so that
\bea
g(w)&=&\frac{h_V(w)}{\sqrt{m_B m_{D^*}} }\nn \\
f(w)&=& \sqrt{m_B m_{D^*}}(1+w) h_{A_1}(w)\nn \\
a_+(w) &=&-\frac{m_{D^*}}{2 \sqrt{m_B m_{D^*}}}  \left( \frac{h_{A_3}(w)}{m_{D^*}}+\frac{h_{A_2}(w)}{m_B} \right)  \\
a_- (w)&=&\frac{m_{D^*}}{2 \sqrt{m_B m_{D^*}}}  \left( \frac{h_{A_3}(w)}{m_{D^*}}-\frac{h_{A_2}(w)}{m_B} \right) \nn \,\,.\eea
The expressions of the helicity amplitudes are:
\bea
H_0&=&\frac{{\cal F}_1(w)}{\sqrt{q^2}} \nn \\
H_\pm&=& f(w) \mp m_B m_{D^*} \sqrt{w^2-1} \, g(w) \,\,\, , \label{HBGL} \eea
with
\be
{\cal F}_1(w)=\sqrt{m_B m_{D^*}}(1+w) \left[(m_B w - m_{D^*})h_{A_1}(w)-m_{D^*}(w-1)h_{A_2}(w)-m_B(w-1) h_{A_3}(w)\right] . \nn
\ee
In the BGL approach, the  observation is used that the $W$ production amplitude of $\bar B \bar D^*$ is related to the $\bar B \to D^*$ form factors by analytic continuation from the semileptonic region
$m_\ell^2 \leq t \leq t_-$  
to the  region $t_+ \leq t$, with $t_\pm=(m_B \pm m_{D^*})^2$ \cite{Boyd:1994tt,Boyd:1995cf,Boyd:1995sq}. In the production region, constraints can be imposed using perturbative QCD, including  quark and gluon condensate corrections. Then, analyticity is exploited. 
The form factors are  written as functions of the conformal variable  $z$ in the form:
 $\displaystyle f(z)=\frac{1}{P_f(z) \phi_f(z)} \sum\limits_{n=0}^N a_n z^n$. 
 The Blatsche factors $P_f(z)$  account for the $t<(m_B+m_{D^*})^2$ poles associated with on-shell production of $\bar c b$ bound states, while $\phi_f(z)$ are outer functions from phase-space integration. The coefficients $a_n$ satisfy unitarity bounds of the type   $\sum\limits_{n=0}^N |a_n|^2 \leq 1$.
For $B \to D^*$, three coefficients $a_n$, with  $n=0,1,2$,   have been fitted for  each form factor  $g$, $f$ and ${\cal F}_1$  \cite{Bigi:2017njr}, and  unitarity bounds have been imposed  \cite{Bigi:2017jbd}. The masses of the $\bar c b$ lowest-lying bound states with suitable $J^P$  quantum numbers are taken from constituent quark models. The resulting values of the parameters are reported in \cite{Bigi:2017njr}: they are used  in our analysis.
\bibliographystyle{JHEP}
\bibliography{refs}

\providecommand{\href}[2]{#2}\begingroup\raggedright\begin{thebibliography}{10}

\bibitem{Amhis:2016xyh}
{\bf HFLAV} Collaboration, Y.~Amhis et~al., {\it {Averages of $b$-hadron,
  $c$-hadron, and $\tau$-lepton properties as of summer 2016}},  {\em Eur.
  Phys. J.} {\bf C77} (2017), no.~12 895,
  [\href{http://arxiv.org/abs/1612.07233}{{\tt arXiv:1612.07233}}].

\bibitem{Lees:2012xj}
{\bf BaBar} Collaboration, J.~P. Lees et~al., {\it {Evidence for an excess of
  $\bar{B} \to D^{(*)} \tau^-\bar{\nu}_\tau$ decays}},  {\em Phys. Rev. Lett.}
  {\bf 109} (2012) 101802, [\href{http://arxiv.org/abs/1205.5442}{{\tt
  arXiv:1205.5442}}].

\bibitem{Lees:2013uzd}
{\bf BaBar} Collaboration, J.~P. Lees et~al., {\it {Measurement of an Excess of
  $\bar{B} \to D^{(*)}\tau^- \bar{\nu}_\tau$ Decays and Implications for
  Charged Higgs Bosons}},  {\em Phys. Rev.} {\bf D88} (2013), no.~7 072012,
  [\href{http://arxiv.org/abs/1303.0571}{{\tt arXiv:1303.0571}}].

\bibitem{Huschle:2015rga}
{\bf Belle} Collaboration, M.~Huschle et~al., {\it {Measurement of the
  branching ratio of $\bar{B} \to D^{(\ast)} \tau^- \bar{\nu}_\tau$ relative to
  $\bar{B} \to D^{(\ast)} \ell^- \bar{\nu}_\ell$ decays with hadronic tagging
  at Belle}},  {\em Phys. Rev.} {\bf D92} (2015), no.~7 072014,
  [\href{http://arxiv.org/abs/1507.03233}{{\tt arXiv:1507.03233}}].

\bibitem{Sato:2016svk}
{\bf Belle} Collaboration, Y.~Sato et~al., {\it {Measurement of the branching
  ratio of $\bar{B}^0 \rightarrow D^{*+} \tau^- \bar{\nu}_{\tau}$ relative to
  $\bar{B}^0 \rightarrow D^{*+} \ell^- \bar{\nu}_{\ell}$ decays with a
  semileptonic tagging method}},  {\em Phys. Rev.} {\bf D94} (2016), no.~7
  072007, [\href{http://arxiv.org/abs/1607.07923}{{\tt arXiv:1607.07923}}].

\bibitem{Hirose:2016wfn}
{\bf Belle} Collaboration, S.~Hirose et~al., {\it {Measurement of the $\tau$
  lepton polarization and $R(D^*)$ in the decay $\bar{B} \to D^* \tau^-
  \bar{\nu}_\tau$}},  {\em Phys. Rev. Lett.} {\bf 118} (2017), no.~21 211801,
  [\href{http://arxiv.org/abs/1612.00529}{{\tt arXiv:1612.00529}}].

\bibitem{Aaij:2015yra}
{\bf LHCb} Collaboration, R.~Aaij et~al., {\it {Measurement of the ratio of
  branching fractions $\mathcal{B}(\bar{B}^0 \to
  D^{*+}\tau^{-}\bar{\nu}_{\tau})/\mathcal{B}(\bar{B}^0 \to
  D^{*+}\mu^{-}\bar{\nu}_{\mu})$}},  {\em Phys. Rev. Lett.} {\bf 115} (2015),
  no.~11 111803, [\href{http://arxiv.org/abs/1506.08614}{{\tt
  arXiv:1506.08614}}]. [Erratum: Phys. Rev. Lett.115,no.15,159901(2015)].

\bibitem{Fajfer:2012vx}
S.~Fajfer, J.~F. Kamenik, and I.~Nisandzic, {\it {On the $B \to D^* \tau \bar
  \nu_{\tau}$ Sensitivity to New Physics}},  {\em Phys. Rev.} {\bf D85} (2012)
  094025, [\href{http://arxiv.org/abs/1203.2654}{{\tt arXiv:1203.2654}}].

\bibitem{Aoki:2016frl}
S.~Aoki et~al., {\it {Review of lattice results concerning low-energy particle
  physics}},  {\em Eur. Phys. J.} {\bf C77} (2017), no.~2 112,
  [\href{http://arxiv.org/abs/1607.00299}{{\tt arXiv:1607.00299}}].

\bibitem{Bigi:2017jbd}
D.~Bigi, P.~Gambino, and S.~Schacht, {\it {$R(D^*)$, $|V_{cb}|$, and the Heavy
  Quark Symmetry relations between form factors}},  {\em JHEP} {\bf 11} (2017)
  061, [\href{http://arxiv.org/abs/1707.09509}{{\tt arXiv:1707.09509}}].

\bibitem{Jaiswal:2017rve}
S.~Jaiswal, S.~Nandi, and S.~K. Patra, {\it {Extraction of $|V_{cb}|$ from
  $B\to D^{(*)}\ell\nu_\ell$ and the Standard Model predictions of
  $R(D^{(*)})$}},  {\em JHEP} {\bf 12} (2017) 060,
  [\href{http://arxiv.org/abs/1707.09977}{{\tt arXiv:1707.09977}}].

\bibitem{Bernlochner:2017jka}
F.~U. Bernlochner, Z.~Ligeti, M.~Papucci, and D.~J. Robinson, {\it {Combined
  analysis of semileptonic $B$ decays to $D$ and $D^*$: $R(D^{(*)})$,
  $|V_{cb}|$, and new physics}},  {\em Phys. Rev.} {\bf D95} (2017), no.~11
  115008, [\href{http://arxiv.org/abs/1703.05330}{{\tt arXiv:1703.05330}}].

\bibitem{Hirose:2017dxl}
{\bf Belle} Collaboration, S.~Hirose et~al., {\it {Measurement of the $\tau$
  lepton polarization and $R(D^*)$ in the decay $\bar{B} \rightarrow D^* \tau^-
  \bar{\nu}_\tau$ with one-prong hadronic $\tau$ decays at Belle}},  {\em Phys.
  Rev.} {\bf D97} (2018), no.~1 012004,
  [\href{http://arxiv.org/abs/1709.00129}{{\tt arXiv:1709.00129}}].

\bibitem{Aaij:2017tyk}
{\bf LHCb} Collaboration, R.~Aaij et~al., {\it {Measurement of the ratio of
  branching fractions
  $\mathcal{B}(B_c^+\,\to\,J/\psi\tau^+\nu_\tau)$/$\mathcal{B}(B_c^+\,\to\,J/\psi\mu^+\nu_\mu)$}},
  \href{http://arxiv.org/abs/1711.05623}{{\tt arXiv:1711.05623}}.

\bibitem{Tran:2018kuv}
C.-T. Tran, M.~A. Ivanov, J.~G. Korner, and P.~Santorelli, {\it {Implications
  of new physics in the decays $B_c \to (J/\psi,\eta_c)\tau\nu$}},
  \href{http://arxiv.org/abs/1801.06927}{{\tt arXiv:1801.06927}}.

\bibitem{Caprini:1997mu}
I.~Caprini, L.~Lellouch, and M.~Neubert, {\it {Dispersive bounds on the shape
  of $\bar B \to D^{(*)}$ lepton anti-neutrino form-factors}},  {\em Nucl.
  Phys.} {\bf B530} (1998) 153--181,
  [\href{http://arxiv.org/abs/hep-ph/9712417}{{\tt hep-ph/9712417}}].

\bibitem{Abdesselam:2017kjf}
{\bf Belle} Collaboration, A.~Abdesselam et~al., {\it {Precise determination of
  the CKM matrix element $\left| V_{cb}\right|$ with $\bar B^0 \to D^{*\,+} \,
  \ell^- \, \bar \nu_\ell$ decays with hadronic tagging at Belle}},
  \href{http://arxiv.org/abs/1702.01521}{{\tt arXiv:1702.01521}}.

\bibitem{Boyd:1994tt}
C.~G. Boyd, B.~Grinstein, and R.~F. Lebed, {\it {Constraints on form-factors
  for exclusive semileptonic heavy to light meson decays}},  {\em Phys. Rev.
  Lett.} {\bf 74} (1995) 4603--4606,
  [\href{http://arxiv.org/abs/hep-ph/9412324}{{\tt hep-ph/9412324}}].

\bibitem{Boyd:1995cf}
C.~G. Boyd, B.~Grinstein, and R.~F. Lebed, {\it {Model independent extraction
  of $|V_{cb}|$ using dispersion relations}},  {\em Phys. Lett.} {\bf B353}
  (1995) 306--312, [\href{http://arxiv.org/abs/hep-ph/9504235}{{\tt
  hep-ph/9504235}}].

\bibitem{Boyd:1995sq}
C.~G. Boyd, B.~Grinstein, and R.~F. Lebed, {\it {Model independent
  determinations of $\bar B \to D$ (lepton), $D^*$ (lepton) anti-neutrino
  form-factors}},  {\em Nucl. Phys.} {\bf B461} (1996) 493--511,
  [\href{http://arxiv.org/abs/hep-ph/9508211}{{\tt hep-ph/9508211}}].

\bibitem{Bigi:2017njr}
D.~Bigi, P.~Gambino, and S.~Schacht, {\it {A fresh look at the determination of
  $|V_{cb}|$ from $B\to D^{*} \ell \nu$}},  {\em Phys. Lett.} {\bf B769} (2017)
  441--445, [\href{http://arxiv.org/abs/1703.06124}{{\tt arXiv:1703.06124}}].

\bibitem{Grinstein:2017nlq}
B.~Grinstein and A.~Kobach, {\it {Model-Independent Extraction of $|V_{cb}|$
  from $\bar{B}\rightarrow D^* \ell \overline{\nu}$}},  {\em Phys. Lett.} {\bf
  B771} (2017) 359--364, [\href{http://arxiv.org/abs/1703.08170}{{\tt
  arXiv:1703.08170}}].

\bibitem{Colangelo:2016ymy}
P.~Colangelo and F.~De~Fazio, {\it {Tension in the inclusive versus exclusive
  determinations of $|V_{cb}|$: a possible role of new physics}},  {\em Phys.
  Rev.} {\bf D95} (2017), no.~1 011701,
  [\href{http://arxiv.org/abs/1611.07387}{{\tt arXiv:1611.07387}}].

\bibitem{Biancofiore:2013ki}
P.~Biancofiore, P.~Colangelo, and F.~De~Fazio, {\it {Anomalous enhancement
  observed in $B \to D^{(*)}\tau{\bar \nu}_\tau$ decays}},  {\em Phys. Rev.}
  {\bf D87} (2013), no.~7 074010, [\href{http://arxiv.org/abs/1302.1042}{{\tt
  arXiv:1302.1042}}].

\bibitem{Duraisamy:2013kcw}
M.~Duraisamy and A.~Datta, {\it {The Full $B \to D^{*} \tau^{-} \bar{\nu_\tau}$
  Angular Distribution and CP violating Triple Products}},  {\em JHEP} {\bf 09}
  (2013) 059, [\href{http://arxiv.org/abs/1302.7031}{{\tt arXiv:1302.7031}}].

\bibitem{Bhattacharya:2015ida}
S.~Bhattacharya, S.~Nandi, and S.~K. Patra, {\it {Optimal-observable analysis
  of possible new physics in $B\to D^{(\ast)}\tau\nu_{\tau}$}},  {\em Phys.
  Rev.} {\bf D93} (2016), no.~3 034011,
  [\href{http://arxiv.org/abs/1509.07259}{{\tt arXiv:1509.07259}}].

\bibitem{Bardhan:2016uhr}
D.~Bardhan, P.~Byakti, and D.~Ghosh, {\it {A closer look at the R$_{D}$ and
  R$_{D^*}$ anomalies}},  {\em JHEP} {\bf 01} (2017) 125,
  [\href{http://arxiv.org/abs/1610.03038}{{\tt arXiv:1610.03038}}].

\bibitem{Alonso:2016gym}
R.~Alonso, A.~Kobach, and J.~Martin~Camalich, {\it {New physics in the
  kinematic distributions of $\bar B\to
  D^{(*)}\tau^-(\to\ell^-\bar\nu_\ell\nu_\tau)\bar\nu_\tau$}},  {\em Phys.
  Rev.} {\bf D94} (2016), no.~9 094021,
  [\href{http://arxiv.org/abs/1602.07671}{{\tt arXiv:1602.07671}}].

\bibitem{Becirevic:2016hea}
D.~Becirevic, S.~Fajfer, I.~Nisandzic, and A.~Tayduganov, {\it {Angular
  distributions of $\bar B \to D^{(\ast)}\ell\bar \nu_\ell$ decays and search
  of New Physics}},  \href{http://arxiv.org/abs/1602.03030}{{\tt
  arXiv:1602.03030}}.

\bibitem{Ligeti:2016npd}
Z.~Ligeti, M.~Papucci, and D.~J. Robinson, {\it {New Physics in the Visible
  Final States of $B\to D^{(*)}\tau\nu$}},  {\em JHEP} {\bf 01} (2017) 083,
  [\href{http://arxiv.org/abs/1610.02045}{{\tt arXiv:1610.02045}}].

\bibitem{Alok:2016qyh}
A.~K. Alok, D.~Kumar, S.~Kumbhakar, and S.~U. Sankar, {\it {$D^{*}$
  polarization as a probe to discriminate new physics in $\bar{B}\to D^{*} \tau
  \bar{\nu}$}},  {\em Phys. Rev.} {\bf D95} (2017), no.~11 115038,
  [\href{http://arxiv.org/abs/1606.03164}{{\tt arXiv:1606.03164}}].

\bibitem{Ivanov:2017mrj}
M.~A. Ivanov, J.~G. Korner, and C.-T. Tran, {\it {Probing new physics in
  $\bar{B}^0 \to D^{(\ast)} \tau^- \bar\nu_{\tau}$ using the longitudinal,
  transverse, and normal polarization components of the tau lepton}},  {\em
  Phys. Rev.} {\bf D95} (2017), no.~3 036021,
  [\href{http://arxiv.org/abs/1701.02937}{{\tt arXiv:1701.02937}}].

\bibitem{Alonso:2017ktd}
R.~Alonso, J.~Martin~Camalich, and S.~Westhoff, {\it {Tau properties in $B\to
  D\tau\nu$ from visible final-state kinematics}},  {\em Phys. Rev.} {\bf D95}
  (2017), no.~9 093006, [\href{http://arxiv.org/abs/1702.02773}{{\tt
  arXiv:1702.02773}}].

\bibitem{Jung:2018lfu}
M.~Jung and D.~M. Straub, {\it {Constraining new physics in $b\to c\ell\nu$
  transitions}},  \href{http://arxiv.org/abs/1801.01112}{{\tt
  arXiv:1801.01112}}.

\bibitem{Bauer:2015knc}
M.~Bauer and M.~Neubert, {\it {Minimal Leptoquark Explanation for the
  R$_{D^{(*)}}$ , R$_K$ , and $(g-2)_g$ Anomalies}},  {\em Phys. Rev. Lett.}
  {\bf 116} (2016), no.~14 141802, [\href{http://arxiv.org/abs/1511.01900}{{\tt
  arXiv:1511.01900}}].

\bibitem{Becirevic:2016yqi}
D.~Becirevic, S.~Fajfer, N.~Kosnik, and O.~Sumensari, {\it {Leptoquark model to
  explain the $B$-physics anomalies, $R_K$ and $R_D$}},  {\em Phys. Rev.} {\bf
  D94} (2016), no.~11 115021, [\href{http://arxiv.org/abs/1608.08501}{{\tt
  arXiv:1608.08501}}].

\bibitem{Crivellin:2017zlb}
A.~Crivellin, D.~MŸller, and T.~Ota, {\it {Simultaneous explanation of
  $R(D^{(*)})$ and $b\to s \mu^+ \mu^-$: the last scalar leptoquarks
  standing}},  {\em JHEP} {\bf 09} (2017) 040,
  [\href{http://arxiv.org/abs/1703.09226}{{\tt arXiv:1703.09226}}].

\bibitem{Buttazzo:2017ixm}
D.~Buttazzo, A.~Greljo, G.~Isidori, and D.~Marzocca, {\it {B-physics anomalies:
  a guide to combined explanations}},  {\em JHEP} {\bf 11} (2017) 044,
  [\href{http://arxiv.org/abs/1706.07808}{{\tt arXiv:1706.07808}}].

\bibitem{Feruglio:2017rjo}
F.~Feruglio, P.~Paradisi, and A.~Pattori, {\it {On the Importance of
  Electroweak Corrections for B Anomalies}},  {\em JHEP} {\bf 09} (2017) 061,
  [\href{http://arxiv.org/abs/1705.00929}{{\tt arXiv:1705.00929}}].

\bibitem{Aaij:2017deq}
{\bf LHCb} Collaboration, R.~Aaij et~al., {\it {Test of Lepton Flavor
  Universality by the measurement of the $B^0 \to D^{*-} \tau^+ \nu_{\tau}$
  branching fraction using three-prong $\tau$ decays}},  {\em Phys. Rev.} {\bf
  D97} (2018), no.~7 072013, [\href{http://arxiv.org/abs/1711.02505}{{\tt
  arXiv:1711.02505}}].

\bibitem{Korner:1989qb}
J.~G. Korner and G.~A. Schuler, {\it {Exclusive Semileptonic Heavy Meson Decays
  Including Lepton Mass Effects}},  {\em Z. Phys.} {\bf C46} (1990) 93.

\bibitem{Gilman:1989uy}
F.~J. Gilman and R.~L. Singleton, {\it {Analysis of Semileptonic Decays of
  Mesons Containing Heavy Quarks}},  {\em Phys. Rev.} {\bf D41} (1990) 142.

\bibitem{Gonzalez-Alonso:2017iyc}
M.~Gonzalez-Alonso, J.~Martin~Camalich, and K.~Mimouni, {\it
  {Renormalization-group evolution of new physics contributions to
  (semi)leptonic meson decays}},  {\em Phys. Lett.} {\bf B772} (2017) 777--785,
  [\href{http://arxiv.org/abs/1706.00410}{{\tt arXiv:1706.00410}}].

\bibitem{Uhlemann:2008pm}
C.~F. Uhlemann and N.~Kauer, {\it {Narrow-width approximation accuracy}},  {\em
  Nucl. Phys.} {\bf B814} (2009) 195--211,
  [\href{http://arxiv.org/abs/0807.4112}{{\tt arXiv:0807.4112}}].

\bibitem{Boyd:1997kz}
C.~G. Boyd, B.~Grinstein, and R.~F. Lebed, {\it {Precision corrections to
  dispersive bounds on form-factors}},  {\em Phys. Rev.} {\bf D56} (1997)
  6895--6911, [\href{http://arxiv.org/abs/hep-ph/9705252}{{\tt
  hep-ph/9705252}}].

\bibitem{Aviles-Casco:2017nge}
A.~Vaquero Avilés-Casco, C.~DeTar, D.~Du, A.~El-Khadra, A.~S. Kronfeld,
  J.~Laiho, and R.~S. Van~de Water, {\it {$\overline{B}\rightarrow
  D^\ast\ell\overline{\nu}$ at Non-Zero Recoil}},  {\em EPJ Web Conf.} {\bf
  175} (2018) 13003, [\href{http://arxiv.org/abs/1710.09817}{{\tt
  arXiv:1710.09817}}].

\bibitem{Bernlochner:2017xyx}
F.~U. Bernlochner, Z.~Ligeti, M.~Papucci, and D.~J. Robinson, {\it {Tensions
  and correlations in $|V_{cb}|$ determinations}},  {\em Phys. Rev.} {\bf D96}
  (2017), no.~9 091503, [\href{http://arxiv.org/abs/1708.07134}{{\tt
  arXiv:1708.07134}}].

\bibitem{Bailey:2014tva}
{\bf Fermilab Lattice, MILC} Collaboration, J.~A. Bailey et~al., {\it {Update
  of $|V_{cb}|$ from the $\bar{B}\to D^*\ell\bar{\nu}$ form factor at zero
  recoil with three-flavor lattice QCD}},  {\em Phys. Rev.} {\bf D89} (2014),
  no.~11 114504, [\href{http://arxiv.org/abs/1403.0635}{{\tt
  arXiv:1403.0635}}].

\bibitem{Sirlin:1981ie}
A.~Sirlin, {\it {Large m(W), m(Z) Behavior of the O(alpha) Corrections to
  Semileptonic Processes Mediated by W}},  {\em Nucl. Phys.} {\bf B196} (1982)
  83--92.

\bibitem{Atwood:1989em}
D.~Atwood and W.~J. Marciano, {\it {Radiative Corrections and Semileptonic $B$
  Decays}},  {\em Phys. Rev.} {\bf D41} (1990) 1736.

\bibitem{Lattice:2015rga}
J.~A. Bailey et~al., {\it {$B \to D$ form factors at nonzero recoil and
  $|V_{cb}|$ from 2+1-flavor lattice QCD}},  {\em Phys. Rev.} {\bf D92} (2015),
  no.~3 034506, [\href{http://arxiv.org/abs/1503.07237}{{\tt
  arXiv:1503.07237}}].

\bibitem{Dungel:2010uk}
{\bf Belle} Collaboration, W.~Dungel et~al., {\it {Measurement of the form
  factors of the decay $B^0 \to D^{*-} \ell^+ \nu$ and determination of the CKM
  matrix element $|V_{cb}|$}},  {\em Phys. Rev.} {\bf D82} (2010) 112007.

\bibitem{Neubert:1993mb}
M.~Neubert, {\it {Heavy quark symmetry}},  {\em Phys. Rept.} {\bf 245} (1994)
  259--396, [\href{http://arxiv.org/abs/hep-ph/9306320}{{\tt hep-ph/9306320}}].

\bibitem{Guido:2017tpe}
{\bf BELLE-II} Collaboration, E.~Guido, {\it {Belle II physics prospects}},
  {\em PoS} {\bf FPCP2017} (2017) 036.

\end{thebibliography}\endgroup
\end{document}